\documentclass[lmodern]{aastex631}
\usepackage{amsmath}
\usepackage{graphicx}   

\usepackage{natbib} 
\PassOptionsToPackage{unicode}{hyperref}
\PassOptionsToPackage{hyphens}{url}
\hypersetup{
	unicode,
	colorlinks,
	breaklinks,
	urlcolor=cyan, 
    linkcolor=black, 
	pdfauthor={Author One, Author Two, Author Three},
	pdfproducer={LaTeX},
	pdfcreator={pdflatex}
}
\definecolor{darkred}{rgb}{0.6, 0, 0}
\definecolor{darkblue}{rgb}{0, 0, 0.6}
\hypersetup{colorlinks=true, linkcolor=black, urlcolor=darkblue, citecolor=darkred}

\usepackage{amsmath,amssymb}
\usepackage{lmodern}
\usepackage{iftex}
\ifPDFTeX
  \usepackage[T1]{fontenc}
  \usepackage[utf8]{inputenc}
  \usepackage{textcomp} 
\else 
  \usepackage{unicode-math}
  \defaultfontfeatures{Scale=MatchLowercase}
  \defaultfontfeatures[\rmfamily]{Ligatures=TeX,Scale=1}
\fi
\IfFileExists{upquote.sty}{\usepackage{upquote}}{}
\IfFileExists{microtype.sty}{
  \usepackage[]{microtype}
  \UseMicrotypeSet[protrusion]{basicmath} 
}{}
\makeatletter
\@ifundefined{KOMAClassName}{
  \IfFileExists{parskip.sty}{%
    \usepackage{parskip}
  }{
    \setlength{\parindent}{0pt}
    \setlength{\parskip}{6pt plus 2pt minus 1pt}}
}{
  \KOMAoptions{parskip=half}}
\makeatother

\makeatletter
\def\@journalinfo{}
\makeatother

\usepackage{xcolor}
\usepackage{longtable,booktabs,array}
\usepackage{multirow}
\usepackage{calc} 
\usepackage{etoolbox}

\IfFileExists{footnotehyper.sty}{\usepackage{footnotehyper}}{\usepackage{footnote}}
\makesavenoteenv{longtable}
\ifLuaTeX
  \usepackage{selnolig}  
\fi
\IfFileExists{bookmark.sty}{\usepackage{bookmark}}{\usepackage{hyperref}}
\IfFileExists{xurl.sty}{\usepackage{xurl}}{} 
\hypersetup{
  hidelinks,
  pdfcreator={LaTeX via pandoc}}
\usepackage{float}
\usepackage{hyperref}

\begin{document}
\title{Potential periodic signals in blazars: significance, forecasting, and deep learning}

\correspondingauthor{M. A. Hashad}
\email{Mohamed.Hashad@bue.edu.eg}
\author{M. A. Hashad}
\affiliation{Department of Basic Science, Modern Academy for Engineering and Technology, Elmokattam 11439, Egypt}
\affiliation{Centre for Theoretical Physics, The British University in Egypt, Sherouk City, Cairo 11837, Egypt}

\author{A. Hammad}
\email{hamed@post.kek.jp}
\affiliation{Theory Center, IPNS, KEK, 1-1 Oho, Tsukuba, Ibaraki 305-0801, Japan}
\affiliation{Center of AI and Natural Science, KIAS, Seoul 02455, Republic of Korea}

\author{ Amr A. EL-Zant}
\affiliation{Centre for Theoretical Physics, The British University in Egypt, Sherouk City, 11837, Cairo, Egypt}
\email{Amr.Elzant@bue.edu.eg}

\begin{abstract}

\noindent Blazars exhibit variable emission on diverse timescales. Some light curves might show signs of quasiperiodic oscillations (QPOs), which may encode clues regarding the physical 
processes behind the emission or point to supermassive binary black holes. We analyzed five blazars with previously reported high significance year-long QPOs, applying four different methods to 
Fermi-LAT data up to early 2025. We furthermore 
examined an additional source (PKS~0139-09), 
where nascent QPO may be present. 
As the light curves showed longer term trends,
we detrended the data 
using an STL decomposition, which often revealed a large 
seasonal component. We find that detrending 
generally leads to an increase in the strength of the QPO signal. However, except for PG 1553+113, 
where a clearly persistent QPO signal is present, we detect
transience on a timescale  of $\lesssim$4000 days.
We then forecast the light curves over the following 
4 yr, using a traditional statistical method as well as 
a transformer-based deep learning (DL) model.  
Applied to a test set, the latter showed significant success in predicting behavior that seems unexpected from 
simple inspection of the past data. 
Analyzing the extended time series suggests a markedly weaker
QPO signals over the coming years in cases where the transient 
behavior appears  near the end of the observational data. 
In contrast, in the nascent candidate QPO source (PKS~0139-09)  
the signal is expected to strengthen significantly.  
These predictions, which may reflect the physical origin of the QPOs, can be tested against future data.

\end{abstract}


\section*{}   
\clearpage    
\vspace{4mm}

\section{Introduction} \label{sec:intro}

Blazars are a prominent type of active galactic nuclei where an accreting supermassive black hole launches a relativistic jet aimed toward the observer \citep{Ulrich_1997, Blandford2019}. The prevailing view holds that emission at X-ray and gamma-ray energies arises through inverse Compton scattering, with  jet electrons boosting photons to high energies by scattering them. The required seed soft photons in the jet plasma may originate from ambient synchrotron photons  (\citealp{Madejski_2016}). Based on optical emission line features, blazars are classified into two subclasses: flat-spectrum radio quasars (FSRQs), showing strong emission lines, and BL Lac objects, showing absent or weak emission lines \citep{Urry1995}. 

A defining feature of blazars is their variable emission across the entire electromagnetic spectrum on diverse timescales, ranging from minutes to years \citep{Urry_2011, Rieger2019, Bhatta2020, Abhir2021, abdollahi2024periodic}. Relatively recently, a growing number of studies have also identified potential quasiperiodic oscillations (QPOs) in the light curves (LCs) of some blazars \citep{2015AckermannPG, Prokhorov2017, Tavani2018, Penil2020, 2021Agarwal, Hashad2023,Penil_2024,abdollahi2024periodic,Hashad2024, 2025penil_SSA}.  
This is of potential significance, as
the existence of QPOs may harbor implications regarding the 
nature of quasar engines, their jets, and 
their emission mechanisms \citep{Mohan2015, Benkhali2020}. Importantly, they may furthermore  
be relevant as probes of supermassive black hole binary (SMBHB) systems  \citep{Begelman1980, 2023ApJL.Agazie, 2023.Orazio, 2025ApJ.Parra, 2025Kiehlmann}, 
where orbital motions  may  
cause jet precessions (e.g., \citealp{Sobacchi-2017}), perturbation by magnetogravitational stress (e.g., \citealt{2017Cavaliere}), or perturbations in the accretion flow (e.g., \citealt{2014ApJFarris, 2015MNRASDo}). This is a possibility that comes with important implications for contemporary cosmology, where the repeated hierarchical merging of dark matter halos---with the galaxies they embed and the supermassive black holes in their centers---is a definite prediction~\citep{Begelman1980, 2003Volonteri, 2024Maiolino}. Indeed, the enumeration of potential SMBHB systems, from observations as well as through cosmological simulations, is currently an active 
area of research (\citealp{Krause_Obs++Rev2025, Supersize_2025ApJS, Habouz_Vo_2025}), and the field is further driven by 
upcoming gravitational wave 
experiments like LISA (\citealp{LISA_2024arXiv240207571C}).

Observed QPO signals in blazar LCs may be of the order of a month or shorter and up to the scale of years. On the shorter end, 
clear transience in the signals tends to be present. 
Such trends may be plausibly associated with ballistic motion in the blazar jet, such as in the curved helical blob model (\citealp{Sarkar_2021, Sharma_2024ApJ, Penil_2025_curved_jet, Tantery_2025}). In this case, the helical motion of a  plasma blob generates periodic variations in the flux, and the curvature of the jet results in a variable 
viewing angle (and hence the apparent decay in  QPO amplitudes). 

For candidate QPOs potentially originating from SMBHB systems, 
the simplest assumption is to 
associate the QPO periods with the orbital timescale of a bound binary system undergoing gravitational wave (GW) ringdown. 
A simple geometric interpretation of QPO in terms of jet 
precession in an SMBHB system predicts a QPO timescale of the 
order of years lest the ringdown becomes too rapid 
for the signal to have a realistic chance of being observed (\citealp{Sobacchi-2017}). 
In some 
models, where orbital motion causes helical (but nonbalistic) motion in the jet, the observed QPO timescale is expected to be shortened by at least an order of magnitude, due to light travel time 
effects (in proportion to the square of the bulk Lorentz factor; \citealp{Rieger2004}). But in such cases, the allowed intrinsic orbital period may also be much larger (in a purely geometric 
model, the maximum timescale is limited by the observable precession angle). Thus, 
any potentially observable QPO signal due to an SMBHB
would be expected to appear on the longest timescale. 
The longest available timescale for 
$\gamma$-ray data is of the order of years.

Confirming or ruling out QPOs with year-long periods 
presents particular difficulties, given that 
the total observing time in the $\gamma$-ray regime 
is not far longer than the purported periodic timescale (\citealp{Penil_2025_Transients,Penil2025P22}; 
for recent studies of longer-term data at other wavelengths see, e.g.,
\citealp{Opt_Rad2025ApJ,Radio2025arXiv251023103M}). 
This is especially the case if the 
signals are accompanied by longer-term trends, including-
longer term periodicities  or transience in the signal. 
Detecting transience, in particular, may also be important
in distinguishing between possible physical origins of 
apparent QPO signals. For example, in purely geometric SMBHB 
models, there is no obvious mechanism to damp the QPO 
signals over observable timescales. 

To test for the persistence of the QPO, one 
may also make predictions by extending LCs based on past behavior, 
which can be compared with future observations. 
An increase in the significance of the QPO signal 
reflects a strong quasi-stationary signal. 
A decrease, on the other hand,  
may point to the presence of transience.  
The results of the predictions encode patterns 
that can  be used to detect and predict  
non-stationarity in the process underlying the signal.
Such predictions can be made using traditional statistical forecasting methods (\citealp{2025penil_SSA}), or through the rapidly developing machine learning techniques. In the present study, we use examples of both.

Recent advances in deep learning (DL) have provided powerful alternatives to traditional statistical forecasting techniques for analyzing complex time series. Classical statistical models, such as autoregressive or decomposition-based methods, typically assume linearity, stationarity or fixed periodicities, which limit their ability to capture the nonlinear, multi-scale variability and noise contamination that characterize astrophysical data. In contrast, DL approaches learn hierarchical feature representations directly from the data, enabling them to model nonlinear correlations and disentangle long- and short-term trends. This makes DL particularly advantageous for forecasting the highly variable emission of gamma-ray blazars, where primary emission processes span diverse physical timescales and are further modulated by jet and environmental effects.
Among DL techniques, transformers are particularly well suited for the task at hand, given their ability to capture both local variability and global periodic trends, and are more adaptable to irregularly sampled or incomplete time series, a common challenge in astrophysical observations.
Unlike other related methods, it does not suffer from shortcomings such as vanishing or exploding gradients, difficulties in learning very long sequences, and sequential processing that hinders parallelization.
transformers overcome these shortcomings through their attention mechanism, which directly models relationships between all time steps without relying on recurrence. 
They can thus make relatively reliable predictions regarding 
the existence and persistence (or otherwise) of QPO signals, 
which may furthermore be tested against future observations. 

Toward these goals,  we reanalyze five blazars previously reported to show strong QPO signals. We use  Fermi-LAT LCs extended up to February 2025. We also examine the blazar PKS 0139-09, which shows a possible long-period QPO along with a strong upward trend in its LC. As such trends constitute a significant complication in quantifying the significance of possible QPO signals of a shorter period, we remove them by detrending the data. We then forecast the LCs 4 yr into the future, testing 
the predictions on a test set and reanalyzing the extended light curves obtained via the transformer method.   

The paper is organized as follows. Section~\ref{LAT Light Curves} describes the Fermi-LAT Gamma-ray data. Gamma-ray QPOs in the observed LCs are analyzed in Section~\ref{gamma-ray periodicity}. Decomposition and detrending of the LCs is presented in Section \ref{lc_detrending}. In Section~\ref{lcForecasting}, we introduce the statistical method (STLForecaster) and a deep learning  (transformer) approach for forecasting the LCs and present the results. We discuss and summarize our findings in Section \ref{Discussion and conclusion}.

\section{Fermi-LAT light curves} 
\label{LAT Light Curves}

Aboard the Fermi satellite, the Fermi Large Area Telescope (Fermi-LAT) is a pair-conversion gamma-ray detector sensitive from $\sim$20 MeV to 500 GeV. Since the start of its operating mode in 2008 August, Fermi-LAT has continuously scanned the full-sky every two orbits (at $\approx$1.6 hr per orbital period) with its wide ($\approx$2.4 sr) field of view. This provides long-coverage observations across diverse timescales, extending from hours to years.
The LAT has a point-spread function <$0.8^\circ$ and the largest effective area of $\sim$8000 $\mathrm{cm^2}$ above 1 GeV \citep{Atwood2009}.

\begin{figure}[!h]
    \centering
    \includegraphics[width=0.7\linewidth]{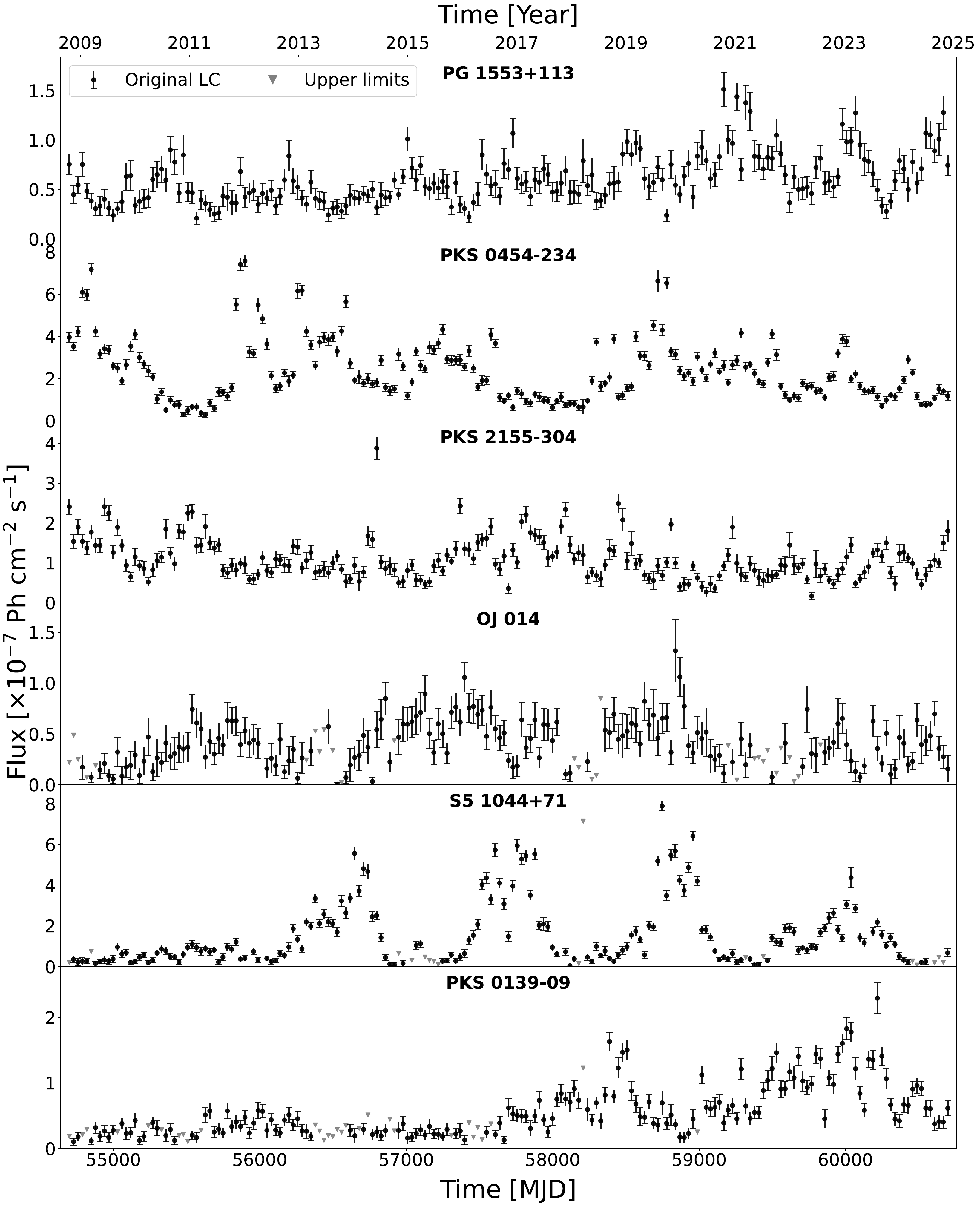}
    \caption{Extended Fermi-LAT $\gamma$-ray integrated flux light curves of the selected blazars. The five previously reported with high-significance periodicity, and the new QPO candidate, PKS 0139-09. The light curves span $\approx$16.5 yr, from 2008 August to 2025 February. The down gray triangles are for bins with upper limits, where their test statistic values <9 ($\approx$3$\sigma$).}
    \label{fig:Original_LCs}
\end{figure}

This work uses LAT observations of six blazars, covering $\approx$16.5 years in the energy range 100 MeV--300 GeV. Five of the  sources were selected because they have been previously reported in the literature to exhibit $\gamma$-ray QPOs with possible high significances (e.g., \citealt{PKS304-2017, Bhatta2020, Wang2022, abdollahi2024periodic, Penil_2024, Penil2025P22}). 
These are PG 1553+113,  PKS 0454-234, PKS 2155-304, OJ 014, and S5 1044+71;
see Table \ref{table1}. 
With the aim of testing for long-term nonstationarity and 
transience, we extend their study using  monthly binned  LCs up to 2025 February.  The LCs were generated using the maximum likelihood technique utilizing fermipy\footnote{\url{http://fermipy.readthedocs.io}}, a Python package for standard \texttt{ScienceTools} implementation (Version 1.2.2; \citealt{Wood_2017}). To further test for nonstationary behavior, 
we do the same for a novel 
candidate QPO source (PKS 0139-09), where we suspect nascent oscillations
to be starting within the observed period. 

To select the photons composing the LCs, the Pass 8 ``SOURCE'' class (EVENT$\_$CLASS = 128) was used from the instrument response function \texttt{P8R3\_SOURCE\_V3}, as well as event type
“FRONT+BACK” (EVENT$\_$TYPE = 3) \citep{2013Atwood}. Regions of interest (ROI) of $15^{\circ}\times 15^{\circ}$ square, centered at the source, were chosen to account for photons within it. A zenith angle cut ($>$90°) was applied to eliminate $\gamma$-ray contamination from the Earth's limb. Standard data quality selections (DATA QUAL$>0$)\&\&(LAT CONFIG$==1$) were also implemented, and time intervals overlapping with LAT-detected $\gamma$-ray bursts and solar flares were excluded. Sources from the 4FGL catalog
\citep{Abdollahi_2022_4FGL-DR3} were modeled using the standard LAT analysis. They were positioned within $20^{\circ}$ from center of the ROI and incorporating the Galactic (\texttt{gll\_iem\_v07.fits}) and isotropic diffuse emissions (\texttt{iso\_P8R3\_SOURCE\_V3\_v1.txt}).

The maximum likelihood analysis was conducted over the entire period of observation, from 2008 August to 2025 February.  
We used eight logarithmic energy bins per decade and 0.1° spatial bins. Sources in the ROI were modeled with their spectral shapes and parameters values set from the catalog. 
The normalization and spectral index of the target source, as well as the normalization of the sources within a $3^{\circ}$ radius from the target position and both diffuse backgrounds, were left free to vary. A good fit quality was obtained (\texttt{fit$\_$quality = 3}) by iteratively running the routines \texttt{gta.optimize()} and \texttt{gta.fit()}. 

In order to generate the LC of each source, we  binned the data into one month bins.  Each bin was then subjected to a full likelihood fit, using the parameters’ values attained from the main stream analysis on the entire LC. During the fit the following parameters converge to their best-fit values: the target’s spectral parameters, the normalizations of sources within $3^{\circ}$ from the center of the ROI, and the normalizations of the diffuse components. The temporal bins with test statistic values (TS) <9 ($\approx$3$\sigma$) were set as upper limits. TS determines the source detection significance (TS $=-2$log$(L_0/L_1)$), where $L_0$ and $L_1$ are the likelihood functions evaluated at the best-fit model parameters without and with the target
source, respectively.
The resulting $\gamma$-ray LCs of the six sources are shown in Fig.~\ref{fig:Original_LCs}.

\begingroup
\small
\setlength{\tabcolsep}{4pt}
\begin{longtable}{@{}
  >{\raggedright\arraybackslash}p{(\columnwidth - 16\tabcolsep) * \real{0.15}}
  >{\centering\arraybackslash}p{(\columnwidth - 16\tabcolsep) * \real{0.15}}
  >{\centering\arraybackslash}p{(\columnwidth - 16\tabcolsep) * \real{0.10}}
  >{\centering\arraybackslash}p{(\columnwidth - 16\tabcolsep) * \real{0.10}}
  >{\centering\arraybackslash}p{(\columnwidth - 16\tabcolsep) * \real{0.06}}
  >{\centering\arraybackslash}p{(\columnwidth - 16\tabcolsep) * \real{0.06}}
  >{\centering\arraybackslash}p{(\columnwidth - 16\tabcolsep) * \real{0.13}}
  >{\centering\arraybackslash}p{(\columnwidth - 16\tabcolsep) * \real{0.20}}
  >{\centering\arraybackslash}p{(\columnwidth - 16\tabcolsep) * \real{0.05}}
}
\caption{Our blazar sample, given by their association name in the point-source Fermi-LAT fourth catalog (4FGL). We list the equatorial coordinates (deg), type, redshift, and $\gamma$-ray QPOs' periods in literature and their significances. The types of AGN: BL Lacertae (bll) and flat-spectrum radio quasar (fsrq). Examples of periods previously reported in the literature. All refer to analysis of $\gamma$-ray data from Fermi-LAT, with the significances obtained through the LSP method.}
\label{table1} \\
\toprule()
Source Name & 4FGL Name & R.A. (J2000) & Decl. (J2000) & Type & $z$ & Reported Period (yr) & Observation Period & Ref. \\
\midrule()
\endfirsthead
\multicolumn{9}{c}{{\tablename\ \thetable{} -- continued from previous page}} \\
\toprule()
Source Name & 4FGL Name & RA (J2000) & Dec (J2000) & Type & $z$ & Historical Period (yr) & Observation Period & Ref. \\
\midrule()
\endhead
\midrule()
\multicolumn{9}{r}{{Continued on next page}} \\
\endfoot
\bottomrule()
\multicolumn{9}{@{}p{\textwidth}@{}}{\textbf{References.} (1) \cite{abdollahi2024periodic}. (2) \cite{Penil2025P22}. (3) \cite{PKS304-2017}. (4) \cite{Wang2022}.} \\
\endlastfoot

PG 1553+113 & J1555.7+1111 & 238.931 & 11.188 & bll & 0.433 & \({2.1}_{4.0\sigma}^{\pm 0.2}\) \({2.1}_{5.1\sigma}^{\pm 0.2}\) &
2008-08--2023-11 2008-08--2020-12 & 1, 2 \\

PKS 0454$-$234 & J0457.0-2324 & 74.261 & -23.415 & fsrq & 1.003 & \({3.6}_{3.2\sigma}^{\pm0.2}\) &
2008-08--2020-12 & 2 \\

PKS 2155$-$304 & J2158.8-3013 & 329.714 & -30.225 & bll & 0.116 & \({1.7}_{4.9\sigma}^{\pm 0.1}\) \({1.7}_{4.1\sigma}^{\pm 0.1}\) &
2008-08--2016-10 2008-08--2020-12 & 3, 2 \\

OJ 014 & J0811.4+0146 & 122.861 & 1.776 & bll & 1.148 & \({4.2}_{3.2\sigma}^{\pm 0.5}\) &
2008-08--2020-12 & 2 \\

S5 1044+71 & J1048.4+7143 & 162.107 & 71.730 & fsrq & 1.150 & \({3.1}_{3.6\sigma}^{\pm 0.4\ }\) &
2012-03--2021-03 & 4 \\

PKS 0139$-$09 & J0141.4-0928 & 25.363 & -9.483 & bll & 0.733 & -\/- & -\/- & -\/- \\
\end{longtable}
\endgroup
 
\section{Search for Gamma-Ray periodicity in the original light curves}
\label{gamma-ray periodicity}

We investigated the $\approx$16.5 yr   
LCs, as obtained above, using the Lomb-Scargle periodogram (LSP) and  weighted wavelet z-transform (WWZ), as we describe below.  

\subsection{Lomb–Scargle periodogram}
\label{sec:LSP}

\subsubsection{Method}
\label{sec:LSP-method}

The LSP is a widely used algorithm for identifying periodicity in unevenly sampled astronomical time series \citep{Lomb1976, Scargle1982}. The standard normalized LSP is equivalent to fitting sinusoidal functions of the form, $y(t)=A{\mathrm{cos}(\omega t)}+B\mathrm{sin}(\omega t)$ to an LC. 
It is defined, for time series $(t_i,y_i)$, as
\[
\begin{aligned}
P\left(\omega \right)=\frac{1}{2}\left\{\frac{{\left(\sum_i{y_i\ \mathrm{cos}\mathrm{\omega }(t_i-\tau )}\right)}^2}{\sum_i{{\mathrm{cos}}^2\mathrm{\omega }(t_i-\tau )}}+\frac{{\left(\sum_i{y_i\ \mathrm{sin}\mathrm{\omega }(t_i-\tau )}\right)}^2}{\sum_i{{\mathrm{sin}}^2\mathrm{\omega }(t_i-\tau )}}\right\},
\end{aligned}
\]
where 
\begin{equation}
\mathrm{\tau }\mathrm{=}\frac{1}{2\omega }\mathrm{{tan}^{-1}}\left(\frac{\sum_i{\mathrm{sin}(2\omega t_i)}}{\sum_i{\mathrm{cos}(2\omega t_i)}}\right).
\end{equation}

We compute the LSP using the periodogram algorithm provided by \cite{Zechmeister2009}\footnote{\url{https://pyastronomy.readthedocs.io/en/latest/pyTimingDoc/pyPeriodDoc/gls.html}}, which includes a constant term (offset) in the fit of the trigonometric function. This approach is advantageous in cases where the data do not provide full phase coverage of the observed signal; as it 
involves a floating-mean, rendering it more robust than the standard LSP \citep{VanderPlas2018}, which typically assumes that the mean of the fitted sinusoid is equal to the mean of the data. The algorithm also accounts for measurement errors; however, no errors were specified in our implementation to ensure a consistent comparison of the periodograms obtained from the original, detrended, and predicted LCs, since the detrended and predicted LCs do not have associated uncertainties. This is further discussed in Sections~\ref{lc_detrending} and ~\ref{lcForecasting}.
To test that the flux uncertainties of the original LCs are approximately homoscedastic, we have calculated the coefficient of variation, $\mathrm{CV} = \sigma_{\mathrm{err}} / \bar{\sigma}_{\mathrm{err}}$, where $\sigma_{\mathrm{err}}$ is the standard deviation of the flux uncertainties and $\bar{\sigma}_{\mathrm{err}}$ is their mean. The resulting CV values were $\lesssim25\%$.
The frequency grid used to evaluate the LSP was chosen following \cite{Scargle1982}, where the minimum frequency, $\nu_{\min}$, and maximum frequency, $\nu_{\max}$, are set to $1/T$ and the Nyquist frequency, respectively. Here, $T$ denotes the total duration of the corresponding light curve. The total number of sampled frequencies can be empirically set as
$N_{\nu} = n_0 T \nu_{\max}$ (see \citealt{VanderPlas2018}). We conservatively adopted $n_0 = 10$ to ensure adequate sampling of periods up to the full duration of the light curve.

Using this procedure, 
we reexamined 
the previously reported QPO periods 
in the selected five blazars after extending their LCs. The period uncertainties were estimated from the half-width at half-maximum of a Gaussian function fitted to the profile at the position of the peak.

\subsubsection{Estimating the significance of periodicities}
\label{sec:significance-estimation}

As blazars, LCs are noisy and can have irregular sampling or contain gaps \citep{Goyal_2017, Goyal_2018, Benkhali2020, Tarnopolski_2020}; spurious peaks could  arise and caution is warranted~\citep{Vaughan2016, VanderPlas2018}. To account for this, we employed simulated LCs, using the \cite{Emmanoulopoulos2013} algorithm to estimate the significance of the peaks. These LCs are statistically equivalent to the original LC in terms of the power spectral density (PSD), probability density function, and data sampling.
We model the PSD of the blazar LCs of our sample  as a simple powerlaw (PL) in the form of $P(\nu) ~\alpha ~\nu^{-\beta}$, where $v$ is the temporal frequency and $\beta$ is the power spectral slope or PL index \citep{Goyal_2017, Bhatta2020}. In general, the observed characteristic slope range in blazars' LCs is approximately 1-3, across timescales from minutes up to decades (e.g., flicker/pink noise for $\beta \sim$1 and red noise for $\beta \sim$2), which differs from the flatter ($\beta \sim$0) PSD signifying uncorrelated white noise \citep{1978-Press, Rieger2019}.
We used $10^6$ simulated LCs utilizing the implementation from \cite{Connolly2015}. It should be noted that, given the number of the simulated LCs, the maximum significance achievable is $\approx$4.8$\sigma$.

\textit{The local and global significance}. In the presence of an a priori expectation of periodicity (e.g., from entirely independent observations or from theory) the statistical significance of the detected period at that position can be directly computed \citep{Benkhali2020, Neill-2022}. This is the so-called local significance (e.g., \citealt{Bell2011}), which corresponds to the probability of recording a periodic signal at the detected period with the same or greater power under the null hypothesis that the signal in the observed LC is due to random red noise.
However, with the lack of a priori, it is statistically more rigorous to compute the global significance \citep{Benkhali2020, 2025penil_SSA}. The global significance accounts for the ``look-elsewhere effect'' (or multiple testing problem), which quantifies the probability of finding aperiodic signal under the above null hypothesis as significant as the one observed at any frequency.

In this work, only the local significance, as defined above, was estimated for the five blazars in our sample where these periods had been reported with possible high significances.
As, for these cases, the period is already specified \textit{a priori}, no correction for the number of independent frequencies is needed.
In contrast, both the local and global significances were estimated for the newly added object 
(PKS 0139-09), 
which was not previously reported as a source of QPO.
The confidence levels shown in the periodogram plots (Fig.~\ref{fig:WWZ-LSP}) are the local significance thresholds computed independently at each frequency. 
We used the 
same procedure outlined in Appendix A of \cite{Neill-2022}. This may be summarized as follows:

\begin{enumerate}
    \item Compute the LSP of the real data. Identify the strongest peak and record its period, \(P_{\text{peak}}\), and its power, \(\mathcal{P}_{\text{peak}}\).
    \item At the specific period \(P_{\text{peak}}\), count all simulations with power \(\mathcal{P}_{\text{sim}} \geq \mathcal{P}_{\text{peak}}\) to estimate the local \(p\)-value, $p_{\mathrm{peak}}$, and the local significance.
    \item For each simulation,
    \begin{enumerate}
        \item compute the LSP at the same frequency grid as the real data;
        \item identify its strongest peak and record its period, \(P_{\text{sim}}\), and its power, \(\mathcal{P}_{\text{sim}}\); and
        \item at period \(P_{\text{sim}}\), calculate the local \(p\)-value for that simulated peak ($p_{\mathrm{sim}}$) using the full ensemble of simulations.
    \end{enumerate}
    \item Count all simulations for which $p_{\mathrm{sim}} \leq p_{\mathrm{peak}}$ to estimate the global \(p\)-value and the global significance.
\end{enumerate}

The ``look-elsewhere effect'' significance correction, which incorporates a specific number of trials $N$ (i.e., the number of blazars in the sample; see, e.g., \citealt{2025penil_SSA}), was not applied. In this work, we search for periodicity signals within a sample of blazars that had already exhibited possible QPOs. Since these sources were previously selected from larger samples by other authors, determining the value of $N$ is not straightforward.

\subsubsection{Application to the light curves}
The best-fit parameters of the power-law model for the six sources, along with their reduced $\chi^2$
values and the results of the LSP applied to the observed light curves of these blazars, are summarized in Table \ref{lsp_origin_lc}. The significance estimates are based on the best-fit PSD model and do not incorporate the uncertainties of the PSD parameters. However, this does not affect the conclusions, as the identified QPOs in this work are, in general, not highly significant. Below we discuss their significance in light of previous work. 

The BL Lac object PG 1553+113 ($z=0.433$; \citealt{2022Dorigo-PGz}) is a remarkable case that showed almost sinusoidal modulation in its LC in the first 7 yr of Fermi-LAT data~(\citealp{Ackermann2015}); with a  period of $2.18\pm0.08$ yr, covering about three oscillation cycles and chance probability of being spurious of $<1\%$. However, QPO candidates should be interpreted with caution where the statistical significance of long-term periodicities may be affected by the limited number of observed cycles, the presence of long term trends, and the underlying red noise variability \citep{Vaughan2016, Goyal_2018, VanderPlas2018, Benkhali2020}.
These initial findings have been supported by several follow-up studies, in $\gamma$-ray as well as optical wavelengths through extended datasets and more analyses (e.g., \citealt{Sobacchi-2017, Tavani2018, Benkhali2020, Ren_2023, abdollahi2024periodic, Chen-2024, Penil_2024}). Recently, ~\cite{abdollahi2024periodic} have used 15 yr of Fermi-LAT data---doubling the observation time of the first claim---to confirm the 2.1 yr-PG 1553+113 $\gamma$-ray QPO. By extending the Fermi-LAT data to 2025 February, and using a monthly binned LC, we find QPO in PG 1553+113 of 2.1 yr period by using LSP at $3.6\sigma$. It should be noted that the period significances we estimated in this study for the five previously reported sources in the literature with possible QPO are local significances (see Section~\ref{sec:significance-estimation}), whereas both local and global significances are estimated for the new source, PKS 0139-09.
This is consistent with their results for the original LC (by LSP; $3.5\sigma$).
After detrending their LC, their reported significance increased to 
$4\sigma$. 
As we will see below (Section \ref{lc_detrending}), we also 
obtain a larger significance ($4.5\sigma$) after detrending.  

In our analysis, the FSRQ PKS 0454-234 showed a QPO with a period of $3.5\pm0.35$ yr at $2.5\sigma$. This period has been previously reported at the higher significance of 3.2$\sigma$ by \cite{Penil2025P22} and by \cite{2025penil_SSA} at the still higher 4.8$\sigma$, both using about 12 yr of the LAT data and using LSP. However, \cite{Penil2025P22} reported also similar periods but at low significances of 2.8$\sigma$, 2.4$\sigma$, and 2.8$\sigma$ using the methods of generalized Lomb–Scargle periodogram (GLSP; \citealt{Zechmeister2009}), phase dispersion minimization (PDM; \citealt{PMW1978}), and continuous wavelet transform (CWT; \citealt{CWT1998}), respectively. 
We also identify 
possible signals with smaller  periods. 
Remarkably, like the previously reported principal peak, which 
turns out to have  a nearby larger period peak, they tend to come in pairs.  
Some are consistent with being harmonic at half, a third, and 
a tenth of the previously reported period, as will be seen below in connection 
to Fig.~\ref{fig:WWZ-LSP}. As we will also see, these secondary peaks are actually detected through the LSP with higher significance than the primary (although they 
are of smaller height, they correspond to more cycles). High-frequency secondary peaks are present in other cases, though only for the present source 
and the nascent QPO candidate (PKS 0139-09)  
do they come with higher significance 
than the principal period. 

\begin{deluxetable}{cccccccccccc}
\tablecaption{Summary of Lomb-Scargle periodogram periods (in years) with their local significance for the original $\gamma$-ray light curves of the six blazars, along with simple power-law (PL) fit parameters and its reduced $\chi^2$. \label{lsp_origin_lc}}
\tablewidth{0pt}
\tablehead{
\multicolumn{2}{c}{PG 1553+113} & 
\multicolumn{2}{c}{PKS 0454$-$234} &
\multicolumn{2}{c}{PKS 2155$-$304} &
\multicolumn{2}{c}{OJ 014} &
\multicolumn{2}{c}{S5 1044+71} &
\multicolumn{2}{c}{PKS 0139$-$09} \\
\tableline
\colhead{PL Par.} & \colhead{Period} &
\colhead{PL Par.} & \colhead{Period} &
\colhead{PL Par.} & \colhead{Period} &
\colhead{PL Par.} & \colhead{Period} &
\colhead{PL Par.} & \colhead{Period} &
\colhead{PL Par.} & \colhead{Period}
}
\startdata
\shortstack{$A=0.003$ \\ $\beta=1.26$ \\ $\chi^2_{\mathrm{red}}=2.8$} & \shortstack{${2.1}_{3.6\sigma}^{\pm 0.13}$ \\ $ $ \\ $ $}&
\shortstack{$A=0.011$ \\ $\beta=1.22$ \\ $\chi^2_{\mathrm{red}}=1.7$} & \shortstack{${3.5}_{2.5\sigma}^{\pm 0.35}$ \\ $ $ \\ $ $}&
\shortstack{$A=0.66$ \\ $\beta=0.83$ \\ $\chi^2_{\mathrm{red}}=1.8$} & \shortstack{${1.7}_{2.6\sigma}^{\pm 0.13}$ \\ $ $ \\ $ $}&
\shortstack{$A=0.027$ \\ $\beta=0.97$ \\ $\chi^2_{\mathrm{red}}=1.3$} & \shortstack{${4.2}_{3.5\sigma}^{\pm 0.57}$ \\ $ $ \\ $ $}&
\shortstack{$A=0.003$ \\ $\beta=1.6$ \\ $\chi^2_{\mathrm{red}}=1.95$} & \shortstack{${3.1}_{4.4\sigma}^{\pm 0.34}$ \\ $ $ \\ $ $}&
\shortstack{$A=0.035$ \\ $\beta=1.02$ \\ $\chi^2_{\mathrm{red}}=1.99$} & \shortstack{${5.2}_{1.6\sigma}^{\pm 0.73}$ \\ $ $ \\ $ $}
\enddata
\end{deluxetable}

The BL Lac object PKS 2155-304 has been reported by \cite{PKS304-2017}, using data from 2008 August to 2016 October, to display a significant (by LSP; $4.9\sigma$) $\gamma$-ray QPO with period $1.7\pm0.13$ yr. This finding is consistent with the results of \cite{Sandrinelli-2014}.
However, in an analysis of the $\sim$10-year $\gamma$-ray LC (from 2008) and the $\sim$14-year optical R-band LC (from 2003), \cite{Sandrinelli2018} found that the significance of these QPOs appears to decrease in recent years, which may be linked to a declining trend in the source's intensity.
\cite{Covino2020} found an $\sim$610 d QPO,
albeit not significant enough. This period has also been noticed in the CWT maps by \cite{Ren_2023}, but after the trial correction it was not significant.
Similar period using LSP has been reported in the optical and $\gamma$-ray using almost 10 years of data by \cite{2019ChevalierOptical}, as well as in the X-ray, UV, and optical bands by \cite{Penil_2024}.
Using almost the first ten years of Fermi-LAT data, \cite{Benkhali2020} reported the same period (1.7 yr) from GLSP at 3.7$\sigma$. The significance of the period decreased to $4.1\sigma$ (by LSP) and $3.3\sigma$ (by GLS and CWT) during the $\gamma$-ray observation period from 2008 August to 2020 December \citep{Penil2025P22}.
Using the decade-long Fermi-LAT data, \cite{Bhatta2020} found two quasiperiodic features for this object in both LSP and WWZ methods, one at $1.7$ yr and the other at $0.7$ yr both at significance of $\gtrsim$99.99$\%$. Moreover, their WWZ analysis showed that over the year, the low-frequency period is gradually shifting toward a slightly higher frequency.
PKS 2155-304 has been identified as a low-significance $\gamma$-ray QPO candidate (2008 August--2021 April) with a period of $0.91\pm0.29$ yr at $2.2\sigma$ using CWT, where it exhibited a time-dependent behavior, with decreasing frequency in time \citep{Ren_2023}.
By extending the observation period to the beginning of 2025 in this study, the period significance from LSP decreased to $2.6\sigma$. 
As we will see, this decrease is partly due to long-term trends and transience;
due to low-frequency modulation and to the principal QPO 
signal itself diminishing with time.

The BL Lac object OJ 014 was reported to exhibit a $\gamma$-ray QPO with period of about 4.2 yr during the period from 2008 August to 2020 December at significance of $3.2\sigma$ (by LSP), $2.6\sigma$ (by GLSP and PDM), and $3.6\sigma$ (by CWT) \citep{Penil2025P22}. The autocorrelation function for twelve years of Fermi-LAT data showed a period of $4.0\pm 0.3$ yr at $2\sigma$ \citep{Penil_2024}. In this study, the LSP for the LC of this object revealed a period of $4.2\pm 0.57$ yr with a small increase in the period's significance to $3.5\sigma$.
 
The FSRQ object S5 1044+71 was reported with QPO of period $3.1\pm0.4$ yr in its $\gamma$-ray LC using LSP during the period from 2012 March to 2021 March at $3.6\sigma$ (covering only three cycles) \citep{Wang2022}, and also during the period from 2008 August to 2020 December at $4.8\sigma$ \citep{2025penil_SSA}.
The same period from the CWT was identified by \cite{Ren_2023} at significance of $4.9\sigma$ using $\gamma$-ray data from 2008 August to 2021 April. 
We confirm the consistency of the QPO in the $\gamma$-ray LC with a potential period of $3.1\pm0.34$ yr (by LSP) at $4.4\sigma$.  

The BL Lac object PKS 0139-09 was not previously reported as a source of QPO. It has been included in this work as it seems an ideal test case for the purpose 
of studying and forecasting nonstationary behavior in QPO. As of now, the LSP for this object revealed two peaks in the PKS 0139-09 $\gamma$-ray LC with two tentative candidate periods at $0.67\pm0.04$ yr and $5.2\pm0.73$ yr, with local significances (global significances) of $2.23\sigma$ ($0.79\sigma$) and $1.63\sigma$ ($0.07\sigma$), respectively. Although the larger period is 
of weaker significance, if this is due to the lack of hitherto observed 
cycles it should be expected to significantly strengthen in the future. 
We thus consider this to be the principal period despite its relatively
small significance (again it seems possible to associate the smaller period, 
detected with higher significance, with  
a high order harmonic; of order $\sim$8).

In summary, examining possible QPO signals in LCs studied 
over different timespans  considered here 
generally leads to similar periods as previously reported 
but with different significances. In addition, we clearly detect 
secondary peaks, which in some cases come with higher significance
than the primary peaks and are not inconsistent with being harmonics. 
In general, the significances of the previously reported 
periodicities (associated with the primary peak) 
do not necessarily increase with the increased timespan considered here. They generally stay approximately the same or  significantly decrease with the increased 
time interval. 
This can be either due to the principal 
QPO signal itself being transient, or because the signal includes superposed low-frequency modes that come with large red noise. We explore these possibilities below. 

\subsection{Weighted wavelet z-transform}
\label{sec:wwz_method}

\begin{figure}[h!]
    \centering
    \begin{minipage}{0.48\textwidth}
        \centering
        \includegraphics[width=\linewidth]{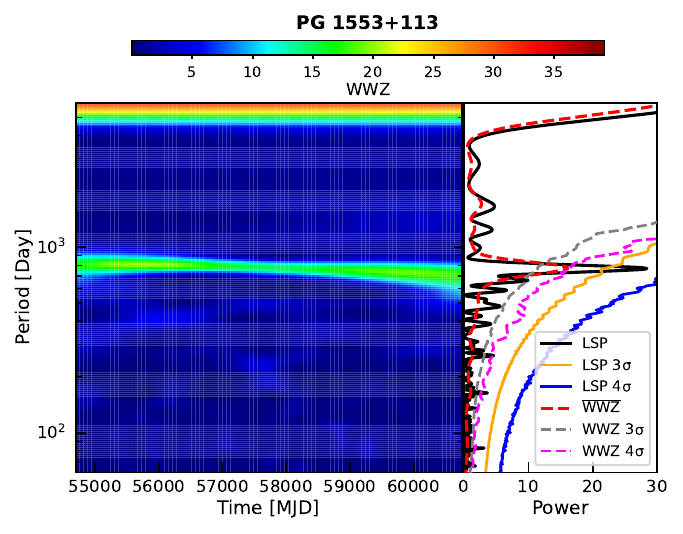}
    \end{minipage}
    \hspace{0.1cm}
    \begin{minipage}{0.48\textwidth}
        \centering
        \includegraphics[width=\linewidth]{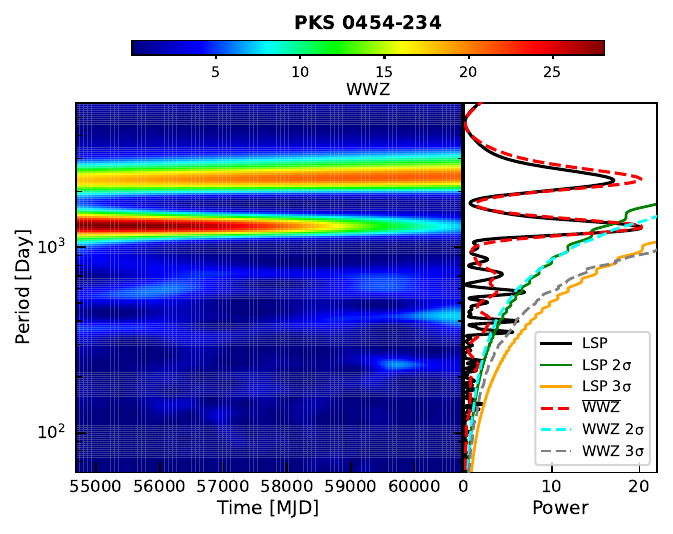}
    \end{minipage}

    \vspace{0.05cm}

    \begin{minipage}{0.48\textwidth}
        \centering
        \includegraphics[width=\linewidth]{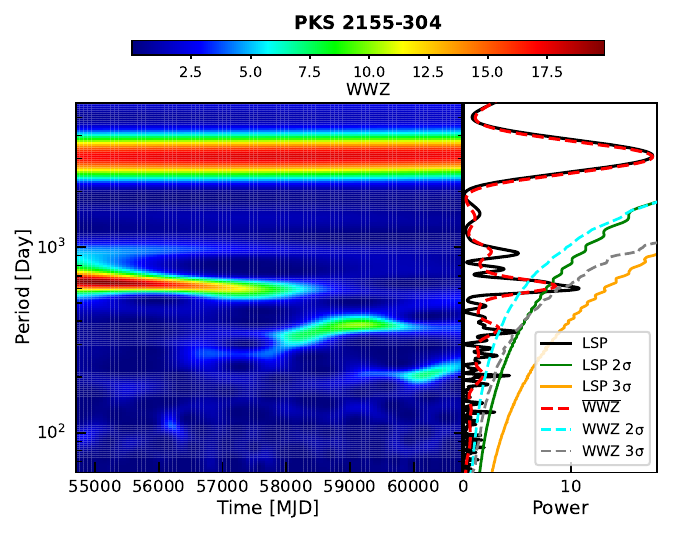}
    \end{minipage}
    \hspace{0.1cm}
    \begin{minipage}{0.48\textwidth}
        \centering
        \includegraphics[width=\linewidth]{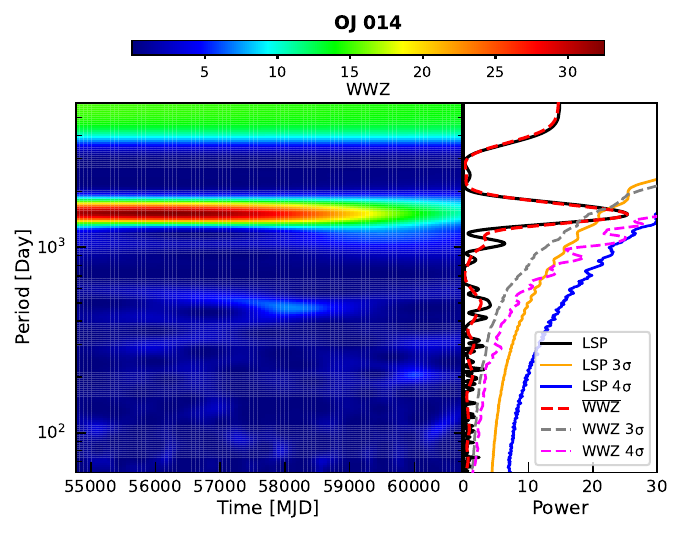}
    \end{minipage}

    \vspace{0.05cm}

    \begin{minipage}{0.48\textwidth}
        \centering
        \includegraphics[width=\linewidth]{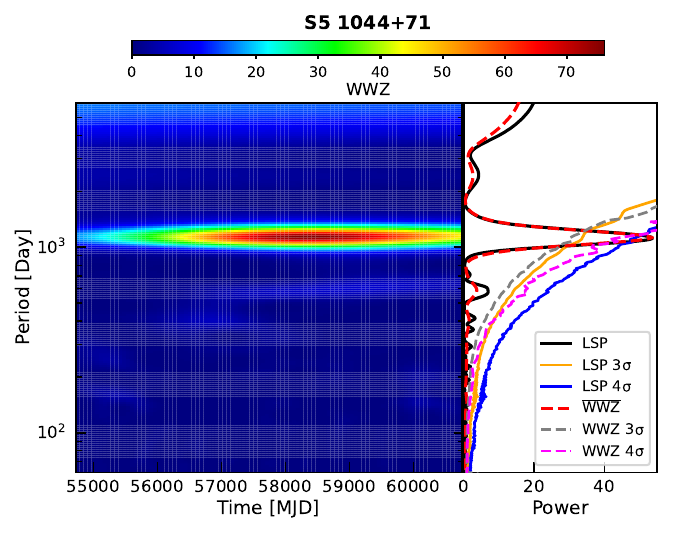}
    \end{minipage}
    \hspace{0.1cm}
    \begin{minipage}{0.48\textwidth}
        \centering
        \includegraphics[width=\linewidth]{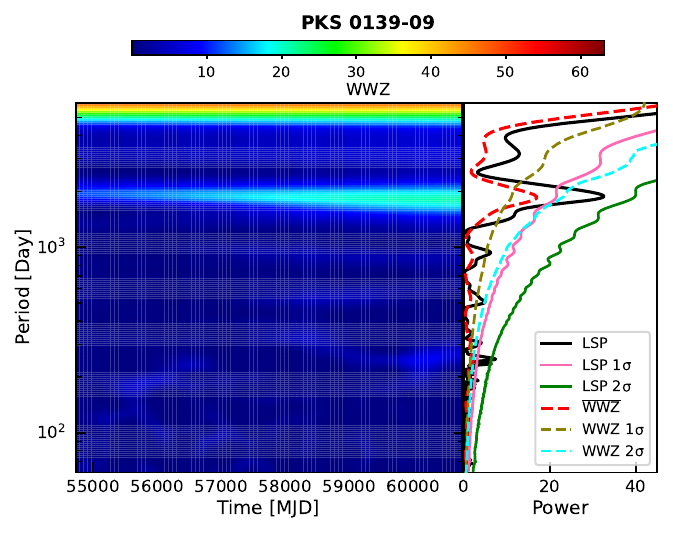}
    \end{minipage}
    \vspace{0.05cm}
    \caption{The 2D scalogram-filled color scale contour plots of the weighted wavelet z-transform for the $\gamma$-ray light curves of the selected sample. Side plots are the corresponding WWZ power spectrum averaged (red dashed line) and the LSP power spectrum (black line), scaled to the WWZ peak. The pink, green, orange, and blue lines denote the  $1 \sigma$, $2 \sigma$, $3 \sigma$, and $4 \sigma$ LSP's confidence levels, respectively. The olive, cyan, gray, and magenta dashed lines represent the corresponding WWZ confidence levels using the \cite{Emmanoulopoulos2013} method.}
    \label{fig:WWZ-LSP}
\end{figure}

To examine the effect of transients 
we use  the WWZ wavelet technique.
While the LSP excels at periodicity detection in unevenly sampled LCs, it cannot optimally accommodate QPOs exhibiting significant frequency or amplitude evolution over time. For such  phenomena  the WWZ proves more effective for detection and quantification. The 
WWZ is based on a concept analogous to the LSP, but with an important difference; it still fits sinusoidal waves to the data, but with wave localization in both the time and frequency domains (e.g., \citealt{Foster1996}). 

In Fig.~\ref{fig:WWZ-LSP}, we show
the 2D contour plots of the WWZ power for the blazar LCs considered. Side plots show the WWZ power spectrum (i.e., averaged along the LC MJD epochs on the x-axis) along with the LSP power spectrum.
To estimate the WWZ significance levels, simulated LCs based on the technique from \cite{Emmanoulopoulos2013}, as described in Section~\ref{sec:significance-estimation}, are used. As the computation of WWZ is demanding, we used only 20,000 simulated LCs, which corresponds to a maximum significance of 3.9$\sigma$. For sources with periods estimated with significances exceeding the 3.9$\sigma$ level, such as PG 1553+113, OJ 014, and S5 1044+71, the number of simulated LCs was increased to $10^6$ to boost the maximum significance to $\approx$4.8$\sigma$.
The major higher significance peaks, associated with the brightly 
shaded strips, correspond to the periods previously inferred  
using the LSP. The principal 
periods obtained through the two methods are  consistent within uncertainties, 
and with broadly similar significances.  
However, the WWZ diagrams show nontrivial  temporal structures. 

Notable nonstationary features may be spotted in 
the WWZ map of PKS 2155-304. This shows a transient pattern with three epochal periods, with the period decreasing by a factor of about 3/5 times from one epoch to the next. 
The first epoch, lasting $\approx$9 years, has a significant period of 1.7 yr (at $2.9 \sigma$); this is followed by a period of 1.0 yr ($2 \sigma$) in the second epoch; and 0.56 yr ($2 \sigma$) in the third.  With the LSP we find an 
additional significant period at $0.35 \pm 0.003$ yr  at $2.2 \sigma$---again consistent with being $3/5$ times the period at 0.56 (which is 
confirmed at $0.56 \pm 0.006$ yr and $2.3 \sigma$)---repeating the aforementioned pattern.  
 
A more stable but still palpably time-dependent 
pattern may also be inferred for PKS 0454-234. Here we 
note even more significant  peaks associated 
with higher frequencies. Like the double peak at the 
main period, these also come in pairs, and some may 
correspond to harmonics. They are particularly clear in the LSP analysis.  
For example, there is a peak  at $1.57 \pm 0.05$ yr,
detected with the LSP at $2.1 \sigma$
significance, which is consistent with being half the principal  period at $3.5\pm0.35$ yr; another peak at $1.1 \pm 0.02$ yr, with significance $2.6 \sigma$, is consistent with being a third of the principal peak period; and another at $0.37 \pm 0.003$ yr, detected with significance $2.6 \sigma$, with period  consistent with being a tenth of the main one. We note that the last two periods have slightly higher significance than the principal peak (at $2.5 \sigma$). Even higher significance comes with the peaks for 0.96 and 0.39 years ($2.7 \sigma$ and $3 \sigma$, respectively), which are ``twinned''
with two of the secondary peaks above (at 1.1 and 0.37 yr).

We note that significant higher frequency peaks are also present
in the case of PG 1553+113. Here they appear, according to the LSP analysis, at $0.45 \pm 0.005$ and $0.71 \pm 0.01$ years (the latter being clearly consistent with having a period of a third of the main one). But they are of small significance relative to the main period at $2.1 \pm 0.1$ yr, detected at  $3.9\sigma$. In contrast to what was found above, in the 
case of PG 1553+113
the strength of the periodic signal is consistent over the whole time-range considered. 

The diagram for S5 1044+71 also exhibits a single clearly defined 
peak at $3.1\pm 0.31$ yr, with no other reaching a significance of $2 \sigma$.
But, unlike with  PG 1553+113, it is time dependent. The corresponding WWZ 
strip  peaks in brightness within the second half of the time interval considered here; it corresponds to the high significance of $4.6\sigma$.
Though high, this is somewhat smaller than that reported by  
\cite{Ren_2023}, who found a period of $3.1 \pm 0.62$ at $>5 \sigma$. 
This may be due to the transient behavior already apparent in the 
brightening and fading of our strip in the panel corresponding to that source in  Fig.~\ref{fig:WWZ-LSP}.

Only in the case of  
PG 1553+113 do we detect a persistent QPO over the whole observed timescale. 
Otherwise, either the signal
fades well within the observational timespan of the data considered here (PKS 0454-234, PKS 2155-304), 
or near its end (OJ 014, S5 1044+71); or (in the case of PKS 0139-09) the apparent
QPOs are nascent and expected to intensify if present. 
These distinctive traits will be important when discussing the 
predicted future light curves (Section~\ref{sec:Forecast}). 

While smaller periods are detected in some of the 
cases, with some consistent with being harmonics, 
in most cases we  furthermore
find  brightly shaded bands at lower frequencies (larger periods) 
than the principal
peak (appearing at the top of the diagrams). These may be  associated 
with longer-term periodic or transient phenomena. 
The exceptions being S5 1044+71, 
where such a band is nearly 
absent, and PKS 0454-234  where the higher period band is just above the principal one. As we will see, such properties will determine the effect of the detrending procedure
on the significance of the principal periods. 

\subsection{ARIMA and SARIMA modeling of the light curves}
\label{arima-sarima}

We used the ARIMA and SARIMA models  to further test whether the variability in the $\gamma$-ray light curves is purely stochastic or may contain additional periodic components. The ARIMA$(p, d, q)$ model comprises three main components: autoregressive (AR), integrated (I), and moving average (MA) processes. The $p, d,$ and $q$ are orders of the model for the number of AR parameters, differences, and MA parameters.
The AR component models the dependence between current and past emissions, where the current flux is expressed as a linear combination of its previous values. The MA component quantifies the dependence of the current emission on recent random shocks \citep{Scargle1981, Feigelson2018}. The integrated component applies differencing to reduce trends and render the time series stationary. Moreover, the Seasonal ARIMA or SARIMA$(p, d, q)\times(P, D, Q)_s$ model is an extension of the ARIMA model that attempts to identify physical periodicity in the LC, where $P, D, Q,$ and $s$ are the seasonal AR order, differencing order, MA order, and seasonal parameter, respectively.
We refer the reader to \citet{HyndmanForecastingBook2021} and \citet{Sharma_2024ApJ} for more details on ARIMA/SARIMA, the latter with an application to a blazar LC.

We fitted both ARIMA and SARIMA models to the monthly binned $\gamma$-ray LC of each source in the sample using \texttt{SARIMAX}\footnote{\url{https://www.statsmodels.org/dev/generated/statsmodels.tsa.statespace.sarimax.SARIMAX.html}} from \texttt{statsmodels} Python Module \citep{seabold2010statsmodels}. Using \texttt{pmdarima.arima.AutoARIMA}\footnote{\url{https://alkaline-ml.com/pmdarima/modules/generated/pmdarima.arima.AutoARIMA.html}} Python class, as an initial step, a stepwise grid search over \citep{Hyndman2008} $p, d, q, P, D,$ and $Q$ is performed to identify the best-fitting model. The seasonal parameter, $s$, is set to the source period of interest from the LSP method. Model selection is performed using the Akaike information criterion ($\mathrm{AIC} = -2 \ln L + 2k$, where $L$ is the likelihood and $k$ is the number of parameters), with the preferred model having the lowest AIC value. The chosen orders are then refit with \texttt{SARIMAX}.

For PG~1553+113, the stochastic modeling using ARIMA and SARIMA yielded the following: the nonseasonal ARIMA(3,~1,~2) model produced an AIC of 804.6, while the seasonal SARIMA(3,~1,~2) $\times$ (3,~0,~0)$_{25}$ model achieved a significantly lower AIC of 526.7.

For S5~1044+71, the nonseasonal ARIMA(3,~0,~3) model resulted in an AIC of 1395.1, whereas the seasonal SARIMA(3,~0,~2)(1,~0,~0)$_{37}$ model gave an AIC of 1160.4.

For the remaining four sources—PKS~0454-234, PKS~2155-304, OJ~014, and PKS~0139-09—the optimal non-seasonal models are ARIMA(1,~0,~0), ARIMA(1,~1,~1), ARIMA(1,~0,~1), and ARIMA(1,~1,~1), with AIC values of 1503.0, 1146.7, 793.4, and 898.4, respectively. For these four LCs, the SARIMA models have $P = 0$, $D = 0$, and $Q = 0$, indicating no significant seasonal component, while the nonseasonal $(p, d, q)$ orders remain the same as those of their ARIMA counterparts.

The results indicate that the LCs of the blazars PG~1553+113 and S5~1044+71 are better represented by a seasonal ARIMA model, whereas the LCs of the remaining four blazars are best described as stochastic processes modeled by nonseasonal ARIMA models.
These modeling findings further support the presence of a possible QPO in PG~1553+113 and S5~1044+71, where they also exhibit the highest QPO significance as measured by both the LSP and the WWZ: 3.6$\sigma$ and 4.4$\sigma$ from the LSP, and 3.9$\sigma$ and 4.6$\sigma$ from the WWZ, respectively.

\subsection{Folded light curve}
\label{folded-lc}

In order to further check the possibility of the presence of the QPO, we run a folding search of each source light curve. The period scan range was 200--2000 days with 15 phase bins and phase zero at MJD corresponds to the start time of LC observation. 
For a given phase bin, the weighted flux is calculated with the formula,

\begin{equation}
\bar{F}_k = \frac{\sum_{i=1}^{N_k} w_i F_i}{\sum_{i=1}^{N_k} w_i}
\end{equation}

where $F_{i}$ and $\sigma_{i}$ are the flux and its associated uncertainty of the \textit{i}th data point within the \textit{k}th phase bin, and $w_{i}$ 
are the statistical weights
($w_i = \frac{1}{\sigma_i^2}$).

\begin{figure}[h!]
    \centering
    \begin{minipage}{0.48\textwidth}
        \centering
        \includegraphics[width=\linewidth]{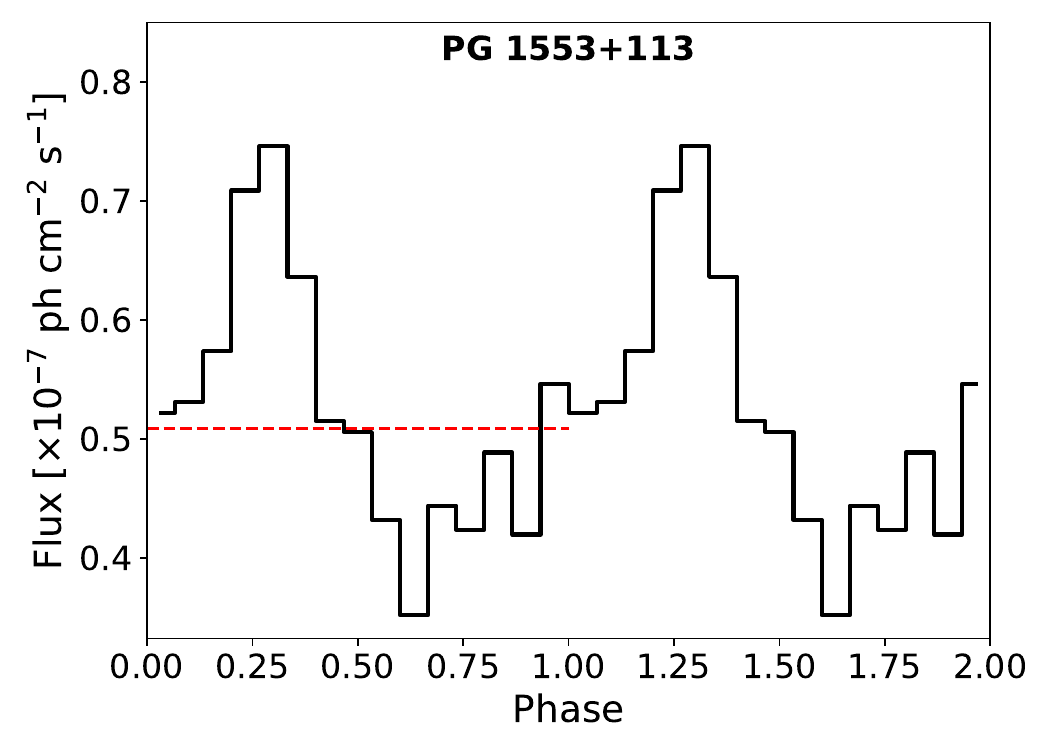}
    \end{minipage}
    \hspace{0.1cm}
    \begin{minipage}{0.48\textwidth}
        \centering
        \includegraphics[width=\linewidth]{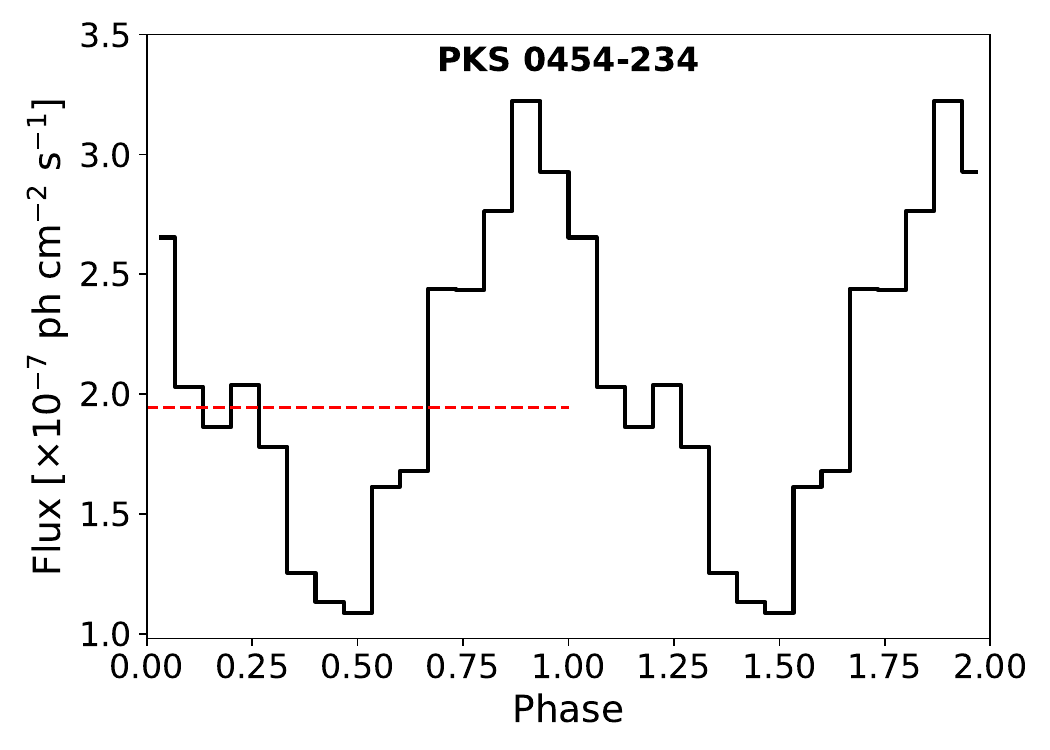}
    \end{minipage}

    \vspace{0.05cm}

    \begin{minipage}{0.48\textwidth}
        \centering
        \includegraphics[width=\linewidth]{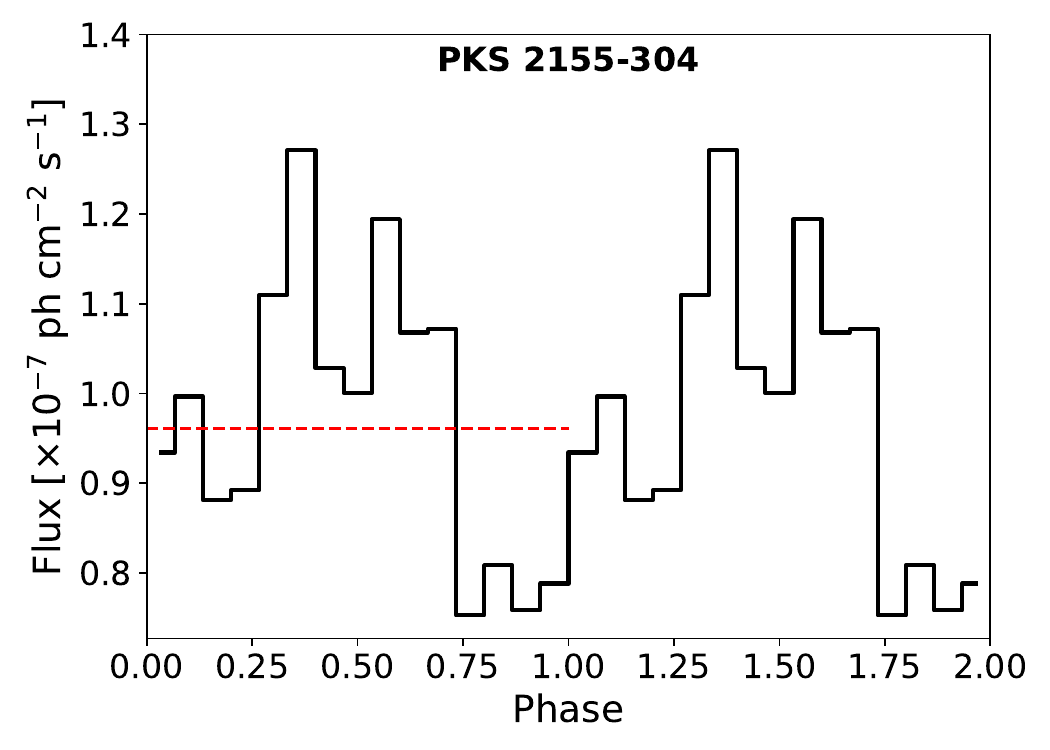}
    \end{minipage}
    \hspace{0.1cm}
    \begin{minipage}{0.48\textwidth}
        \centering
        \includegraphics[width=\linewidth]{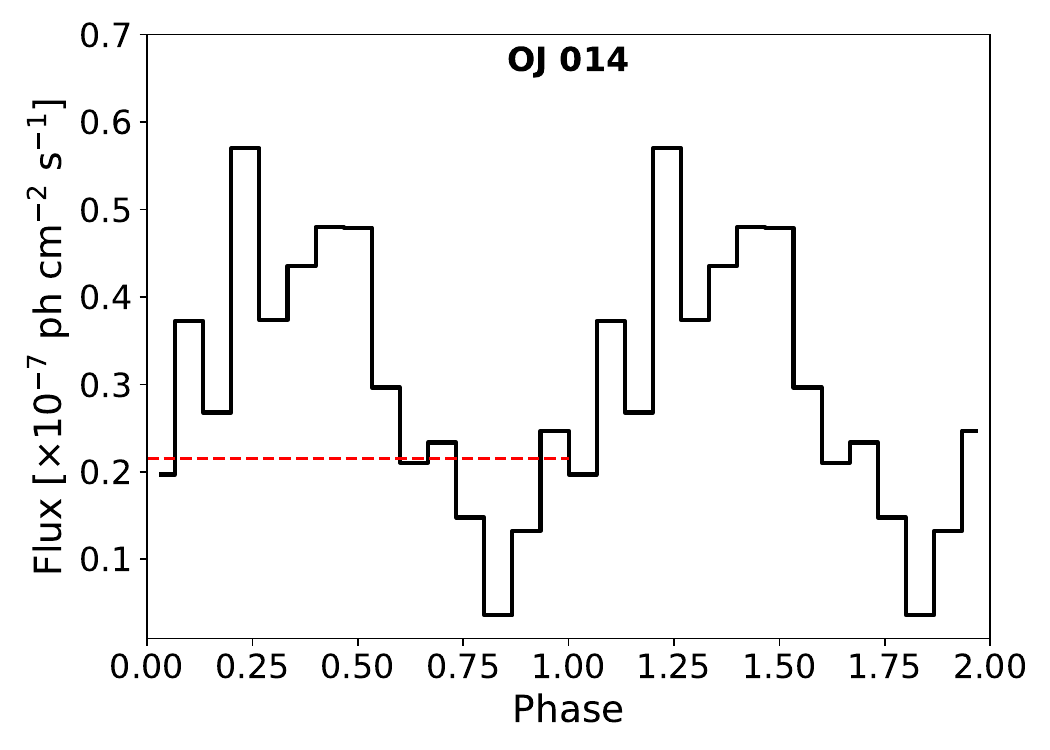}
    \end{minipage}

    \vspace{0.05cm}

    \begin{minipage}{0.48\textwidth}
        \centering
        \includegraphics[width=\linewidth]{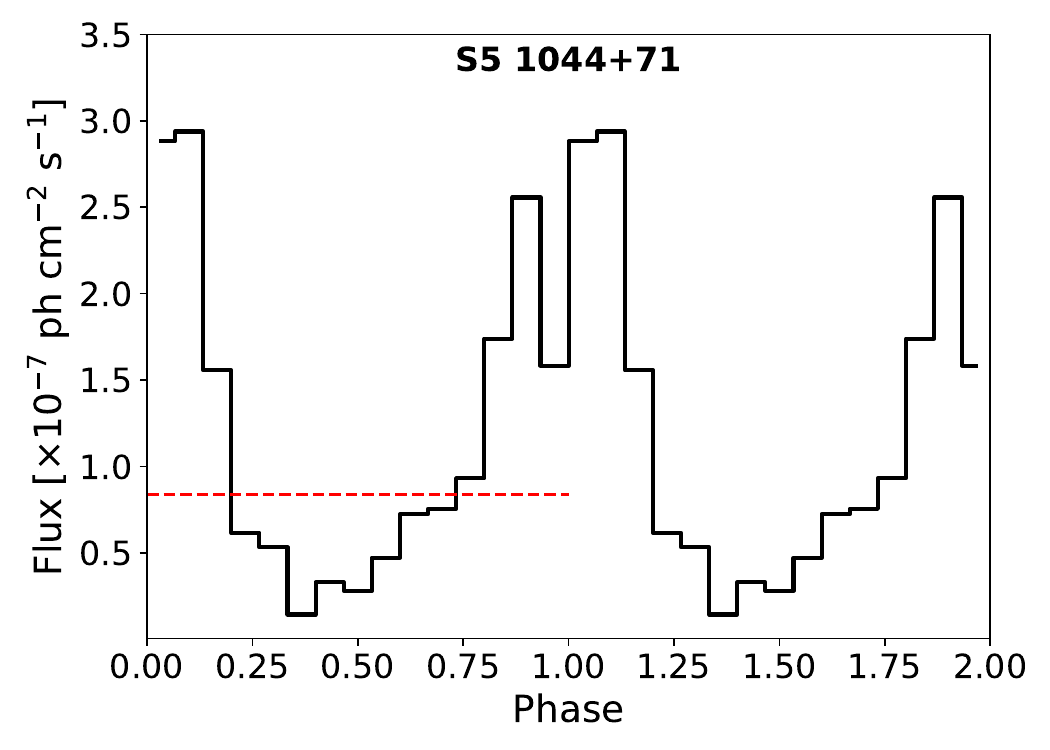}
    \end{minipage}
    \hspace{0.1cm}
    \begin{minipage}{0.48\textwidth}
        \centering
        \includegraphics[width=\linewidth]{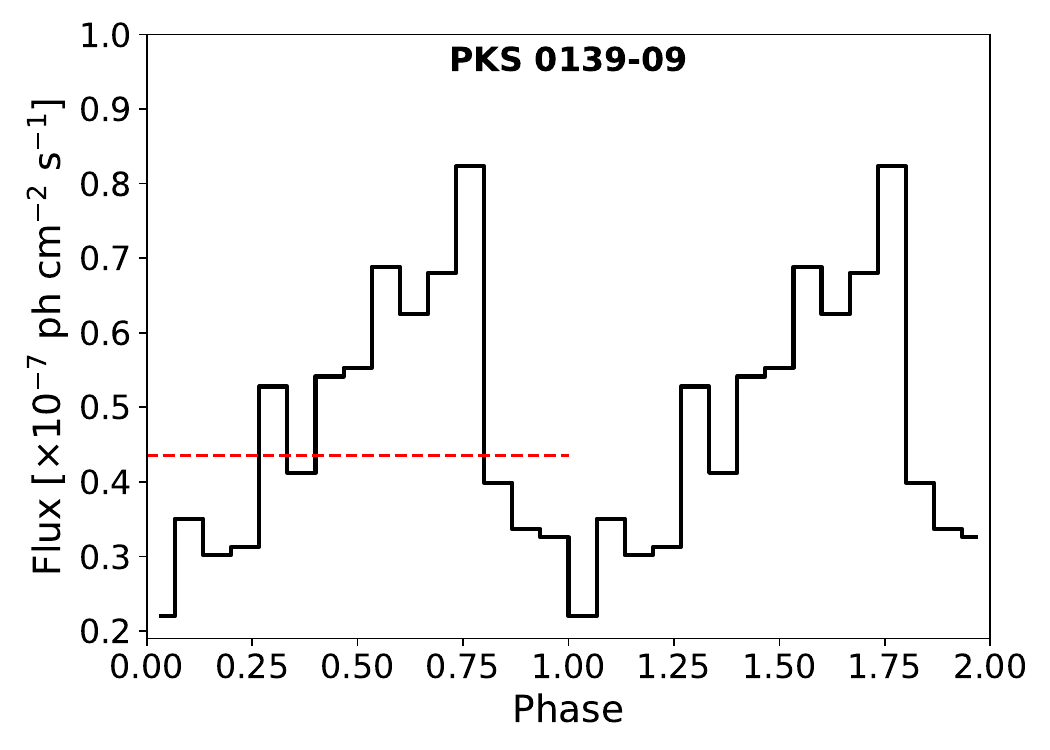}
    \end{minipage}
    \vspace{0.05cm}
    \caption{Folded light curves of the six sources. The folded light curves use 15 phase ranges with phase zero at MJD corresponds to the start time of LC observation. The periods were set to that identified by the LS method. The dashed red lines represent the weighted mean values. Two cycles are displayed for clarity.}
    \label{fig:folded-LCs}
\end{figure}

For all sources, the detected period was within the period uncertainty identified by the LS method. Moreover, for PG~1553+113 one more period is detected (4.2 yr) at comparable $\chi^2_{\mathrm{red}}$ value to the period of interest (2.1 yr) as well as PKS~2155-304 which showed few higher periods at little higher $\chi^2_{\mathrm{red}}$ values and one shorter period (0.5 yr) at the same $\chi^2_{\mathrm{red}}$ value of the period of interest (1.7 yr). Fig.~\ref{fig:folded-LCs} shows the folded LCs of the six sources with period set to that identified by the LS method. Two period cycles are displayed for clarity.

The variation with the phase appears clear in the sources PKS 0454-234 ($\chi^2_{\mathrm{red}}=168.7$; from the fitting of the folded LC with a constant) and S5 1044+71 ($\chi^2_{\mathrm{red}}=330.1$) and moderate in PG 1553+113 ($\chi^2_{\mathrm{red}}=15.1$), OJ 014 ($\chi^2_{\mathrm{red}}=30.3$), and PKS 0139-09 ($\chi^2_{\mathrm{red}}=30.4$). The PKS 2155-304 showed the lowest variation ($\chi^2_{\mathrm{red}}=13.5$).

\section{Light-curve decomposition and detrending}
\label{lc_detrending}

\subsection{Method}
\label{lc_detrending_method}

Simple inspection (of Fig.~\ref{fig_decomposed_LCs}) reveals that some blazar LCs  display long-term trends alongside possible periodic signals with 
relatively short periods. This is an effect that was also present 
in previous studies (e.g., \citealt{2022-Rueda, abdollahi2024periodic, Penil_2024, 2025penil_SSA}).
As trends distort the embedded periodic signals, they affect any inferred statistical significance regarding the presence of periodicity \citep{2022-Rueda}. They may also reflect QPO modulation on longer timescales (with larger errors). The inclusion of such trends in the fit of a source’s behavior may thus be essential in estimating the significance of any inferred periodicity on characteristic timescales shorter than the trend \citep{Welsh1999, 2025penil_SSA}.

With this in mind, we decompose our LC time series into components that correspond to distinct temporal variations \citep{makridakis1998}. Namely, 

\begin{enumerate}
    \item \textit{A Trend Component}. Reflecting low-frequency or systematic long-term variation in the data amplitudes.
    
    \item \textit{A Seasonal Component}. Capturing cyclical variation aligned with smaller periodic frequencies, such as daily, weekly, or annual patterns.
    
    \item \textit{A Remainder Component}. Representing variation unaccounted for by trend or seasonality, which may contain noise or latent informational patterns.
\end{enumerate}

Traditional decomposition frameworks often assume mutual independence among components, formalized by decomposition of an observed 
time series $Y_t$ with $n$ data points through an additive model:
\begin{equation}
    Y_t = T_t + S_t + R_t, \quad t = 1, \dots, n.
\end{equation}
Here $T_t$, $S_t$, and $R_t$ denote the long-term trend, the seasonality, and the remainder (or irregularities), respectively.
For this purpose, we used the STLForecaster\footnote{\textcolor{black}{\url{https://www.sktime.net/en/stable/api_reference/auto_generated/sktime.forecasting.trend.STLForecaster.html}}}; a statistical model that employs a decomposition-based framework to generate predictions for both LC detrending and forecasting. 
Rooted in the principles of time series decomposition, this method applies the filtering procedure STL (seasonal-trend decomposition via LOESS), introduced by \cite{stl}, to decompose a given time series to the aforementioned components (trend, seasonal, and residual). The detrended LC is computed by subtracting the trend component from the original LC. Any detrending technique, including STL decomposition, can introduce correlations between neighboring data points through the smoothing process. However, in this work, the STL procedure is used to remove the slowly varying long-term trend from the light curves, whereas our analysis focuses on year-like periodicities. Therefore, the induced correlations are expected to mainly affect low-frequency variations rather than the year-scale signals investigated here. The predicted time series is then derived by independently modeling each component. It is also possible  to recompose their projections (as described in Section \ref{STL_forecasting}).

The STL algorithm iteratively applies LOESS (locally estimated scatterplot smoothing). This is a nonparametric local regression smoothing technique. It is used to isolate seasonal and trend-cycle patterns while accounting for local variations. Unlike rigid parametric models (e.g., the parametric model to search for periods in LCs of high-energy $\gamma$-ray sources;  \citealt{2022-Rueda}), the STL’s modular architecture comprises sequential smoothing iterations. It thus ensures robustness in handling diverse temporal structures, including nonlinear trends and evolving seasonality \citep{THEODOSIOU2011STL}.
It derives its  strength from the ability to specify the number of observations per cycle of seasonal components; resilience  to short-term anomalies or transient behavior in the data; and compatibility with time series including gaps \citep{stl}. This is needed, as our sample includes three sources with gaps in their LCs---OJ 014, S5 1044+71, and PKS 0139-09. One of these has many missing values, mostly during source's low states ($\approx$16$\%$).
To overcome this issue, we interpolate its LC employing a linear interpolation method.
This is further discussed in Section~\ref{STL_forecasting}. The STL algorithm provides the three decomposed components without associated uncertainties. Consequently, the propagation of uncertainties to the detrended LC data points introduced by the detrending procedure is not taken into account.

For each point $x$ in the time series, LOESS implements a weighted regression on a selected nearest neighbor window. The weights of neighbors $x_i$ follow a weighting function $W$, e.g., Gaussian or tri-cube function, where the weight decreases as with the distance $x_i$  from $x$.
The neighborhood weight for any $x_i$ is then 
\begin{equation} \label{eq:neu}
    \nu_i(x) = W\left(\frac{\lvert x_i - x \rvert}{\lambda(x)}\right),
\end{equation}
where $\lambda(x)$ is the distance amplitude from $x$.

The essential outputs of the procedure are the strengths of the trend and seasonality in the LCs \citep{Wang_2006}. The trend strength is defined as \cite{makridakis1998}
\begin{equation}
    F_{\rm{T}} = \max \left( 0,\  1 - \frac{\operatorname{Var}(R_t)}{\operatorname{Var}(T_t + R_t)} \right),
\end{equation}
where Var($T_t + R_t$) and Var($R_t$) are the variances of the deseasonalized series and the series after trend and seasonality adjusting, respectively. In time series exhibiting a strong trend, the seasonally adjusted series demonstrates significantly greater variability than the remainder. Consequently, the variance ratio Var($R_t$)/Var($T_t + R_t$) will be substantially reduced. Conversely, data with minimal or absent trend exhibit approximately equivalent variances between the remainder and the seasonally adjusted series. Similarly, the seasonality strength is defined by,
\begin{equation}
    F_{\rm{S}} = \max \left( 0,\  1 - \frac{\operatorname{Var}(R_t)}{\operatorname{Var}(S_t + R_t)} \right),
\end{equation}
where Var($S_t+R_t$) is the variance of the detrended series. A series with strong seasonality gives $F_{\rm{S}}$ close to unity, as Var($S_t+R_t$) will be much greater than Var($R_t$), while a series with almost no seasonality gives $F_S$ close to zero. The remainder strength, $F_{\rm{R}}$, is computed by replacing Var($R_t$) in the previous equation by Var($S_t$).

\subsection{Results of the detrending procedure}

\begin{figure}[h!]
    \centering
    \begin{minipage}{0.48\textwidth}
        \centering
        \includegraphics[width=\linewidth]{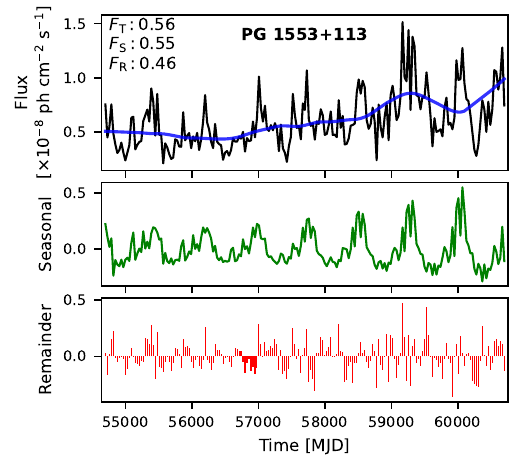}
    \end{minipage}
    \hspace{0.1cm}
    \begin{minipage}{0.48\textwidth}
        \centering
        \includegraphics[width=\linewidth]{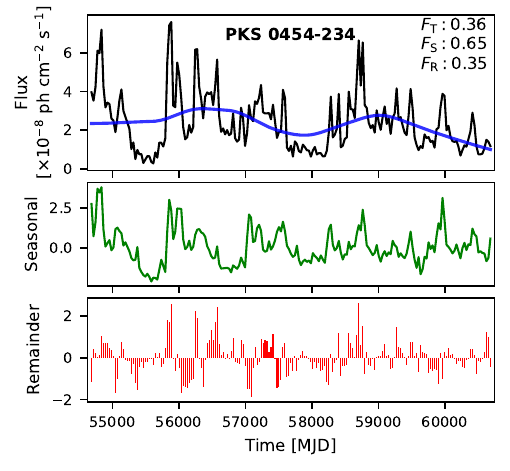}
    \end{minipage}

    \vspace{0.05cm}

    \begin{minipage}{0.48\textwidth}
        \centering
        \includegraphics[width=\linewidth]{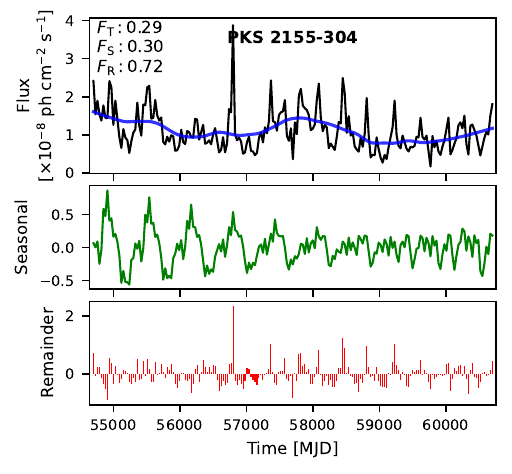}
    \end{minipage}
    \hspace{0.1cm}
    \begin{minipage}{0.48\textwidth}
        \centering
        \includegraphics[width=\linewidth]{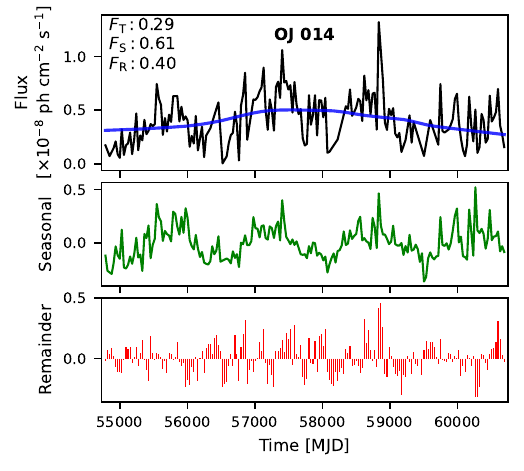}
    \end{minipage}

    \vspace{0.05cm}

    \begin{minipage}{0.48\textwidth}
        \centering
        \includegraphics[width=\linewidth]{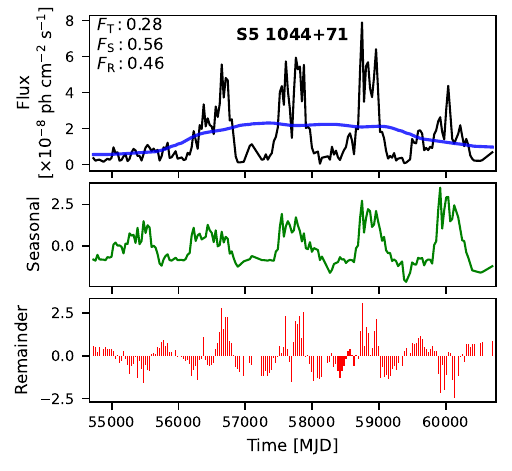}
    \end{minipage}
    \hspace{0.1cm}
    \begin{minipage}{0.48\textwidth}
        \centering
        \includegraphics[width=\linewidth]{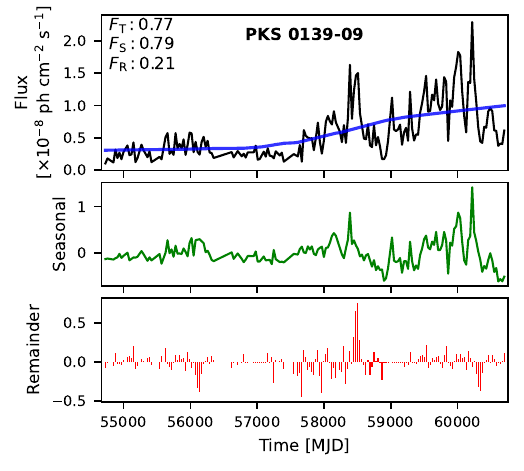}
    \end{minipage}
    \caption{Decomposition of the monthly $\gamma$-ray light curves using the STL method. For each source, the trend (blue curve) component is depicted in the top panel with source's original light curve (black curve). The seasonal and remainder components are displayed in the middle and bottom panels, respectively. The strength of the components are as denoted in the panels.}
    \label{fig_decomposed_LCs}
\end{figure}

Fig.~\ref{fig_decomposed_LCs} shows the decomposition of the monthly $\gamma$-ray LCs using the STL method. In the case of PG 1553+113
the procedure reveals a general increasing trend over time, followed by apparent oscillations. 
We note that in previous studies, the trend for this source was modeled as a linear increase during the first $\approx$12-15 yr of Fermi-LAT data \citep{abdollahi2024periodic, Penil_2024, 2025penil_SSA}. However, by extending the LC with currently available Fermi-LAT data up to 2025 February, it is evident that the flux has followed a saturating path rather than continuing to monotonically increase. The STL method captures this reversal 
that was not accounted for in previous work.
In other cases the trend appears to exhibit clear time modulation, 
particularly in the case 
of PKS 0454-234 (where we recall had a larger period peak close to the main one in its 
periodogram; Fig.~\ref{fig:WWZ-LSP}).

In general, all blazars in the sample exhibited long-term trends, albeit with varying strength and timescales. These  were  captured in the  contour plots of the WWZ, where they are represented as bands or peaks referring to tentative low-frequency  signals (Fig.~\ref{fig:WWZ-LSP}). In addition to significant trend components,  all sources exhibited seasonality values above $0.55$, with the exception of PKS 2155-304. For this source, the strength of the seasonality fades after the first $\approx$9 yr of observations, which is consistent with the transient behavior also inferred from the WWZ in Fig.~\ref{fig:WWZ-LSP}. In contrast, in the case of S5 1044+71 the seasonal component increases with time, but it comes with a noisy residual that appears correlated with it 
over the first three peaks of the original LC and then anticorrelated with the fourth. We finally note the strong seasonal component for PKS 0454-234  despite the low-significance level of the main period at  3.5 yr. This further suggests 
that the smaller period peaks found in Fig.~\ref{fig:WWZ-LSP} 
may indeed be effectively harmonics that contribute
toward a strong overall seasonal quasiperiodic (but nonsinusoidal) signal.

\subsection{Possible quasiperiodic signals in the detrended light curves}

We now repeat the LSP and WWZ analysis on the detrended LCs. The detrending removes the low-frequency signals previously found with low significance, where they are 
represented by the bands at the upper end of the 
panels of Fig.~\ref{fig:WWZ-LSP}. As may be expected, major differences 
in QPO signal significances are found when these are prominent. 

The results of periods and their local significances, along with simple power-law fit parameters and its reduced $\chi^2$ are summarized in Table~\ref{LSP_WWZ}. 
The main periods identified for the detrended LCs are the same as, or within the uncertainties of, those from the original LCs. The statistical significances, 
on the other hand, increase significantly in all cases; with  exceptions 
for PKS 0454-234, where the major band at lower frequency was previously found 
to be of similar period to the main QPO signal (Fig.~~\ref{fig:WWZ-LSP}), 
and the case of S5 1044+71, where a clear low-frequency band is absent. 
In the latter case, although a significant 
trend is present, it seems to correspond to a more diffuse 
band of lower frequencies.

\clearpage
\begin{longtable}{@{} 
  >{\raggedright\arraybackslash}p{2.5cm} 
  >{\centering\arraybackslash}p{1.6cm}  
  >{\centering\arraybackslash}p{1.6cm}  
  >{\centering\arraybackslash}p{1.6cm}  
  >{\centering\arraybackslash}p{1.6cm}  
  >{\centering\arraybackslash}p{1.6cm}  
  >{\centering\arraybackslash}p{1.6cm}@{} 
}

\caption{Summary of the periods (in years) and their local significances, as inferred from the data, for both original and detrended $\gamma$-ray light curves of the six blazars, along with simple power-law (PL) fit parameters and its reduced $\chi^2$.} \\
\toprule
\multirow{2}{*}{\raggedright Source Name} &
\multicolumn{3}{c}{Original LC} & 
\multicolumn{3}{c}{Detrended LC} \\
\cmidrule(lr){2-4} \cmidrule(lr){5-7}
& PL Par. & LSP & WWZ & PL Par. & LSP & WWZ \\
\midrule
\endfirsthead

\toprule
\multirow{2}{*}{\raggedright Source Name} &
\multicolumn{3}{c}{Original LC} & 
\multicolumn{3}{c}{Detrended LC} \\
\cmidrule(lr){2-4} \cmidrule(lr){5-7}
& PL Par. & LSP & WWZ & PL Par. & LSP & WWZ \\
\midrule
\endhead

\bottomrule
\endlastfoot

PG 1553+113 &
$A = 0.003$ \newline $\Gamma = 1.26$ \newline $\chi^2_{\text{red}} = 2.8$ &
${2.1}_{3.6\sigma}^{\pm 0.13}$ & ${2.1}_{3.9\sigma}^{\pm 0.19}$ &
$A = 0.0081$ \newline $\Gamma = 1.17$ \newline $\chi^2_{\text{red}} = 2.84$ &
${2.1}_{4.5\sigma}^{\pm 0.12}$ & ${2.1}_{4.0\sigma}^{\pm 0.18}$ \\
\addlinespace[0.2cm]

PKS 0454$-$234 &
$A = 0.011$ \newline $\Gamma = 1.22$ \newline $\chi^2_{\text{red}} = 1.7$ &
${3.5}_{2.5\sigma}^{\pm 0.35}$ & ${3.5}_{2.2\sigma}^{\pm 0.42}$ &
$A = 0.015$ \newline $\Gamma = 1.19$ \newline $\chi^2_{\text{red}} = 1.75$ &
${3.6}_{2.6\sigma}^{\pm 0.4}$ & ${3.6}_{2.4\sigma}^{\pm 0.44}$ \\
\addlinespace[0.2cm]

PKS 2155$-$304 &
$A = 0.66$ \newline $\Gamma = 0.83$ \newline $\chi^2_{\text{red}} = 1.8$ &
${1.7}_{2.6\sigma}^{\pm 0.13}$ & ${1.7}_{2.9\sigma}^{\pm 0.17}$ &
$A = 0.23$ \newline $\Gamma = 0.62$ \newline $\chi^2_{\text{red}} = 2.47$ &
${1.7}_{3.3\sigma}^{\pm 0.13}$ & ${1.7}_{3.3\sigma}^{\pm 0.16}$ \\
\addlinespace[0.2cm]

OJ 014 &
$A = 0.027$ \newline $\Gamma = 0.97$ \newline $\chi^2_{\text{red}} = 1.3$ &
${4.2}_{3.5\sigma}^{\pm 0.57}$ & ${4.2}_{3.6\sigma}^{\pm 0.53}$ &
$A = 0.51$ \newline $\Gamma = 0.53$ \newline $\chi^2_{\text{red}} = 1.37$ &
${4.2}_{4.2\sigma}^{\pm 0.53}$ & ${4.2}_{4.3\sigma}^{\pm 0.55}$ \\
\addlinespace[0.2cm]

S5 1044+71 &
$A = 0.003$ \newline $\Gamma = 1.6$ \newline $\chi^2_{\text{red}} = 1.95$ &
${3.1}_{4.4\sigma}^{\pm 0.34}$ & ${3.1}_{4.6\sigma}^{\pm 0.31}$ &
$A = 0.0036$ \newline $\Gamma = 1.6$ \newline $\chi^2_{\text{red}} = 1.98$ &
${3.1}_{4.3\sigma}^{\pm 0.37}$ & ${3.1}_{4.4\sigma}^{\pm 0.35}$ \\
\addlinespace[0.2cm]

PKS 0139$-$09 &
$A = 0.035$ \newline $\Gamma = 1.02$ \newline $\chi^2_{\text{red}} = 1.99$ &
${5.2}_{1.6\sigma}^{\pm 0.73}$ & ${5.1}_{1.7\sigma}^{\pm 0.71}$ &
$A = 0.067$ \newline $\Gamma = 1.0$ \newline $\chi^2_{\text{red}} = 2.27$ &
${5.1}_{2.0\sigma}^{\pm 0.96}$ & ${5.0}_{2.0\sigma}^{\pm 0.80}$
\label{LSP_WWZ}
\end{longtable}

In general, the postdetrending increase in the strength of the QPO signal  
was larger in the case of the LSP-inferred periods than with the local in time WWZ method. But it was still not always 
large enough to account for the possible high-significance levels in previous work, where the LCs were studied on a shorter time interval.
This is notably the case with PKS 2155-304,  which is also 
the case in which the transient nature of the QPO 
signal seems clearest. Thus, in addition to the contamination 
by slow time-dependent modulation that may be removed by detrending,
intrinsic transience in the signal strength at the principal period 
may affect the significance of the QPO signal at that period over time. 
Except for the case  PG 1553+113, clear transience is present in 
the WWZ diagrams of the detrended data as was the case 
with the original LCs. 

\section{Light-curve forecasting} 
\label{lcForecasting} 

\subsection{Motivation and general framework}

Forecasting future signals is important in our context, as it enables us to make testable predictions that may ascertain or rule out sustained QPO signals, which may contain important clues to their strength and physical origin. Persistence of QPOs in extrapolated curves 
reflect their stregth of an existing QPO and  makes testable predictions that may be compared with data from future observational campaigns.  

For this purpose, we will use both the traditional statistical 
method of the previous section, in addition to a more recent machine learning based technique described below.   

\subsection{Statistical learning forecasting: STLForecaster}
\label{STL_forecasting}

So far we have used the detrending procedure  
in order to isolate a periodic signal from a general trend and residual of an existing time series. But the method is also commonly used for forecasting future signals.
The associated 
STLForecaster framework enhances forecasting accuracy through the  decomposition of the time series into its constituent components, as in the previous section. Forecasts are then generated by extrapolating the deterministic trend and seasonal cycles into future intervals and combining these projections with stochastic residual estimates. This component-wise approach ensures that systematic patterns and random fluctuations are modeled independently. This leads to improvement in overall predictive reliability.

The LC is split into training and test datasets, with the test set comprising approximately quarter of the complete observation period of the LC to evaluate the model’s performance. To optimize the hyperparameters involved, we performed a grid search over predefined search spaces using the machine learning method \texttt{ForecastingGridSearchCV}
(from \texttt{sktime.forecasting.model$\_$selection}) with scoring metric mean squared error (MSE). The mean absolute percentage error (MAPE), which measures the deviation of predictions from the actual (ground-truth) observations, was also computed for the comparison with the transformer (Section~\ref{Transformer}).
The critical hyperparameter seasonal period \texttt{sp}, which defines the length of the seasonal period passed to \texttt{statsmodels} STL, is then assigned a range of values centered around the period given by the LSP.

\begin{figure}[!h]
    \centering
    \includegraphics[width=0.8\linewidth]{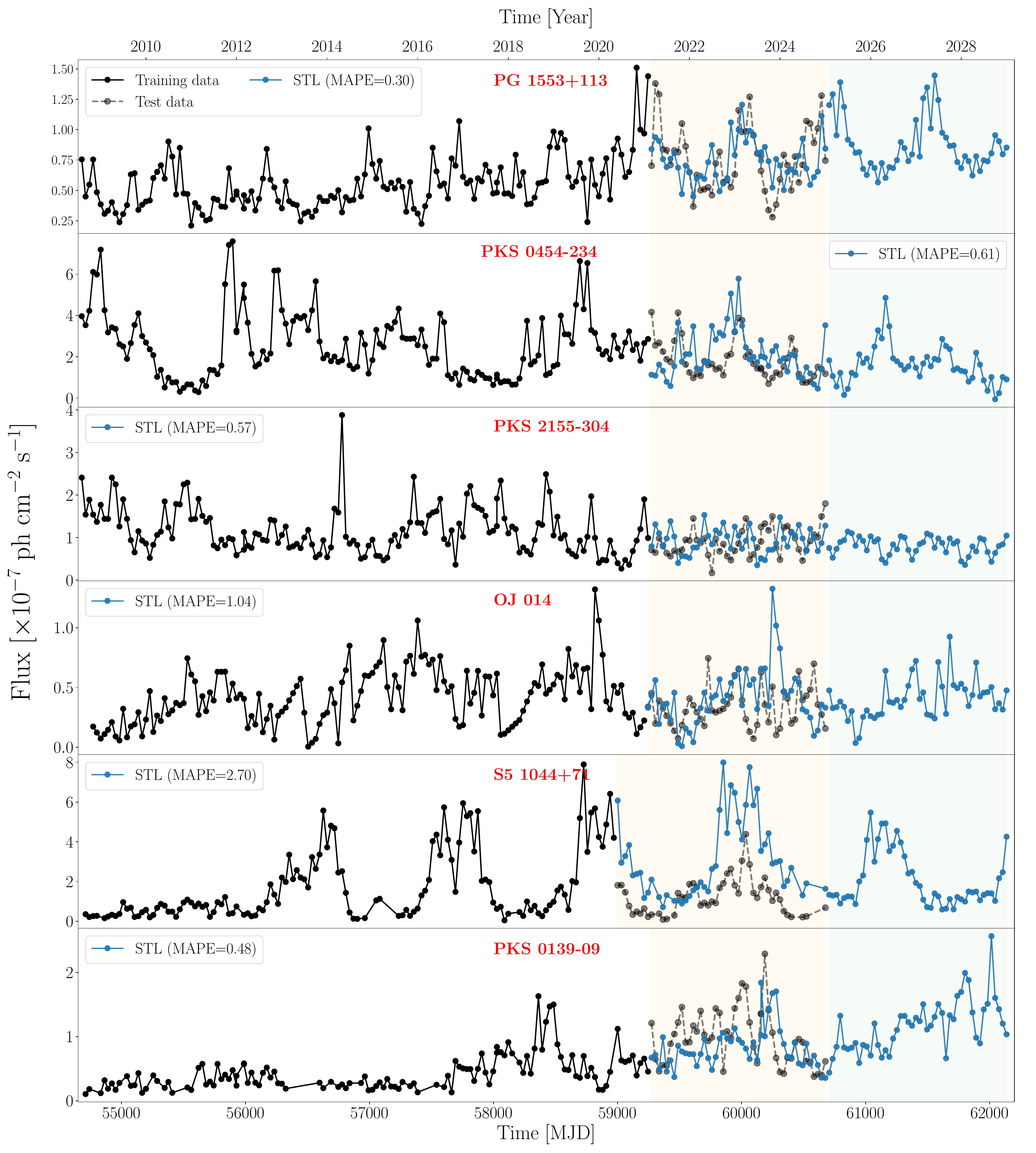}
    \caption{Current observed monthly $\gamma$-ray light curves and their forecast for the following 4 yr. The blue line represents the light-curve predictions on the test set (pink highlighted region) and the future epoch (light blue highlighted region) using STLForecaster. The mean absolute percentage error (MAPE) for the test set predictions is presented. The prediction on the future epoch is based on the training on the entire observed light curve.}
    \label{fig:forecasting_STL}
\end{figure}

Another important hyperparameter is \texttt{forecaster$\_$trend}. The trend component extracted via STL decomposition is fitted to a polynomial function to project future trend. For modeling the seasonal component, the \texttt{NaiveForecaster(strategy=“last")} outperformed different tested forecasters. 

The residual of the future period is set to the mean of the residual component of the STL decomposition. The prediction is then the sum of the predictions of trend, seasonal, and residual components. The results of this procedure are presented in Fig.~\ref{fig:forecasting_STL}. They will be compared to those obtained by the transformer technique described below.

\subsection{Deep learning forecasting: transformers} 
\label{Transformer}

\subsubsection{Motivation and overview}

Compared to classical statistical approaches, such as those described in the preceding subsection (STL), transformers offer clear advantages for gamma-ray blazar forecasting. STL relies on additive decomposition into smooth seasonal, trend, and residual components, which makes it effective for regular, stationary time series but poorly suited for the highly irregular and nonstationary variability of blazars. In particular, STL assumes fixed periodicities and smooth variations, limiting its ability to model sudden flaring events or quasiperiodic oscillations. Moreover, while STL focuses on short-term extrapolation of trends, it cannot naturally account for long-range temporal dependencies or multiscale variability. 

Unlike traditional sequential models that propagate information step by step, the transformer attends to all time steps simultaneously. This makes it especially suitable for forecasting \citep{wen2022transformers,shin2025enhancing}.
For forecasting, the transformer first processes sequential flux measurements from the existing LCs. The temporal structure of these sequences contains both rapid short-term flaring activity and long-range quasiperiodic variability, and the attention mechanism allows the model to capture both simultaneously. 
This is critical when forecasting over a horizon of 48 time steps.

At the core of transformer architectures lies the \textit{attention mechanism}, which computes weighted interactions among sequence elements. These weights reflect how strongly different time steps influence each other. This is learned adaptively through training. 

The input to the transformer is a tensor $X \in \mathbb{R}^{T \times F}$, where $T$ denotes the number of observed past time steps and $F$ the number of features per step. In the simplest case, $F=1$ corresponds to flux values alone; however, additional contextual variables (e.g., flux uncertainties, spectral indices, or external observational flags) may also be incorporated. Each input vector $X_{t,*}$ is projected into a higher-dimensional latent space through an embedding layer, which enriches the representation and enables the network to capture nonlinear dynamics beyond the raw flux signal.

The model then generates outputs $\hat{Y} \in \mathbb{R}^{H \times F}$, where $H=48$ is the forecast horizon. Each predicted time step $\hat{Y}_h$ corresponds to a forecasted flux value (or set of features) $h$ steps ahead of the most recent observation. This direct mapping from past context to future trajectory is what distinguishes transformers from statistical decomposition-based approaches.
We emphasize that, although transformer architectures provide a flexible framework for capturing complex multiscale temporal correlations through self-attention mechanisms, their forecasting performance remains fundamentally dependent on the temporal stability and representativeness of the underlying data.
\subsubsection{The attention mechanism}
Self-attention in transformers allows the model to capture dependencies across all time steps simultaneously, unlike traditional recurrent models that process data sequentially. This global view enables the network to identify long-term patterns and seasonal trends in time series data more effectively. Self-attention also provides dynamic weighting of past observations, so the model can focus on the most relevant historical points for each prediction. Furthermore, it is highly parallelizable, making training on large datasets faster and more efficient compared to sequential architectures. Overall, self-attention enhances the model’s ability to handle complex, nonlinear temporal relationships, improving the accuracy and robustness of forecasts.

The embedded sequence, $X^{\rm embed}$, of length $T$ is transformed into queries ($Q$), keys ($K$), and values ($V$) by learnable linear projections~\citep{vaswani2017attention}
\begin{equation}
Q^{T \times d} = X^{\rm embed} W^Q, \hspace{6mm} 
K^{T \times d} = X^{\rm embed} W^K, \hspace{6mm} 
V^{T \times d} = X^{\rm embed} W^V,
\end{equation}
where $d$ is the latent dimensionality encoding the 
minimal effective number of degrees of through which the data 
may be represented and $W^Q$, $W^Q$, and $W^Q$ are their learnable weight matrices.

Attention scores quantify the relative importance of different time steps in the input light curve when constructing the prediction at a given epoch. In this context, they measure how strongly past flux states influence the current forecast, enabling the model to dynamically focus on the most informative temporal patterns while down-weighting irrelevant or noisy observations. This adaptive weighting allows the network to capture both short- and long-range temporal correlations. The attention scores are computed as

\begin{equation}
\alpha^{T \times T} = \text{softmax}\left(\frac{Q K^T}{\sqrt{d}}\right),
\end{equation}
where $\alpha_{ij}$ measures the contribution of time step $j$ when updating step $i$ and $K^T$ 
denotes the transpose of the key matrix. In blazar LCs, this enables the model to focus on temporally distant but correlated variability, capturing flaring episodes and long-memory effects.

The aggregated output is:
\begin{equation}
\mathcal{Z} = \alpha V,
\end{equation}
where $\mathcal{Z}$ represents a context-aware reconstruction of the sequence. Multihead attention extends this mechanism by learning multiple relational subspaces in parallel: some heads specialize in short-term fluctuations, while others capture broader, slowly evolving trends.

The outputs from all attention heads are concatenated and linearly projected back to the model’s embedding dimension by the learned weight matrix $W^O$:
\begin{equation}
\mathcal{O} = \text{concat}(\mathcal{Z}^1, \ldots, \mathcal{Z}^n) W^O,
\end{equation}
and combined with the original embedding through a residual connection:
\begin{equation}
\widetilde{X} = X^{\rm embed} + \mathcal{O},
\end{equation}
which stabilizes training and preserves temporal structure. Stacking multiple transformer blocks allows progressive refinement of temporal dependencies across scales.

The self-attention mechanism in transformers directly captures both short- and long-term correlations within the LC, enabling the model to simultaneously learn rapid flux fluctuations and slowly evolving patterns. This makes it a powerful forecasting tool that adapts dynamically to the complexity of blazar variability, providing accurate predictions over horizons as large as 48 steps.
As a result, the transformer architecture provides a flexible, data-driven alternative to statistical decomposition methods. Its ability to learn multiscale temporal dependencies and adapt to irregular variability makes it particularly well suited for forecasting the complex and dynamic behavior of gamma-ray blazars.

\subsubsection{Network structure}
The forecasting network is constructed upon a transformer encoder tailored to capture temporal dependencies in multivariate time series. Each input sequence consists of $96$ past observations. Prior to embedding, the raw input series $x_t$ is decomposed into two components, a slowly varying trend and a rapidly fluctuating residual, according to
\begin{equation}
    x_t = T_t + R_t.
\end{equation}
Here $T_t$ denotes the trend component, representing smooth long-term variations; $R_t$ denotes the residual, capturing short-term fluctuations, irregularities, and seasonal patterns. Importantly, the two components are processed independently: $T_t$ and $R_t$ are each passed through separate linear embedding layers and then through parallel stacks of transformer encoders. In this way, each encoder is specialized in learning the temporal dependencies of either the long-term trend or the high-frequency residual. The resulting encoded representations are concatenated at the output of the encoder stage, providing the forecasting head with complementary information from both structural components of the time series. This decomposition and parallel encoding stabilize optimization by reducing interference between long- and short-range patterns, while also improving interpretability.

To preserve temporal ordering, positional encodings are added to the embeddings, as the self-attention mechanism is inherently permutation-invariant \citep{Hammad_2024cae, Hammad_2023sbd}. In this work, sinusoidal positional encodings are used, defined as
\begin{align}
    \text{PE}_{(pos,2i)}   = \sin\!\left(\frac{pos}{10000^{2i/d_{\text{model}}}}\right), \hspace{4mm}  \text{PE}_{(pos,2i+1)} = \cos\!\left(\frac{pos}{10000^{2i/d_{\text{model}}}}\right),
\end{align}
where pos is the position index, $i$ indexes the embedding dimensions, and $d_{\text{model}}=256$ is the embedding size. 
Positional encoding introduces explicit information about the position of each element, allowing the model to distinguish values at different time steps and capture temporal dependencies. Without it, the transformer would interpret the input as an unordered set, preventing effective learning of sequential patterns.
Positional encoding is essential for transformers in time series forecasting because self-attention  does not encode the order of sequence elements. In time series, temporal order is critical, as the relative timing of observations carries important information for forecasting.

Each of the trend and residual encoder branches consists of four transformer encoder layers. Each layer contains a multihead self-attention mechanism with $h=8$ heads, followed by a position-wise feed-forward network. Given an input representation $X \in \mathbb{R}^{L \times d_{\text{model}}}$, queries $Q$, keys $K$, and values $V$ are computed through learned linear projections. The scaled dot-product attention is then expressed as
\begin{equation}
    \text{Attention}(Q,K,V) = \text{softmax}\!\left(\frac{QK^\top}{\sqrt{d_k}}\right)V,
\end{equation}
where $d_k = d_{\text{model}}/h$ is the dimension per head. Multihead attention concatenates the outputs of the $h$ attention heads and applies a linear transformation, enabling the model to attend to different representation subspaces in parallel.

Each attention block is followed by a position-wise feed-forward network, implemented as
\begin{equation}
    \text{FFN}(x) = \text{ReLU}(xW_1 + b_1)W_2 + b_2,
\end{equation}
where $W_1, W_2$ are learnable weight matrices, $b_1$ and $b_2$ are biases, and ReLU introduces nonlinearity. Residual connections and layer normalization are applied after both the attention and feed-forward sublayers, ensuring stable gradient propagation. A dropout rate of $0.1$ is applied throughout to improve generalization.

\begin{figure}[!h]
    \centering
    \includegraphics[width=0.8\linewidth]{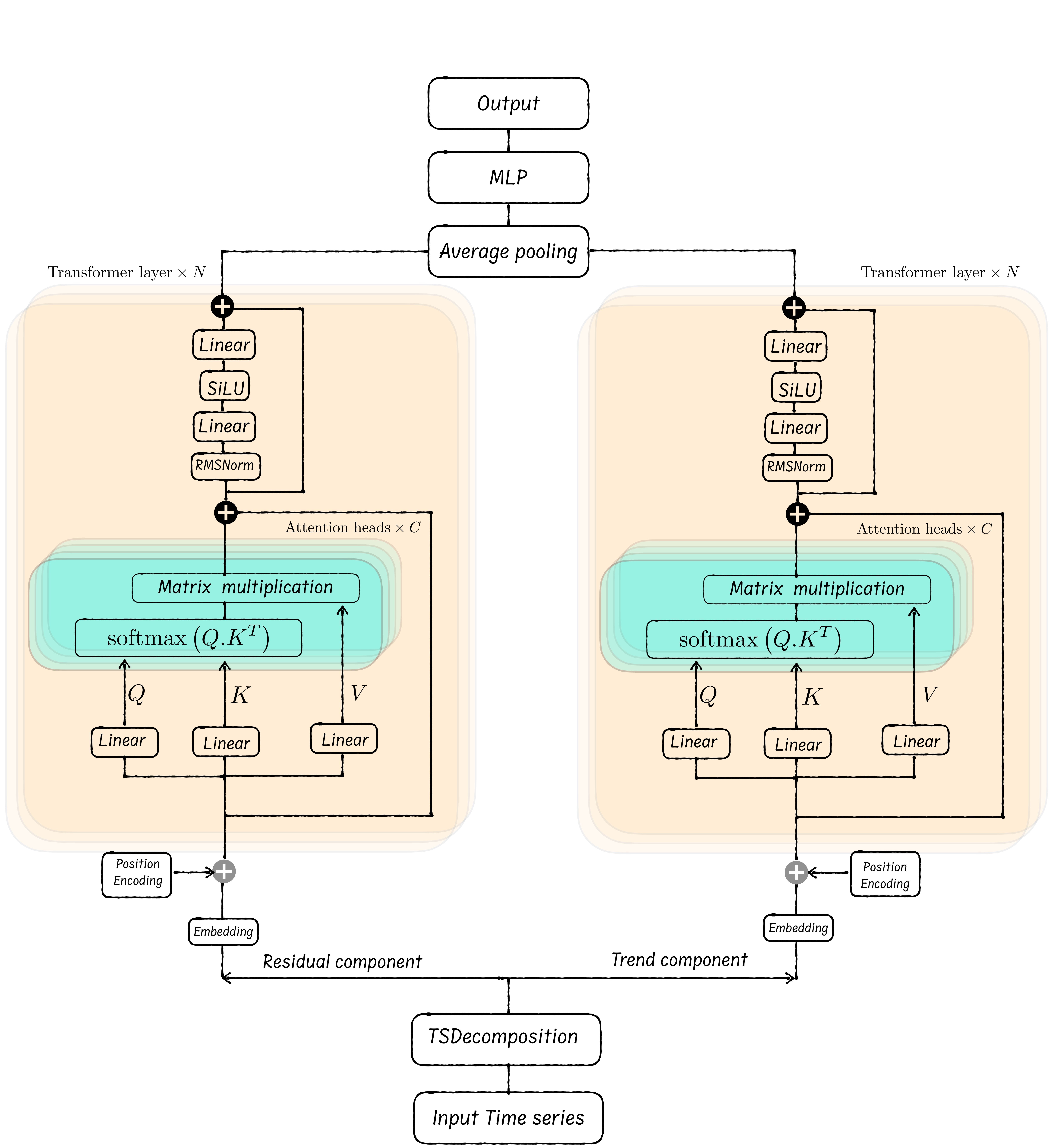}
    \caption{Schematic illustration of the used transformer network architecture.}
    \label{fig:network}
\end{figure}

At the output of the encoder stage, the encoded trend and residual representations are concatenated and passed through a projection head to produce forecasts for the $48$-step horizon. This fusion ensures that both long-term dependencies captured in the trend and localized variations preserved in the residual are exploited in the final prediction. An illustration of the network architecture is shown in Fig.~\ref{fig:network}. 

\subsubsection{Training}

Training is performed end-to-end using the PyTorch Lightning framework. The model is trained with mini-batches of $64$ sequences, and $10\%$ of the dataset is held out for validation. The objective is to minimize the MSE between the predicted horizon and the ground-truth sequence. Optimization is carried out using the Adam algorithm with an initial learning rate of $5 \times 10^{-6}$. To enhance convergence stability, a step learning rate scheduler decays the learning rate by a factor of $0.9$ every $20$ epochs. Training proceeds for up to $100$ epochs, although an early stopping criterion with patience of five epochs halts training if no improvement in validation loss is observed. Model checkpoints are saved based on the lowest validation loss, ensuring that the best-performing parameters are retained for evaluation and deployment.

\begin{figure}[!h]
    \centering
    \includegraphics[width=0.8\linewidth]{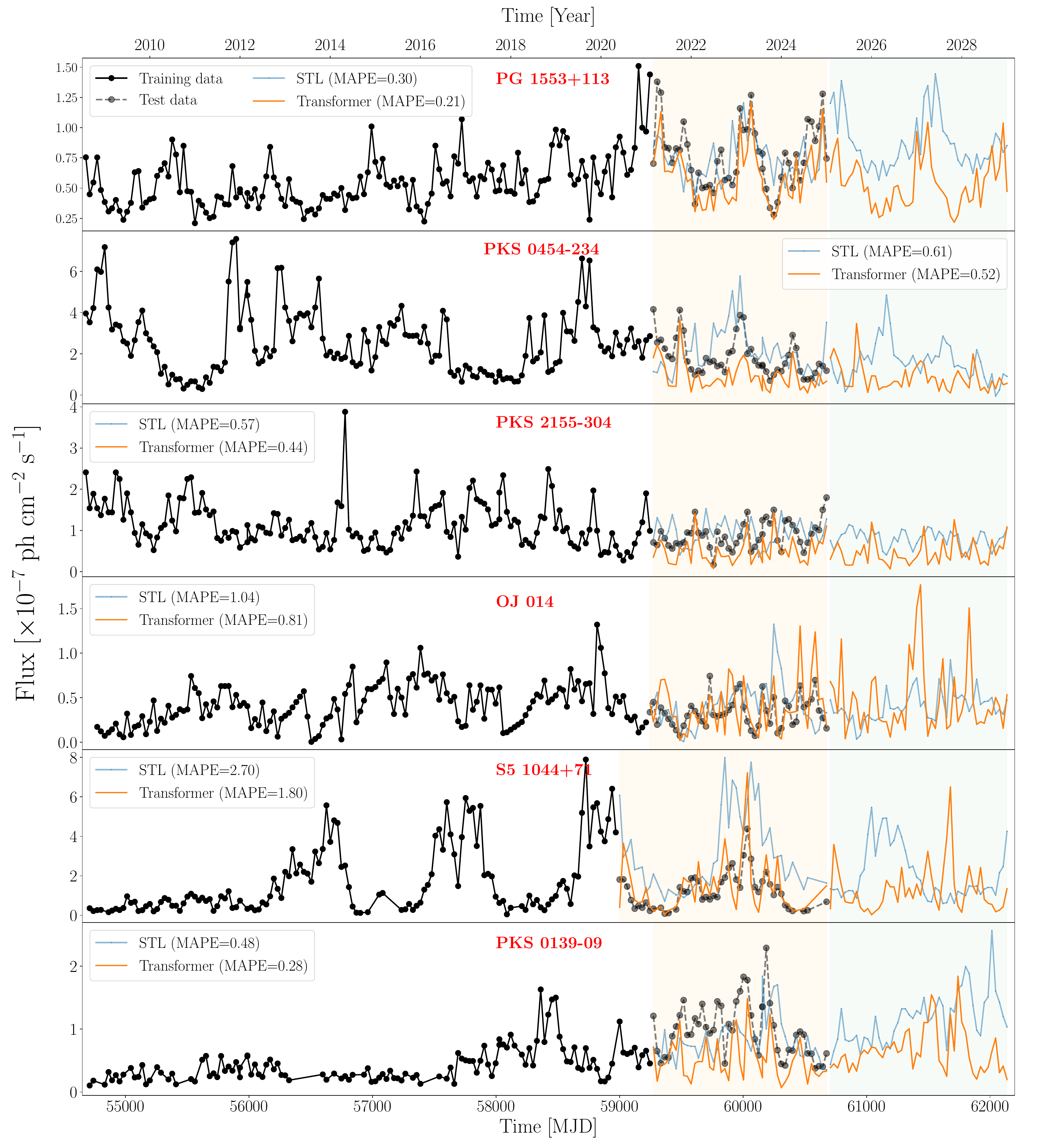}
    \caption{Current observed monthly $\gamma$-ray light curves and their forecasting for the following 4 yr. The blue and yellow lines represent the light-curve predictions on the test set (pink highlighted region) and the future epoch (light blue highlighted region) using STLForecaster and transformer, respectively. The MAPE for the test set predictions is presented for both methods, the transformer and the STLForecaster. The prediction on the future epoch is based on the training on the entire observed light curve.}
    \label{fig:results}
\end{figure}

The future behavior of the six sources in our sample, as predicted by both the transformer and STLForecaster methods, is shown in the light blue highlighted region of Fig.~\ref{fig:results}. 

We start by comparing the predictions on the test set, represented  by the pink highlighted region. The mean absolute percentage error results for the  demonstrate the general superiority of the deep learning transformer method over the statistical STLForecaster method in learning the complexity of the LC and accurately predicting the sources' behavior. This is particularly apparent  in the cases of PKS 1553+113. S5 1044+71 and PKS 0139-09. 
Qualitatively, in the case of PG 1553+113, the transformer successfully captured the long-term trend, whereas the LOESS method overestimates it; in the case 
of S5 1044+71, the detail of the peak at about 60000 MJD is somewhat counterintuitively captured, as one may have expected a naive continuation of the previous pattern apparent by simple inspection of 
 the training set to lead to the broader peaked behavior that is indeed predicted by LOESS; the details of the peaks in the test set are similarly much better represented by the STLForecaster in the case of PKS 0139-09. 

\subsubsection{Forecast and predictions}
\label{sec:Forecast}

Given the encouraging results on the test set, we make predictions 
regarding future behavior. In general, the transformer predictions 
include a weaker future trend. For PG 1553+113, for instance, 
the QPO cycle starting around 2020 is expected to repeat without further enhancement of the mean signal. 
This comes in agreement with the hint of a 21.8 $\pm$ 4.7 yr oscillation found by \citealt{Penil2024-period22yr} and expected peak of the 22 yr oscillation expected to
occur around 2025 July.
A similar conclusion can be made regarding PKS 0454-234, with cycle starting from 2023.  In the case of PKS 2155-304, the 
initial transient high amplitude periodicities (that could be identified 
on the WWZ diagram in Fig.~\ref{fig:WWZ-LSP}), are expected to 
completely disappear,  with the smaller amplitude oscillations persisting. 
In the case of OJ 014, on the other hand, large flaring that appears to embody  a stochastic component is predicted.   For S5 1044+71, the aforementioned unexpected single peak, previously predicted for the test set, is expected to repeat. A similar expectation is predicted for 
PKS 0139-09, with the double peak starting at $\gtrsim$6000 MJD repeating.  

The expected future periods and their local significances from the application of the LSP method, along with simple power-law fit parameters and its reduced $\chi^2$, are summarized in Table \ref{LSP_orig_future}. This is for both original and detrended $\gamma$-ray future LCs of the six blazars considered, where ``future'' refers to the observed LC plus  future data as  predicted by the transformer technique. 

The periods of the inferred QPOs are compatible with the observed LCs. Without detrending we 
find only two cases where QPO signal's significance increases: the case of PG 1553+113, where the QPO signal appears exceptionally strong and is predicted to persist with high significance; and the new source studied here (PKS 0139-09), where the tentative nascent QPO is expected to strengthen significantly. The local and global significances of this latter source were $2.01\sigma$ and $0.48 \sigma$, respectively, which are comparable to those of the original-detrended LC. The future behavior of the LC of PKS 0139-09, in particular, will represent a definite test case for the existence and predictability of QPO.

As with the case of the observed light curves, detrending the extended 
LCs generally enhances the statistical significance of the periods. 
In particular, the two cases just mentioned (PG 1553+113 and PKS 0139-09), the local significance of the QPO increased from $3.5 \sigma$ and $1.6 \sigma$  (global significance of $0.07 \sigma$) in the original LC to >4.8 $\sigma$ and $3.2 \sigma$ (global significance of $2.02 \sigma$), respectively, in the detrended and predicted LC. 

The general features of our predictions can be 
outlined as follows. In the case of PG 1553+113, 
we expect persistent, strong QPO to continue. 
For cases that showed early transient behavior, 
namely PKS 0454-234 and PKS 2155-304, a suppressed QPO signal is expected 
to persist for the extrapolated light curve. 
On the other hand, we predict significant suppression of the QPO
signal, relative to that inferred from the observational data alone, 
in cases where  transience 
appeared near the end of the period covered by the 
observational data (OJ 014 and S5 1044+71). 
Finally, our results suggest significant strengthening of the QPO signal in significance in the case where a nascent QPO signal is apparent 
(PKS 0139-09).

\begingroup
\setlength{\tabcolsep}{2pt} 
\begin{longtable}{@{} l c c c c c c @{}}

\caption{Summary of periods and local significances inferred from observed and projected data, using the LSP, for both original and detrended light curves, along with simple power-law (PL) fit parameters and its reduced $\chi^2$.}\\

\toprule()
\multirow{2}{*}{\begin{minipage}[t]{1.5cm}\raggedright
Source Name
\end{minipage}} &
\multicolumn{3}{c}{Present Period (yr)} &
\multicolumn{3}{c}{Future Period (yr)} \\
\cmidrule(lr){2-4} \cmidrule(lr){5-7}
&
\ \ \ \ \ \ \ \ PL Par. &
Original LC & 
Detrended LC &
\ \ \ \ \ \ \ \ PL Par. &
Original LC & Detrended LC \\
\midrule()
\endhead

\bottomrule()
\endlastfoot

PG 1553+113 &
\centering
\begin{tabular}{@{}c@{}} $A = 0.003$ \\ $\Gamma = 1.26$ \\ $\chi^2_{\text{red}} = 2.8$ \end{tabular} &
${2.1}_{3.6\sigma}^{\pm 0.13}$ & ${2.1}_{4.5\sigma}^{\pm 0.12}$ &
\centering
\begin{tabular}{@{}c@{}} $A = 0.00901$ \\ $\Gamma = 1.10$ \\ $\chi^2_{\text{red}} = 2.8$ \end{tabular} &
${2.1}_{4.2\sigma}^{\pm0.12}$ & ${2.1}_{>4.8\sigma}^{\pm0.11}$ \\
\addlinespace[0.05cm]

PKS 0454$-$234 &
\centering
\begin{tabular}{@{}c@{}} $A = 0.011$ \\ $\Gamma = 1.22$ \\ $\chi^2_{\text{red}} = 1.7$ \end{tabular} &
${3.5}_{2.5\sigma}^{\pm 0.35}$ & ${3.6}_{2.6\sigma}^{\pm 0.4}$ &
\centering
\begin{tabular}{@{}c@{}} $A = 0.0207$ \\ $\Gamma = 1.11$ \\ $\chi^2_{\text{red}} = 2.5$ \end{tabular} &
${3.6}_{2.3\sigma}^{\pm0.34}$ & ${3.6}_{2.9\sigma}^{\pm0.73}$ \\
\addlinespace[0.05cm]

PKS 2155$-$304 &
\centering
\begin{tabular}{@{}c@{}} $A = 0.66$ \\ $\Gamma = 0.83$ \\ $\chi^2_{\text{red}} = 1.8$ \end{tabular} &
${1.7}_{2.6\sigma}^{\pm 0.13}$ & ${1.7}_{3.3\sigma}^{\pm 0.13}$ &
\centering
\begin{tabular}{@{}c@{}} $A = 0.568$ \\ $\Gamma = 0.49$ \\ $\chi^2_{\text{red}} = 1.6$ \end{tabular} &
${1.7}_{2.7\sigma}^{\pm0.18}$ & ${1.7}_{3.4\sigma}^{\pm0.18}$ \\
\addlinespace[0.05cm]

OJ 014 &
\centering
\begin{tabular}{@{}c@{}} $A = 0.027$ \\ $\Gamma = 0.97$ \\ $\chi^2_{\text{red}} = 1.3$ \end{tabular} &
${4.2}_{3.5\sigma}^{\pm 0.57}$ & ${4.2}_{4.2\sigma}^{\pm 0.53}$ &
\centering
\begin{tabular}{@{}c@{}} $A = 0.370$ \\ $\Gamma = 0.66$ \\ $\chi^2_{\text{red}} = 1.6$ \end{tabular} &
${4.1}_{2.9\sigma}^{\pm0.38}$ & ${4.0}_{2.9\sigma}^{\pm0.39}$ \\
\addlinespace[0.05cm]

S5 1044+71 &
\centering
\begin{tabular}{@{}c@{}} $A = 0.003$ \\ $\Gamma = 1.6$ \\ $\chi^2_{\text{red}} = 2.0$ \end{tabular} &
${3.1}_{4.4\sigma}^{\pm 0.34}$ & ${3.1}_{4.3\sigma}^{\pm 0.37}$ &
\centering
\begin{tabular}{@{}c@{}} $A = 0.0396$ \\ $\Gamma = 1.11$ \\ $\chi^2_{\text{red}} = 1.7$ \end{tabular} &
${3.2}_{3.3\sigma}^{\pm0.53}$ & ${3.2}_{3.6\sigma}^{\pm0.47}$ \\
\addlinespace[0.05cm]

PKS 0139$-$09 &
\centering
\begin{tabular}{@{}c@{}} $A = 0.035$ \\ $\Gamma = 1.02$ \\ $\chi^2_{\text{red}} = 2.0$ \end{tabular} &
${5.2}_{1.6\sigma}^{\pm 0.73}$ & ${5.1}_{2.0\sigma}^{\pm 0.96}$ &
\centering
\begin{tabular}{@{}c@{}} $A = 0.0200$ \\ $\Gamma = 1.07$ \\ $\chi^2_{\text{red}} = 3.8$ \end{tabular} &
${4.9}_{2.0\sigma}^{\pm0.75}$ & ${5.0}_{3.2\sigma}^{\pm0.81}$ 
\label{LSP_orig_future}
\end{longtable}
\endgroup

For more specific predictions, in the two cases where 
the QPO signal is expected to increase in 
significance---namely, PG 1553+113 and PKS 0139-09---we expect 
the upcoming periodic flaring to occur in 2027 April for PG 1553+113   
and  twin peaks in 2027 May and December in the case of PKS 0139-09. 
In the two cases where transience was detected
near the end of the observational period---namely, 
OJ 014 and S5 1044+71---we expect the novel behavior to lead to flaring 
in 2027 February and 2028 March in the case of OJ 014 and in 2027 October in the case of S5 1044+71.
The two remaining cases are expected to remain relatively 
quiescent, with smaller peaks at timescales 
consistent with the inferred periods.  

\section{Conclusion} 
\label{Discussion and conclusion}

We conducted a comprehensive temporal analysis of the $\gamma$-ray light curves of five blazars that were previously reported to exhibit high-significance (>$3 \sigma$)
quasiperiodic oscillations on year-long timescales.   
We extended previous analyses with further data (up to 2025 February), 
examining the persistence of the QPO signals, the existence of 
long-term trends, and their effect on the strengths of the signals. 
We furthermore forecast the light curves into the future, enabling 
predictions that may be tested against future data. 
The above procedures were also 
applied to one new candidate, the BL Lac PKS 0139-09, where 
nascent QPOs may be appearing.  

To search for periodicities, we initially 
employed the LSP (Section~\ref{sec:LSP}). 
The inferred periods are consistent with, or within the uncertainties of those previously reported in the literature. 
However, in general,  results for our extended timespan do not show a significant  increase  
in the QPO signal's strength, but more 
often the reverse. The clear exception 
being the case of S5 1044+71, where we find a QPO 
significance of 4.4$\sigma$, which is larger 
than the value 3.6$\sigma$ previously found by~\cite{Wang2022}. 

The observed LCs were also subjected to ARIMA/SARIMA (Section~\ref{arima-sarima}) and folded light curve (Section~\ref{folded-lc}) analyses. The ARIMA/SARIMA results indicate that the LCs of PG~1553+113 and S5~1044+71 are better represented by a seasonal ARIMA model, whereas those of the remaining four blazars are best described as stochastic processes modeled by nonseasonal ARIMA models. The folded light-curve analysis further shows that, for all sources, the detected period lies within the period uncertainty identified by the LSP method, with S5~1044+71 exhibiting the highest variation ($\chi^2_{\mathrm{red}} = 330.1$).

The general lack of consolidation of 
QPO signal confidences may be due to the effect of long-term trends in the data, which may themselves reflect longer-term periodicities, or  
due to intrinsic transience in the principal QPO signal. 
We studied both effects by detrending through  
an STL decomposition (Section~\ref{lc_detrending_method}), and by using the WWZ wavelet technique to look for transients 
(Section~\ref{sec:wwz_method}).  

Long-term trends  
are associated with significant red noise, 
which can suppress the detectability of the principal periodic signals.
A key aspect of our analysis addressed its confounding effects, 
with the central finding being that detrending  can in some cases significantly improve the significance of the principal QPO periods
(Table~\ref{LSP_WWZ}). The exceptions to this were the sources 
S5 1044+71, where the low-frequency signal associated with the trend is very small compared to the 
principal peak; 
and PKS 0454-234, where the lower-frequency peak is close to the 
principal peak (Fig.~\ref{fig:WWZ-LSP}). A notably 
remarkable feature of this latter case is the repetition 
of this double peaked structure at smaller periods, some of which 
are consistent with being harmonics of the principal period. These peaks are ` `secondary'' in the sense that they are of smaller height. But they  can, in fact, come with higher significance, as they correspond to more oscillatory cycles. Significant secondary 
peaks are also 
present in other cases (PG 1553+113, PKS 2155-304, and PKS 0139-09). 
A first harmonic has recently been 
confirmed in the case of PKS J1309+1154  by~\cite{RedheadHarmonic2026ApJ}. 

If harmonic peaks may indeed be detected by analyzing LCs, 
it seems possible that the principal peaks themselves, which are usually associated with the characteristic times of the physical process leading to the QPO, are 
in reality harmonics of a longer principal period. The long-term
 modulations represented by the inferred trends suggest 
that larger periods may indeed be present. 
The presence of significant harmonics (or near harmonics) may 
contribute to an  
elevated seasonal component in the STL 
decomposition in some of the light curves (Fig.~\ref{fig_decomposed_LCs}), 
even when the primary peak significance is relatively 
weak.  

In a couple of cases (PKS 0454-234 and  PKS 2155-304), even  with detrending, the QPO signal 
significances are markedly lower than found in previous studies. In these cases, clear transients are present in 
the data early on in the observing period.  
Indeed, examination of the WWZ diagrams 
reveals clear transient behavior
even for the detrended data; 
with  significant weakening found 
well within the observed timespan, 
on a timescale of $\lesssim$4000
days for PKS 0454-234, and  within $\sim$3000 days for 
PKS 2155-304. 

In fact, except for 
PG 1553+113,
which displays a persistently strong QPO signal, all sources considered here show transient behavior on 
a timescale $\lesssim$4000 days. 
While transience in the two sources discussed above 
results in the suppression of the QPO 
signal well within the observed timescale, 
in two other cases (OJ 014 and S5 1044+71),
the signal is attenuated near 
the end of that timespan. In the case
S5 1044+71 the QPO signal appears weaker not 
only near the end of the observational period 
but also at its start.
A weak start is also apparent in the case of 
the new source, PKS 0139-09. But 
here the significance of the QPO 
steadily increases within the whole
observational period. 
In this case, one may thus expect the peak associated with  the 
$\lesssim$4000 day QPO mean life to appear in the 
coming years. To make such statements 
more precise, and make testable predictions, we introduced future state forecasting. 

We used  a statistical model, the STLForecaster, as well as  a deep learning transformer architecture, to project the blazars’ data four years into the future (Section~\ref{lcForecasting}). 
Applying the methods on a test set we find that 
the  transformer outperformed the statistical STLForecaster in prediction accuracy. It is successful in capturing complex, non-linear long-term trends; as evidenced, for example, by its successful prediction of a trend saturation in PG 1553+113 (in agreement with the hint of a 21.8 $\pm$ 4.7 yr oscillation found by \citealt{Penil2024-period22yr}), contrary to the recent historical ($\approx$10 yr) upward trend, as well as unexpected flaring in the cases of S5 1044+71
and PKS 0139-09 (Fig.~\ref{lcForecasting}). 
Table~\ref{LSP_orig_future}
lists the values and expected significance of the principal periods.   

Sustained oscillations are 
predicted to continue in the future in the case of  PG 1553+113.
Indeed, here the significance of the 
2.1 year period increased from $3.6 \sigma$ in the original light curve 
to >4.8$ \sigma$, when the data was forecast into the future and then detrended. Also, the principal candidate signal in the new BL Lac object PKS 0139-09, which was only marginally detectable (local significance $1.6\sigma$/global significance $0.07 \sigma$) in the original data showed its local significance rise to $3.2 \sigma$ (global significance $2.02 \sigma$) in the detrended extrapolated curve.
Thus, a solid testable prediction  for this new source is the strengthening of the nascent QPO signal in the coming years. 

On the other hand, for the cases where the QPO signal was significantly attenuated near the end of the observed data timescale---namely, in the cases of OJ 014 and S5 1044+71---the late-time suppression is expected to persist. This is  reflected in the smaller significance of the QPOs in 
the time-extrapolated light curves compared 
to the originals. The persistence of late-time behavior  
is also expected for the cases where the QPO signal 
significantly died down well within  
the observed timespan (PKS 0454-234 and PKS 2155-304).  
As the transient behavior here occurs early on, the expectation is for the QPO significance to remain  at a  similar level in the extrapolated curves. This reflects the  
prediction of a new quasi-stationary state that persists 
through the extrapolation period.  

In addition to these generic expected trends, we make specific 
predictions regarding the future behavior, 
particularly regarding the dates of upcoming flarings, 
in the light curves of the objects studied here. 
These can potentially guide future observations and data 
analyses, against which they may be tested.

Our inferences and predictions regarding the 
persistence of year-long periodic signals have
implications for their theoretical interpretations. 
While no single widely accepted theory yet exists to describe the physical mechanisms that may be responsible for QPOs in blazars, a variety of models  have 
been proposed to account for apparent year-long periodicities detected in the data
(e.g., \citealp{abdollahi2024periodic} Section~4.1, for a recent summary).  
These may differ especially in terms 
of the longer-term trends and transient behavior studied here.  For example, as mentioned in the introduction,  
ballistic blobs in the curved jet model should lead to transient 
decaying signals. On the other hand, purely geometric models invoking 
jet precession in a binary  
on a circular orbit 
(e.g., \citealp{Sobacchi-2017}) may not account for the transient behavior inferred 
in this study. 
Likewise with purely rigid precession of a 
jet driven by torques (\citealp{KatzPrecess1997ApJ}) 
or direct modulation of the accretion flow 
rate on the SMBH due to torques from a companion, as originally 
proposed by \cite{SinapoaSMBHBOrig1988ApJ}. 
Nevertheless, as  they show, 
the inclusion of a disk potential leads
to orbital precession that can be associated with
longer term modulations, 
possibly compatible with the more complex patterns 
displayed by most systems considered here.
Nontrivial time dependence  may also ensue from  
tidal interactions between SMBB disks. 
Additional flaring, with possible signature signals in 
specific wavelengths, can result from passage of an SMBH through the accretion
disks of the other (\citealp{optflarepassagedisk2000ApJ}). 
Double jetted models with interactions between the jets 
may furthermore lead to double peaked periodic patterns (\citealp{DoubleJetted1998MNRAS}).   
In addition to simple precession, perturbations
of the jet in an SMBHB  can lead to furthermore instigate 
complex magnetohydrodynamic processes 
(\citealp{2017Cavaliere, Precesssimul2023ApJ}).
Such precession and associated QPOs may be present even in the absence 
of a secondary SMBH, notably through  
Lense-Thirring precession of tilted accretion disks (\citealp{Bardeen-Peterson1975, FragileLenseT2007ApJ}). 

Thus, except possibly for the case 
of PG 1553+113, the simplest 
geometric models appear ruled out 
as sources of QPO by our analysis 
of the past light curves and predictions for the next 
years. The time dependencies inferred in this study 
favor richer scenarios, where further physical 
processes are triggered (by the presence of a
companion SMBH or otherwise). 
Detailed simulations 
coupled with 
Multiwavelength observations may serve to distinguish between such scenarios. Integrating multifrequency observations will also help increase the accuracy of the neural network in forecasting future behavior.

\section*{Acknowledgements}
The authors thanks Waleed Esmail for the fruitful discussion about the DL analysis. A.H. is funded by grant No. 22H05113, ``Foundation of Machine Learning Physics,'' Grant in Aid for Transformative Research Areas and 22K03626, Grant-in-Aid for Scientific Research (C). A.H. is partially supported by the Science, Technology and Innovation Funding Authority (STDF) under grant No. 50806.

\section*{Software}
Fermitools-conda, DELCgen-Simulating light curves \citep{Connolly2015}, NumPy \citep{Harris2020},
SciPy \citep{SciPy2020},
Matplotlib \citep{Hunter2007},
Statsmodels \citep{seabold2010statsmodels},
sktime \citep{2019SKtime},
PyTorch \citep{Pytorch2017},
astroML \citep{Ivezic2014}. 
For the purpose of open science, all code and analysis pipelines are available from a public GitHub repository\footnote{\href{https://github.com/mohamed-hashad/Potential-periodic-signals-in-blazars-significance-forecasting-and-deep-learning}{GitHub repository}} and Zenodo \citep{Hashad2026Zenodo}.

\bibliography{biblo}

@article{Abhir2021,
   author = {Abhir, Jayant and Joseph, Jophin and Patel, Sonal R and Bose, Debanjan},
   title = {Multi-frequency temporal and spectral variability study of blazar PKS 1424-418},
   journal = {Monthly Notices of the Royal Astronomical Society},
   volume = {501},
   number = {2},
   pages = {2504-2511},
   ISSN = {0035-8711},
   year = {2021},
   type = {Journal Article},
   url={https://doi.org/10.1093/mnras/staa3639
}
}

@article{Ackermann2015,
   author = {Ackermann, M and Ajello, M and Albert, A and Atwood, WB and Baldini, LUCA and Ballet, J and Barbiellini, G and Bastieri, D and Gonzalez, J Becerra and Bellazzini, R},
   title = {Multiwavelength evidence for quasi-periodic modulation in the gamma-ray blazar PG 1553+ 113},
   journal = {The Astrophysical Journal Letters},
   volume = {813},
   number = {2},
   pages = {L41},
   ISSN = {2041-8205},
   year = {2015},
   type = {Journal Article}
}

@article{Atwood2009,
   author = {Atwood, WB and Abdo, Aous A and Ackermann, Markus and Althouse, W and Anderson, B and Axelsson, M and Baldini, Luca and Ballet, J and Band, DL and Barbiellini, Guido},
   title = {The large area telescope on the Fermi gamma-ray space telescope mission},
   journal = {The Astrophysical Journal},
   volume = {697},
   number = {2},
   pages = {1071},
   ISSN = {0004-637X},
   year = {2009},
   type = {Journal Article}
}

@article{Begelman1980,
   author = {Begelman, Mitchell C and Blandford, Roger D and Rees, Martin J},
   title = {Massive black hole binaries in active galactic nuclei},
   journal = {Nature},
   volume = {287},
   number = {5780},
   pages = {307-309},
   ISSN = {1476-4687},
   year = {1980},
   type = {Journal Article},
   url = {https://doi.org/10.1038/287307a0}
}

@article{Benkhali2020,
   author = {Benkhali, F Ait and Hofmann, W and Rieger, FM and Chakraborty, Nachiketa},
   title = {Evaluating quasi-periodic variations in the Gamma-ray light curves of Fermi-LAT blazars},
   journal = {Astronomy \& Astrophysics},
   volume = {634},
   pages = {A120},
   ISSN = {0004-6361},
   year = {2020},
   type = {Journal Article},
   url = {https://doi.org/10.1051/0004-6361/201935117}
}

@article{Bhatta2020,
   author = {Bhatta, Gopal and Dhital, Niraj},
   title = {The nature of $\gamma$-ray variability in blazars},
   journal = {The Astrophysical Journal},
   volume = {891},
   number = {2},
   pages = {120},
   ISSN = {0004-637X},
   year = {2020},
   type = {Journal Article},
   url = {https://doi.org/10.3847/1538-4357/ab7455}
}

@article{Blandford2019,
   author = {Blandford, Roger and Meier, David and Readhead, Anthony},
   title = {Relativistic jets from active galactic nuclei},
   journal = {Annual Review of Astronomy and Astrophysics},
   volume = {57},
   pages = {467-509},
   ISSN = {0066-4146},
   year = {2019},
   type = {Journal Article},
   url = {https://doi.org/10.1146/annurev-astro-081817-051948}
}

@article{Connolly2015,
   author = {Connolly, SD},
   title = {A Python Code for the Emmanoulopoulos et al.[arXiv: 1305.0304] Light Curve Simulation Algorithm},
   journal = {arXiv e-prints},
   pages = {arXiv: 1503.06676},
   year = {2015},
   type = {Journal Article}
}

@article{Emmanoulopoulos2013,
   author = {Emmanoulopoulos, D and McHardy, IM and Papadakis, IE},
   title = {Generating artificial light curves: revisited and updated},
   journal = {Monthly Notices of the Royal Astronomical Society},
   volume = {433},
   number = {2},
   pages = {907-927},
   ISSN = {1365-2966},
   year = {2013},
   type = {Journal Article},
   url = {https://doi.org/10.1093/mnras/stt764}
}

@article{Foster1996,
   author = {Foster, Grant},
   title = {Wavelets for period analysis of unevenly sampled time series},
   journal = {The Astronomical Journal},
   volume = {112},
   pages = {1709-1729},
   ISSN = {0004-6256},
   year = {1996},
   type = {Journal Article},
   url = {https://doi.org/10.1086/118137}
}

@article{Harris2020,
   author = {Harris, Charles R and Millman, K Jarrod and Van Der Walt, Stéfan J and Gommers, Ralf and Virtanen, Pauli and Cournapeau, David and Wieser, Eric and Taylor, Julian and Berg, Sebastian and Smith, Nathaniel J},
   title = {Array programming with NumPy},
   journal = {Nature},
   volume = {585},
   number = {7825},
   pages = {357-362},
   ISSN = {1476-4687},
   year = {2020},
   type = {Journal Article}
}

@article{Hashad2023,
   author = {Hashad, MA and EL-Zant, Amr A and Abdou, Y},
   title = {Quasi-periodic variability in the $\gamma$-Ray blazar PKS 0426–380},
   journal = {Advances in Space Research},
   volume = {72},
   number = {8},
   pages = {3538-3549},
   ISSN = {0273-1177},
   year = {2023},
   type = {Journal Article},
   url = {https://doi.org/10.1016/j.asr.2023.06.042}
}

@article{Hunter2007,
   author = {Hunter, John D},
   title = {Matplotlib: A 2D graphics environment},
   journal = {Computing in science \& engineering},
   volume = {9},
   number = {03},
   pages = {90-95},
   ISSN = {1521-9615},
   year = {2007},
   type = {Journal Article},
   url = {https://doi.org/10.1109/MCSE.2007.55}
}

@inbook{Ivezic2014,
   author = {Ivezić, zeljko and Connolly, Andrew J and VanderPlas, Jacob T and Gray, Alexander},
   title = {Statistics, data mining, and machine learning in astronomy},
   booktitle = {Statistics, Data Mining, and Machine Learning in Astronomy},
   publisher = {Princeton University Press},
   ISBN = {1400848911},
   pages = {403-468},
   year = {2014},
   type = {Book Section},
   url = {https://doi.org/10.23943/princeton/9780691151687.001.0001}
}

@article{Lomb1976,
   author = {Lomb, Nicholas R},
   title = {Least-squares frequency analysis of unequally spaced data},
   journal = {Astrophysics and space science},
   volume = {39},
   number = {2},
   pages = {447-462},
   ISSN = {1572-946X},
   year = {1976},
   type = {Journal Article},
   url = {https://doi.org/10.1007/BF00648343}
}

@article{Mohan2015,
   author = {Mohan, Prashanth and Mangalam, A},
   title = {Kinematics of and emission from helically orbiting blobs in a relativistic magnetized jet},
   journal = {The Astrophysical Journal},
   volume = {805},
   number = {2},
   pages = {91},
   ISSN = {0004-637X},
   year = {2015},
   type = {Journal Article},
   url = {https://doi.org/10.1088/0004-637X/805/2/91}
}

@article{Penil2020,
   author = {Peñil, P and Domínguez, A and Buson, S and Ajello, M and Otero-Santos, J and Barrio, JA and Nemmen, R and Cutini, S and Rani, B and Franckowiak, A},
   title = {Systematic search for $\gamma$-ray periodicity in active galactic nuclei detected by the Fermi Large Area Telescope},
   journal = {The Astrophysical Journal},
   volume = {896},
   number = {2},
   pages = {134},
   ISSN = {0004-637X},
   year = {2020},
   type = {Journal Article},
   url = {https://doi.org/10.3847/1538-4357/ab910d}
}

@article{Prokhorov2017,
   author = {Prokhorov, DA and Moraghan, A},
   title = {A search for cyclical sources of $\gamma$-ray emission on the period range from days to years in the Fermi-LAT sky},
   journal = {Monthly Notices of the Royal Astronomical Society},
   volume = {471},
   number = {3},
   pages = {3036-3042},
   ISSN = {0035-8711},
   year = {2017},
   type = {Journal Article},
   url = {https://doi.org/10.1093/mnras/stx1742}
}

@article{Rieger2004,
   author = {Rieger, Frank M},
   title = {On the geometrical origin of periodicity in blazar-type sources},
   journal = {The Astrophysical Journal},
   volume = {615},
   number = {1},
   pages = {L5},
   ISSN = {0004-637X},
   year = {2004},
   type = {Journal Article},
   url = {https://doi.org/10.1086/426018}
}

@article{Rieger2019,
   author = {Rieger, Frank M},
   title = {Gamma-Ray Astrophysics in the Time Domain},
   journal = {Galaxies},
   volume = {7},
   number = {1},
   pages = {28},
   year = {2019},
   type = {Journal Article},
   url = {https://doi.org/10.3390/galaxies701002}
}

@article{Sandrinelli2018,
   author = {Sandrinelli, Angela and Covino, Stefano and Treves, Aldo and Holgado, A Miguel and Sesana, Alberto and Lindfors, Elina and Ramazani, V Fallah},
   title = {Quasi-periodicities of BL Lacertae objects},
   journal = {Astronomy \& Astrophysics},
   volume = {615},
   pages = {A118},
   ISSN = {0004-6361},
   year = {2018},
   type = {Journal Article},
   url = {https://doi.org/10.1051/0004-6361/201732550}
}

@article{Scargle1982,
   author = {Scargle, Jeffrey D},
   title = {Studies in astronomical time series analysis. II-Statistical aspects of spectral analysis of unevenly spaced data},
   journal = {The Astrophysical Journal},
   volume = {263},
   pages = {835-853},
   ISSN = {0004-637X},
   year = {1982},
   type = {Journal Article},
   url = {https://doi.org/10.1086/160554}
}

@ARTICLE{Sobacchi-2017,
       author = {{Sobacchi}, Emanuele and {Sormani}, Mattia C. and {Stamerra}, Antonio},
        title = "{A model for periodic blazars}",
      journal = {\mnras},
         year = 2017,
        month = feb,
       volume = {465},
       number = {1},
        pages = {161-172},
          doi = {10.1093/mnras/stw2684},
archivePrefix = {arXiv},
       eprint = {1610.04709},
 primaryClass = {astro-ph.HE},
       adsurl = {https://ui.adsabs.harvard.edu/abs/2017MNRAS.465..161S}
}

@ARTICLE{Bardeen-Peterson1975,
       author = {{Bardeen}, James M. and {Petterson}, Jacobus A.},
        title = "{The Lense-Thirring Effect and Accretion Disks around Kerr Black Holes}",
      journal = {\apjl},
         year = 1975,
        month = jan,
       volume = {195},
        pages = {L65},
          doi = {10.1086/181711},
       adsurl = {https://ui.adsabs.harvard.edu/abs/1975ApJ...195L..65B}
}

@ARTICLE{KatzPrecess1997ApJ,
       author = {{Katz}, J.~I.},
        title = "{A Precessing Disk in OJ 287?}",
      journal = {\apj},
         year = 1997,
        month = mar,
       volume = {478},
       number = {2},
        pages = {527-529},
          doi = {10.1086/303811},
       adsurl = {https://ui.adsabs.harvard.edu/abs/1997ApJ...478..527K}
}

@ARTICLE{Precesssimul2023ApJ,
       author = {{Nolting}, Chris and {Ball}, Jay and {Nguyen}, Tri M.},
        title = "{Simulations of Precessing Jets and the Formation of X-shaped Radio Galaxies}",
      journal = {\apj},
         year = 2023,
        month = may,
       volume = {948},
       number = {1},
          eid = {25},
        pages = {25},
          doi = {10.3847/1538-4357/acc652},
archivePrefix = {arXiv},
       eprint = {2301.04343},
 primaryClass = {astro-ph.HE},
       adsurl = {https://ui.adsabs.harvard.edu/abs/2023ApJ...948...25N}
}

@ARTICLE{SinapoaSMBHBOrig1988ApJ,
       author = {{Sillanpaa}, A. and {Haarala}, S. and {Valtonen}, M.~J. and {Sundelius}, B. and {Byrd}, G.~G.},
        title = "{OJ 287: Binary Pair of Supermassive Black Holes}",
      journal = {\apj},
         year = 1988,
        month = feb,
       volume = {325},
        pages = {628},
          doi = {10.1086/166033},
       adsurl = {https://ui.adsabs.harvard.edu/abs/1988ApJ...325..628S}
}

@ARTICLE{FragileLenseT2007ApJ,
       author = {{Fragile}, P. Chris and {Blaes}, Omer M. and {Anninos}, Peter and {Salmonson}, Jay D.},
        title = "{Global General Relativistic Magnetohydrodynamic Simulation of a Tilted Black Hole Accretion Disk}",
      journal = {\apj},
         year = 2007,
        month = oct,
       volume = {668},
       number = {1},
        pages = {417-429},
          doi = {10.1086/521092},
archivePrefix = {arXiv},
       eprint = {0706.4303},
 primaryClass = {astro-ph},
       adsurl = {https://ui.adsabs.harvard.edu/abs/2007ApJ...668..417F}
}

@ARTICLE{Chen-2024,
       author = {{Chen}, Yutong and {Yi}, Tingfeng and {Chen}, Junping and {Lu}, He and {Shen}, Yuncai and {Wang}, Junjie and {Wang}, Liang and {Zhang}, Shun and {Mao}, Lisheng and {Dong}, Liang},
        title = "{Revisiting the quasi-periodic oscillations in blazar PG 1553＋113 with multi-wavebands data}",
      journal = {\na},
         year = 2024,
        month = may,
       volume = {108},
          eid = {102186},
        pages = {102186},
          doi = {10.1016/j.newast.2023.102186},
       adsurl = {https://ui.adsabs.harvard.edu/abs/2024NewA..10802186C}
}

@ARTICLE{Sandrinelli-2014,
       author = {{Sandrinelli}, A. and {Covino}, S. and {Treves}, A.},
        title = "{Quasi-periodicities of the BL Lacertae Object PKS 2155-304}",
      journal = {\apjl},
         year = 2014,
        month = sep,
       volume = {793},
       number = {1},
          eid = {L1},
        pages = {L1},
          doi = {10.1088/2041-8205/793/1/L1},
archivePrefix = {arXiv},
       eprint = {1408.0015},
 primaryClass = {astro-ph.HE},
       adsurl = {https://ui.adsabs.harvard.edu/abs/2014ApJ...793L...1S}
}

@article{Tavani2018,
   author = {Tavani, Marco and Cavaliere, Alfonso and Munar-Adrover, Pere and Argan, Andrea},
   title = {The blazar PG 1553+ 113 as a binary system of supermassive black holes},
   journal = {The Astrophysical Journal},
   volume = {854},
   number = {1},
   pages = {11},
   ISSN = {0004-637X},
   year = {2018},
   type = {Journal Article},
   url = {https://doi.org/10.3847/1538-4357/aaa3f4}
}

@article{Urry1995,
   author = {Urry, C Megan and Padovani, Paolo},
   title = {Unified schemes for radio-loud active galactic nuclei},
   journal = {Publications of the Astronomical Society of the Pacific},
   volume = {107},
   number = {715},
   pages = {803},
   ISSN = {1538-3873},
   year = {1995},
   type = {Journal Article},
   url = {https://doi.org/10.1086/133630}
}

@article{Ulrich_1997, title={VARIABILITY OF ACTIVE GALACTIC NUCLEI}, volume={35}, ISSN={1545-4282}, url={http://dx.doi.org/10.1146/annurev.astro.35.1.445}, DOI={10.1146/annurev.astro.35.1.445}, number={1}, journal={Annual Review of Astronomy and Astrophysics}, publisher={Annual Reviews}, author={Ulrich, Marie-Helene and Maraschi, Laura and Urry, C. Megan}, year={1997}, month=sep, pages={445–502} }

@article{Madejski_2016, title={Gamma-Ray Observations of Active Galactic Nuclei}, volume={54}, ISSN={1545-4282}, url={http://dx.doi.org/10.1146/annurev-astro-081913-040044}, DOI={10.1146/annurev-astro-081913-040044}, number={1}, journal={Annual Review of Astronomy and Astrophysics}, publisher={Annual Reviews}, author={Madejski, Grzegorz (Greg) and Sikora, Marek}, year={2016}, month=sep, pages={725–760} }

@article{Urry_2011, title={Gamma-Ray and Multiwavelength Emission from Blazars}, volume={32}, ISSN={0973-7758}, url={http://dx.doi.org/10.1007/s12036-011-9072-x}, DOI={10.1007/s12036-011-9072-x}, number={1–2}, journal={Journal of Astrophysics and Astronomy}, publisher={Springer Science and Business Media LLC}, author={Urry, Meg}, year={2011}, month=jun, pages={139–145} }

@article{VanderPlas2018,
   author = {VanderPlas, Jacob T},
   title = {Understanding the lomb–scargle periodogram},
   journal = {The Astrophysical Journal Supplement Series},
   volume = {236},
   number = {1},
   pages = {16},
   ISSN = {0067-0049},
   year = {2018},
   type = {Journal Article},
   url = {https://doi.org/10.3847/1538-4365/aab766}
}

@article{Vaughan2016,
   author = {Vaughan, S and Uttley, P and Markowitz, AG and Huppenkothen, D and Middleton, MJ and Alston, WN and Scargle, JD and Farr, WM},
   title = {False periodicities in quasar time-domain surveys},
   journal = {Monthly Notices of the Royal Astronomical Society},
   volume = {461},
   number = {3},
   pages = {3145-3152},
   ISSN = {1365-2966},
   year = {2016},
   type = {Journal Article},
   url = {https://doi.org/10.1093/mnras/stw1412}
}

@article{Wang2022,
   author = {Wang, GG and Cai, JT and Fan, JH},
   title = {A Possible 3 yr Quasi-periodic Oscillation in $\gamma$-Ray Emission from the FSRQ S5 1044+ 71},
   journal = {The Astrophysical Journal},
   volume = {929},
   number = {2},
   pages = {130},
   ISSN = {0004-637X},
   year = {2022},
   type = {Journal Article},
   url = {https://doi.org/10.3847/1538-4357/ac5b08}
}

@article{Zechmeister2009,
   author = {Zechmeister, M and Kürster, M},
   title = {The generalised Lomb-Scargle periodogram-a new formalism for the floating-mean and Keplerian periodograms},
   journal = {Astronomy \& Astrophysics},
   volume = {496},
   number = {2},
   pages = {577-584},
   ISSN = {0004-6361},
   year = {2009},
   type = {Journal Article},
   url = {https://doi.org/10.1051/0004-6361:200811296}
}

@article{Ren_2023, title={Quasi-periodic oscillations in the Gamma-ray light curves of bright active galactic nuclei}, volume={672}, ISSN={1432-0746}, url={http://dx.doi.org/10.1051/0004-6361/202244754}, DOI={10.1051/0004-6361/202244754}, journal={Astronomy \& amp; Astrophysics}, publisher={EDP Sciences}, author={Ren, Helena X. and Cerruti, Matteo and Sahakyan, Narek}, year={2023}, month=apr, pages={A86} }

@ARTICLE{PKS304-2017,
       author = {{Zhang}, Peng-fei and {Yan}, Da-hai and {Liao}, Neng-hui and {Wang}, Jian-cheng},
        title = "{Revisiting Quasi-periodic Modulation in {\ensuremath{\gamma}}-Ray Blazar PKS 2155-304 with Fermi Pass 8 Data}",
      journal = {\apj},
         year = 2017,
        month = feb,
       volume = {835},
       number = {2},
          eid = {260},
        pages = {260},
          doi = {10.3847/1538-4357/835/2/260},
archivePrefix = {arXiv},
       eprint = {1611.04354},
 primaryClass = {astro-ph.HE},
       adsurl = {https://ui.adsabs.harvard.edu/abs/2017ApJ...835..260Z}
}

@inproceedings{Wood_2017, series={ICRC2017}, title={Fermipy: An open-source Python package for analysis of Fermi-LAT Data}, url={http://dx.doi.org/10.22323/1.301.0824}, DOI={10.22323/1.301.0824}, booktitle={Proceedings of 35th International Cosmic Ray Conference — PoS(ICRC2017)}, publisher={Sissa Medialab}, author={Wood, Matthew and Caputo, Regina and Charles, Eric and Di Mauro, Mattia and Magill, Jeffrey and Perkins, Jeremy S.}, year={2017}, month=aug, collection={ICRC2017} }

@article{Abdollahi_2022_4FGL-DR3,
doi = {10.3847/1538-4365/ac6751},
url = {https://dx.doi.org/10.3847/1538-4365/ac6751},
year = {2022},
month = {jun},
publisher = {The American Astronomical Society},
volume = {260},
number = {2},
pages = {53},
author = {S. Abdollahi and F. Acero and L. Baldini and J. Ballet and D. Bastieri and R. Bellazzini and B. Berenji and A. Berretta and E. Bissaldi and R. D. Blandford and E. Bloom and R. Bonino and A. Brill and R. J. Britto and P. Bruel and T. H. Burnett and S. Buson and R. A. Cameron and R. Caputo and P. A. Caraveo and D. Castro and S. Chaty and C. C. Cheung and G. Chiaro and N. Cibrario and S. Ciprini and J. Coronado-Blázquez and M. Crnogorcevic and S. Cutini and F. D’Ammando and S. De Gaetano and S. W. Digel and N. Di Lalla and F. Dirirsa and L. Di Venere and A. Domínguez and V. Fallah Ramazani and S. J. Fegan and E. C. Ferrara and A. Fiori and H. Fleischhack and A. Franckowiak and Y. Fukazawa and S. Funk and P. Fusco and G. Galanti and V. Gammaldi and F. Gargano and S. Garrappa and D. Gasparrini and F. Giacchino and N. Giglietto and F. Giordano and M. Giroletti and T. Glanzman and D. Green and I. A. Grenier and M.-H. Grondin and L. Guillemot and S. Guiriec and M. Gustafsson and A. K. Harding and E. Hays and J. W. Hewitt and D. Horan and X. Hou and G. Jóhannesson and C. Karwin and T. Kayanoki and M. Kerr and M. Kuss and D. Landriu and S. Larsson and L. Latronico and M. Lemoine-Goumard and J. Li and I. Liodakis and F. Longo and F. Loparco and B. Lott and P. Lubrano and S. Maldera and D. Malyshev and A. Manfreda and G. Martí-Devesa and M. N. Mazziotta and I. Mereu and M. Meyer and P. F. Michelson and N. Mirabal and W. Mitthumsiri and T. Mizuno and A. A. Moiseev and M. E. Monzani and A. Morselli and I. V. Moskalenko and M. Negro and E. Nuss and N. Omodei and M. Orienti and E. Orlando and D. Paneque and Z. Pei and J. S. Perkins and M. Persic and M. Pesce-Rollins and V. Petrosian and R. Pillera and H. Poon and T. A. Porter and G. Principe and S. Rainò and R. Rando and B. Rani and M. Razzano and S. Razzaque and A. Reimer and O. Reimer and T. Reposeur and M. Sánchez-Conde and P. M. Saz Parkinson and L. Scotton and D. Serini and C. Sgrò and E. J. Siskind and D. A. Smith and G. Spandre and P. Spinelli and K. Sueoka and D. J. Suson and H. Tajima and D. Tak and J. B. Thayer and D. J. Thompson and D. F. Torres and E. Troja and J. Valverde and K. Wood and G. Zaharijas},
title = {Incremental Fermi Large Area Telescope Fourth Source Catalog},
journal = {The Astrophysical Journal Supplement Series}}

@article{Penil_2024, title={Multiwavelength variability analysis of Fermi-LAT blazars}, volume={529}, ISSN={1365-2966}, url={http://dx.doi.org/10.1093/mnras/stae594}, DOI={10.1093/mnras/stae594}, number={2}, journal={Monthly Notices of the Royal Astronomical Society}, publisher={Oxford University Press (OUP)}, author={Peñil, P and Otero-Santos, J and Ajello, M and Buson, S and Domínguez, A and Marcotulli, L and Torres−Albà, N and González, J Becerra and Acosta-Pulido, J A}, year={2024}, month=feb, pages={1365–1385} }

@ARTICLE{stl,
       author = {{Cleveland}, B},
        title = "{STL: A seasonal-trend decomposition procedure based on loess}",
      journal = {J Off Stat},
         year = 1990,
       volume = {6},
        pages = {3-73},
}

@ARTICLE{2022-Rueda,
       author = {{Rueda}, H{\'e}ctor and {Glicenstein}, Jean-Fran{\c{c}}ois and {Brun}, Fran{\c{c}}ois},
        title = "{Search for Periodicities in High Energy AGNs with a Time Domain Approach}",
      journal = {\apj},
         year = 2022,
        month = jul,
       volume = {934},
       number = {1},
          eid = {6},
        pages = {6},
          doi = {10.3847/1538-4357/ac771c},
archivePrefix = {arXiv},
       eprint = {2206.07614},
 primaryClass = {astro-ph.HE},
       adsurl = {https://ui.adsabs.harvard.edu/abs/2022ApJ...934....6R}
}

@article{Tarnopolski_2020, title={A Comprehensive Power Spectral Density Analysis of Astronomical Time Series. I. The Fermi-LAT Gamma-Ray Light Curves of Selected Blazars}, volume={250}, ISSN={1538-4365}, url={http://dx.doi.org/10.3847/1538-4365/aba2c7}, DOI={10.3847/1538-4365/aba2c7}, number={1}, journal={The Astrophysical Journal Supplement Series}, publisher={American Astronomical Society}, author={Tarnopolski, Mariusz and Żywucka, Natalia and Marchenko, Volodymyr and Pascual-Granado, Javier}, year={2020}, month=aug, pages={1} }

@article{Goyal_2018, title={A Comparative Study of Multiwavelength Blazar Variability on Decades to Minutes Timescales}, volume={6}, ISSN={2075-4434}, url={http://dx.doi.org/10.3390/galaxies6010034}, DOI={10.3390/galaxies6010034}, number={1}, journal={Galaxies}, publisher={MDPI AG}, author={Goyal, Arti}, year={2018}, month=mar, pages={34} }

@article{Goyal_2017, title={Multiwavelength Variability Study of the Classical BL Lac Object PKS 0735+178 on Timescales Ranging from Decades to Minutes}, volume={837}, ISSN={1538-4357}, url={http://dx.doi.org/10.3847/1538-4357/aa6000}, DOI={10.3847/1538-4357/aa6000}, number={2}, journal={The Astrophysical Journal}, publisher={American Astronomical Society}, author={Goyal, Arti and Stawarz, Lukasz and Ostrowski, Michał and Larionov, Valeri and Gopal-Krishna and Wiita, Paul J. and Joshi, Santosh and Soida, Marian and Agudo, Iván}, year={2017}, month=mar, pages={127} }

@ARTICLE{1978-Press,
       author = {{Press}, W.~H.},
        title = "{Flicker noises in astronomy and elsewhere.}",
      journal = {Comments on Astrophysics},
         year = 1978,
        month = jan,
       volume = {7},
       number = {4},
        pages = {103-119},
       adsurl = {https://ui.adsabs.harvard.edu/abs/1978ComAp...7..103P}
}

@book{makridakis1998,
  author    = {Makridakis, Spyros and Wheelwright, Steven C. and Hyndman, Rob J.},
  title     = {Forecasting Methods and Applications},
  year      = {1998},
  publisher = {John Wiley \& Sons},
  address   = {New York},
  edition   = {3rd}
}

@article{Wang_2006, title={Characteristic-Based Clustering for Time Series Data}, volume={13}, ISSN={1573-756X}, url={http://dx.doi.org/10.1007/s10618-005-0039-x}, DOI={10.1007/s10618-005-0039-x}, number={3}, journal={Data Mining and Knowledge Discovery}, publisher={Springer Science and Business Media LLC}, author={Wang, Xiaozhe and Smith, Kate and Hyndman, Rob}, year={2006}, month=may, pages={335–364} }

@article{abdollahi2024periodic,
  title={Periodic gamma-ray modulation of the blazar PG 1553+ 113 confirmed by Fermi-LAT and multiwavelength observations},
  author={Abdollahi, S and Baldini, L and Barbiellini, G and Bellazzini, R and Berenji, B and Bissaldi, E and Blandford, RD and Bonino, R and Bruel, P and Buson, S and others},
  journal={The Astrophysical Journal},
  volume={976},
  number={2},
  pages={203},
  year={2024},
  publisher={IOP Publishing}
}

@ARTICLE{2023ApJL.Agazie,
       author = {{Agazie}, Gabriella and {Anumarlapudi}, Akash and {Archibald}, Anne M. and {Arzoumanian}, Zaven and {Baker}, Paul T. and {B{\'e}csy}, Bence and {Blecha}, Laura and {Brazier}, Adam and {Brook}, Paul R. and {Burke-Spolaor}, Sarah and {Case}, Robin and {Casey-Clyde}, J. Andrew and {Charisi}, Maria and {Chatterjee}, Shami and {Cohen}, Tyler and {Cordes}, James M. and {Cornish}, Neil J. and {Crawford}, Fronefield and {Cromartie}, H. Thankful and {Crowter}, Kathryn and {Decesar}, Megan E. and {Demorest}, Paul B. and {Digman}, Matthew C. and {Dolch}, Timothy and {Drachler}, Brendan and {Ferrara}, Elizabeth C. and {Fiore}, William and {Fonseca}, Emmanuel and {Freedman}, Gabriel E. and {Garver-Daniels}, Nate and {Gentile}, Peter A. and {Glaser}, Joseph and {Good}, Deborah C. and {G{\"u}ltekin}, Kayhan and {Hazboun}, Jeffrey S. and {Hourihane}, Sophie and {Jennings}, Ross J. and {Johnson}, Aaron D. and {Jones}, Megan L. and {Kaiser}, Andrew R. and {Kaplan}, David L. and {Kelley}, Luke Zoltan and {Kerr}, Matthew and {Key}, Joey S. and {Laal}, Nima and {Lam}, Michael T. and {Lamb}, William G. and {Lazio}, T. Joseph W. and {Lewandowska}, Natalia and {Liu}, Tingting and {Lorimer}, Duncan R. and {Luo}, Jing and {Lynch}, Ryan S. and {Ma}, Chung-Pei and {Madison}, Dustin R. and {McEwen}, Alexander and {McKee}, James W. and {McLaughlin}, Maura A. and {McMann}, Natasha and {Meyers}, Bradley W. and {Meyers}, Patrick M. and {Mingarelli}, Chiara M.~F. and {Mitridate}, Andrea and {Ng}, Cherry and {Nice}, David J. and {Ocker}, Stella Koch and {Olum}, Ken D. and {Pennucci}, Timothy T. and {Perera}, Benetge B.~P. and {Petrov}, Polina and {Pol}, Nihan S. and {Radovan}, Henri A. and {Ransom}, Scott M. and {Ray}, Paul S. and {Romano}, Joseph D. and {Sardesai}, Shashwat C. and {Schmiedekamp}, Ann and {Schmiedekamp}, Carl and {Schmitz}, Kai and {Shapiro-Albert}, Brent J. and {Siemens}, Xavier and {Simon}, Joseph and {Siwek}, Magdalena S. and {Stairs}, Ingrid H. and {Stinebring}, Daniel R. and {Stovall}, Kevin and {Susobhanan}, Abhimanyu and {Swiggum}, Joseph K. and {Taylor}, Jacob and {Taylor}, Stephen R. and {Turner}, Jacob E. and {Unal}, Caner and {Vallisneri}, Michele and {van Haasteren}, Rutger and {Vigeland}, Sarah J. and {Wahl}, Haley M. and {Witt}, Caitlin A. and {Young}, Olivia and {Nanograv Collaboration}},
        title = "{The NANOGrav 15 yr Data Set: Bayesian Limits on Gravitational Waves from Individual Supermassive Black Hole Binaries}",
      journal = {\apjl},
         year = 2023,
        month = jul,
       volume = {951},
       number = {2},
          eid = {L50},
        pages = {L50},
          doi = {10.3847/2041-8213/ace18a},
archivePrefix = {arXiv},
       eprint = {2306.16222},
 primaryClass = {astro-ph.HE},
       adsurl = {https://ui.adsabs.harvard.edu/abs/2023ApJ...951L..50A}
}

@ARTICLE{2023.Orazio,
       author = {{D'Orazio}, Daniel J. and {Charisi}, Maria},
        title = "{Observational Signatures of Supermassive Black Hole Binaries}",
      journal = {arXiv e-prints},
         year = 2023,
        month = oct,
          eid = {arXiv:2310.16896},
        pages = {arXiv:2310.16896},
          doi = {10.48550/arXiv.2310.16896},
archivePrefix = {arXiv},
       eprint = {2310.16896},
 primaryClass = {astro-ph.HE},
       adsurl = {https://ui.adsabs.harvard.edu/abs/2023arXiv231016896D}
}

@ARTICLE{2025Kiehlmann,
       author = {{Kiehlmann}, S. and {de la Parra}, P.~V. and {Sullivan}, A.~G. and {Synani}, A. and {Liodakis}, I. and {Mr{\'o}z}, P. and {N{\ae}ss}, S.~K. and {Readhead}, A.~C.~S. and {Begelman}, M.~C. and {Blandford}, R.~D. and {Chatziioannou}, K. and {Ding}, Y. and {Graham}, M.~J. and {Harrison}, F. and {Homan}, D.~C. and {Hovatta}, T. and {Kulkarni}, S.~R. and {Lister}, M.~L. and {Maiolino}, R. and {Max-Moerbeck}, W. and {Molina}, B. and {O'Dea}, C.~P. and {Pavlidou}, V. and {Pearson}, T.~J. and {Aller}, M.~F. and {Lawrence}, C.~R. and {Lazio}, T.~J.~W. and {O'Neill}, S. and {Prince}, T.~A. and {Ravi}, V. and {Reeves}, R.~A. and {Tassis}, K. and {Vallisneri}, M. and {Zensus}, J.~A.},
        title = "{PKS 2131‑021{\textemdash}Discovery of Strong Coherent Sinusoidal Variations from Radio to Optical Frequencies: Compelling Evidence for a Blazar Supermassive Black Hole Binary}",
      journal = {\apj},
         year = 2025,
        month = may,
       volume = {985},
       number = {1},
          eid = {59},
        pages = {59},
          doi = {10.3847/1538-4357/adc567},
archivePrefix = {arXiv},
       eprint = {2407.09647},
 primaryClass = {astro-ph.HE},
       adsurl = {https://ui.adsabs.harvard.edu/abs/2025ApJ...985...59K}
}

@ARTICLE{2025ApJ.Parra,
       author = {{de la Parra}, P.~V. and {Kiehlmann}, S. and {Mr{\'o}z}, P. and {Readhead}, A.~C.~S. and {Synani}, A. and {Begelman}, M.~C. and {Blandford}, R.~D. and {Ding}, Y. and {Harrison}, F. and {Liodakis}, I. and {Max-Moerbeck}, W. and {Pavlidou}, V. and {Reeves}, R. and {Vallisneri}, M. and {Aller}, M.~F. and {Graham}, M.~J. and {Hovatta}, T. and {Lawrence}, C.~R. and {Lazio}, T.~J.~W. and {Mahabal}, A.~A. and {Molina}, B. and {O'Neill}, S. and {Pearson}, T.~J. and {Ravi}, V. and {Tassis}, K. and {Zensus}, J.~A.},
        title = "{PKS J0805-0111: A Second Owens Valley Radio Observatory Blazar Showing Highly Significant Sinusoidal Radio Variability{\textemdash}The Tip of the Iceberg}",
      journal = {\apj},
         year = 2025,
        month = jul,
       volume = {987},
       number = {2},
          eid = {191},
        pages = {191},
          doi = {10.3847/1538-4357/addc60},
archivePrefix = {arXiv},
       eprint = {2408.02645},
 primaryClass = {astro-ph.HE},
       adsurl = {https://ui.adsabs.harvard.edu/abs/2025ApJ...987..191D}
}

@ARTICLE{Hashad2024,
       author = {{Hashad}, M.~A. and {EL-Zant}, Amr A. and {Abdou}, Y. and {Badran}, H.~M.},
        title = "{Quasiperiodic {\ensuremath{\gamma}}-Ray Modulations in the Blazars PKS 2155-83 and PKS 2255-282}",
      journal = {\apj},
         year = 2024,
        month = nov,
       volume = {975},
       number = {2},
          eid = {164},
        pages = {164},
          doi = {10.3847/1538-4357/ad7a6e},
archivePrefix = {arXiv},
       eprint = {2409.10622},
 primaryClass = {astro-ph.HE},
       adsurl = {https://ui.adsabs.harvard.edu/abs/2024ApJ...975..164H}
}

@ARTICLE{2015AckermannPG,
       author = {{Ackermann}, M. and {Ajello}, M. and {Albert}, A. and {Atwood}, W.~B. and {Baldini}, L. and {Ballet}, J. and {Barbiellini}, G. and {Bastieri}, D. and {Becerra Gonzalez}, J. and {Bellazzini}, R. and {Bissaldi}, E. and {Blandford}, R.~D. and {Bloom}, E.~D. and {Bonino}, R. and {Bottacini}, E. and {Bregeon}, J. and {Bruel}, P. and {Buehler}, R. and {Buson}, S. and {Caliandro}, G.~A. and {Cameron}, R.~A. and {Caputo}, R. and {Caragiulo}, M. and {Caraveo}, P.~A. and {Cavazzuti}, E. and {Cecchi}, C. and {Chekhtman}, A. and {Chiang}, J. and {Chiaro}, G. and {Ciprini}, S. and {Cohen-Tanugi}, J. and {Conrad}, J. and {Cutini}, S. and {D'Ammando}, F. and {de Angelis}, A. and {de Palma}, F. and {Desiante}, R. and {Di Venere}, L. and {Dom{\'\i}nguez}, A. and {Drell}, P.~S. and {Favuzzi}, C. and {Fegan}, S.~J. and {Ferrara}, E.~C. and {Focke}, W.~B. and {Fuhrmann}, L. and {Fukazawa}, Y. and {Fusco}, P. and {Gargano}, F. and {Gasparrini}, D. and {Giglietto}, N. and {Giommi}, P. and {Giordano}, F. and {Giroletti}, M. and {Godfrey}, G. and {Green}, D. and {Grenier}, I.~A. and {Grove}, J.~E. and {Guiriec}, S. and {Harding}, A.~K. and {Hays}, E. and {Hewitt}, J.~W. and {Hill}, A.~B. and {Horan}, D. and {Jogler}, T. and {J{\'o}hannesson}, G. and {Johnson}, A.~S. and {Kamae}, T. and {Kuss}, M. and {Larsson}, S. and {Latronico}, L. and {Li}, J. and {Li}, L. and {Longo}, F. and {Loparco}, F. and {Lott}, B. and {Lovellette}, M.~N. and {Lubrano}, P. and {Magill}, J. and {Maldera}, S. and {Manfreda}, A. and {Max-Moerbeck}, W. and {Mayer}, M. and {Mazziotta}, M.~N. and {McEnery}, J.~E. and {Michelson}, P.~F. and {Mizuno}, T. and {Monzani}, M.~E. and {Morselli}, A. and {Moskalenko}, I.~V. and {Murgia}, S. and {Nuss}, E. and {Ohno}, M. and {Ohsugi}, T. and {Ojha}, R. and {Omodei}, N. and {Orlando}, E. and {Ormes}, J.~F. and {Paneque}, D. and {Pearson}, T.~J. and {Perkins}, J.~S. and {Perri}, M. and {Pesce-Rollins}, M. and {Petrosian}, V. and {Piron}, F. and {Pivato}, G. and {Porter}, T.~A. and {Rain{\`o}}, S. and {Rando}, R. and {Razzano}, M. and {Readhead}, A. and {Reimer}, A. and {Reimer}, O. and {Schulz}, A. and {Sgr{\`o}}, C. and {Siskind}, E.~J. and {Spada}, F. and {Spandre}, G. and {Spinelli}, P. and {Suson}, D.~J. and {Takahashi}, H. and {Thayer}, J.~B. and {Thompson}, D.~J. and {Tibaldo}, L. and {Torres}, D.~F. and {Tosti}, G. and {Troja}, E. and {Uchiyama}, Y. and {Vianello}, G. and {Wood}, K.~S. and {Wood}, M. and {Zimmer}, S. and {Berdyugin}, A. and {Corbet}, R.~H.~D. and {Hovatta}, T. and {Lindfors}, E. and {Nilsson}, K. and {Reinthal}, R. and {Sillanp{\"a}{\"a}}, A. and {Stamerra}, A. and {Takalo}, L.~O. and {Valtonen}, M.~J.},
        title = "{Multiwavelength Evidence for Quasi-periodic Modulation in the Gamma-Ray Blazar PG 1553+113}",
      journal = {\apjl},
         year = 2015,
        month = nov,
       volume = {813},
       number = {2},
          eid = {L41},
        pages = {L41},
          doi = {10.1088/2041-8205/813/2/L41},
archivePrefix = {arXiv},
       eprint = {1509.02063},
 primaryClass = {astro-ph.HE},
       adsurl = {https://ui.adsabs.harvard.edu/abs/2015ApJ...813L..41A}
}

@ARTICLE{2021Agarwal,
       author = {{Agarwal}, A. and {Mihov}, B. and {Andruchow}, I. and {Cellone}, S.~A. and {Anupama}, G.~C. and {Agrawal}, V. and {Zola}, S. and {Slavcheva-Mihova}, L. and {{\"O}zd{\"o}nmez}, A. and {Ege}, Erg{\"u}n and {Raj}, A. and {Mammana}, L. and {Zibecchi}, L. and {Fern{\'a}ndez-Laj{\'u}s}, E.},
        title = "{Multi-band behaviour of the TeV blazar PG 1553+113 in optical range on diverse timescales. Flux and spectral variations}",
      journal = {\aap},
         year = 2021,
        month = jan,
       volume = {645},
          eid = {A137},
        pages = {A137},
          doi = {10.1051/0004-6361/202039301},
archivePrefix = {arXiv},
       eprint = {2011.04074},
 primaryClass = {astro-ph.HE},
       adsurl = {https://ui.adsabs.harvard.edu/abs/2021A&A...645A.137A}
}

@ARTICLE{2025penil_SSA,
       author = {{Rico}, A. and {Dom{\'\i}nguez}, A. and {Pe{\~n}il}, P. and {Ajello}, M. and {Buson}, S. and {Adhikari}, S. and {Movahedifar}, M.},
        title = "{Singular spectrum analysis of Fermi-LAT blazar light curves: A systematic search for periodicity and trends in the time domain}",
      journal = {\aap},
         year = 2025,
        month = may,
       volume = {697},
          eid = {A35},
        pages = {A35},
          doi = {10.1051/0004-6361/202452495},
archivePrefix = {arXiv},
       eprint = {2412.05812},
 primaryClass = {astro-ph.HE},
       adsurl = {https://ui.adsabs.harvard.edu/abs/2025A&A...697A..35R}
}

@ARTICLE{2003Volonteri,
       author = {{Volonteri}, Marta and {Haardt}, Francesco and {Madau}, Piero},
        title = "{The Assembly and Merging History of Supermassive Black Holes in Hierarchical Models of Galaxy Formation}",
      journal = {\apj},
         year = 2003,
        month = jan,
       volume = {582},
       number = {2},
        pages = {559-573},
          doi = {10.1086/344675},
archivePrefix = {arXiv},
       eprint = {astro-ph/0207276},
 primaryClass = {astro-ph},
       adsurl = {https://ui.adsabs.harvard.edu/abs/2003ApJ...582..559V}
}

@ARTICLE{2024Maiolino,
       author = {{Maiolino}, Roberto and {Scholtz}, Jan and {Curtis-Lake}, Emma and {Carniani}, Stefano and {Baker}, William and {de Graaff}, Anna and {Tacchella}, Sandro and {{\"U}bler}, Hannah and {D'Eugenio}, Francesco and {Witstok}, Joris and {Curti}, Mirko and {Arribas}, Santiago and {Bunker}, Andrew J. and {Charlot}, St{\'e}phane and {Chevallard}, Jacopo and {Eisenstein}, Daniel J. and {Egami}, Eiichi and {Ji}, Zhiyuan and {Jones}, Gareth C. and {Lyu}, Jianwei and {Rawle}, Tim and {Robertson}, Brant and {Rujopakarn}, Wiphu and {Perna}, Michele and {Sun}, Fengwu and {Venturi}, Giacomo and {Williams}, Christina C. and {Willott}, Chris},
        title = "{JADES: The diverse population of infant black holes at 4 < z < 11: Merging, tiny, poor, but mighty}",
      journal = {\aap},
         year = 2024,
        month = nov,
       volume = {691},
          eid = {A145},
        pages = {A145},
          doi = {10.1051/0004-6361/202347640},
archivePrefix = {arXiv},
       eprint = {2308.01230},
 primaryClass = {astro-ph.GA},
       adsurl = {https://ui.adsabs.harvard.edu/abs/2024A&A...691A.145M}
}

@ARTICLE{optflarepassagedisk2000ApJ,
       author = {{Valtaoja}, E. and {Ter{\"a}sranta}, H. and {Tornikoski}, M. and {Sillanp{\"a}{\"a}}, A. and {Aller}, M.~F. and {Aller}, H.~D. and {Hughes}, P.~A.},
        title = "{Radio Monitoring of OJ 287 and Binary Black Hole Models for Periodic Outbursts}",
      journal = {\apj},
         year = 2000,
        month = mar,
       volume = {531},
       number = {2},
        pages = {744-755},
          doi = {10.1086/308494},
       adsurl = {https://ui.adsabs.harvard.edu/abs/2000ApJ...531..744V}
}

@ARTICLE{DoubleJetted1998MNRAS,
       author = {{Villata}, M. and {Raiteri}, C.~M. and {Sillanpaa}, A. and {Takalo}, L.~O.},
        title = "{A beaming model for the OJ 287 periodic optical outbursts}",
      journal = {\mnras},
         year = 1998,
        month = jan,
       volume = {293},
       number = {1},
        pages = {L13-L16},
          doi = {10.1046/j.1365-8711.1998.01244.x},
       adsurl = {https://ui.adsabs.harvard.edu/abs/1998MNRAS.293L..13V}
}

@ARTICLE{2017Cavaliere,
       author = {{Cavaliere}, A. and {Tavani}, M. and {Vittorini}, V.},
        title = "{Blazar Jets Perturbed by Magneto-gravitational Stresses in Supermassive Binaries}",
      journal = {\apj},
         year = 2017,
        month = feb,
       volume = {836},
       number = {2},
          eid = {220},
        pages = {220},
          doi = {10.3847/1538-4357/836/2/220},
archivePrefix = {arXiv},
       eprint = {1701.05350},
 primaryClass = {astro-ph.HE},
       adsurl = {https://ui.adsabs.harvard.edu/abs/2017ApJ...836..220C}
}

@ARTICLE{2013Atwood,
       author = {{Atwood}, W. and {Albert}, A. and {Baldini}, L. and {Tinivella}, M. and {Bregeon}, J. and {Pesce-Rollins}, M. and {Sgr{\`o}}, C. and {Bruel}, P. and {Charles}, E. and {Drlica-Wagner}, A. and {Franckowiak}, A. and {Jogler}, T. and {Rochester}, L. and {Usher}, T. and {Wood}, M. and {Cohen-Tanugi}, J. and {Zimmer}, S.},
        title = "{Pass 8: Toward the Full Realization of the Fermi-LAT Scientific Potential}",
      journal = {arXiv e-prints},
         year = 2013,
        month = mar,
          eid = {arXiv:1303.3514},
        pages = {arXiv:1303.3514},
          doi = {10.48550/arXiv.1303.3514},
archivePrefix = {arXiv},
       eprint = {1303.3514},
 primaryClass = {astro-ph.IM},
       adsurl = {https://ui.adsabs.harvard.edu/abs/2013arXiv1303.3514A}
}

@ARTICLE{Krause_Obs++Rev2025,
       author = {{Krause}, Martin Gustav Heinrich and {Bourne}, Martin Albert and {Britzen}, Silke and {Foord}, Adi and {Greene}, Jenny and {Habouzit}, Melanie and {Horton}, Maya and {Mayer}, Lucio and {Middleton}, Hannah and {Nealon}, Rebecca and {Sisk-Reyn{\'e}s}, Julia and {Reynolds}, Christopher and {Sijacki}, Debora},
        title = "{Evidence for supermassive black hole binaries}",
      journal = {\pasa},
         year = 2025,
        month = nov,
       volume = {42},
          eid = {e162},
        pages = {e162},
          doi = {10.1017/pasa.2025.10120},
archivePrefix = {arXiv},
       eprint = {2510.07534},
 primaryClass = {astro-ph.HE},
       adsurl = {https://ui.adsabs.harvard.edu/abs/2025PASA...42..162K}
}

@ARTICLE{LISA_2024arXiv240207571C,
       author = {{Colpi}, Monica and {Danzmann}, Karsten and {Hewitson}, Martin and {Holley-Bockelmann}, Kelly and {Jetzer}, Philippe and {Nelemans}, Gijs and {Petiteau}, Antoine and {Shoemaker}, David and {Sopuerta}, Carlos and {Stebbins}, Robin and {Tanvir}, Nial and {Ward}, Henry and {Weber}, William Joseph and {Thorpe}, Ira and {Daurskikh}, Anna and {Deep}, Atul and {Fern{\'a}ndez N{\'u}{\~n}ez}, Ignacio and {Garc{\'\i}a Marirrodriga}, C{\'e}sar and {Gehler}, Martin and {Halain}, Jean-Philippe and {Jennrich}, Oliver and {Lammers}, Uwe and {Larra{\~n}aga}, Jonan and {Lieser}, Maike and {L{\"u}tzgendorf}, Nora and {Martens}, Waldemar and {Mondin}, Linda and {Piris Ni{\~n}o}, Ana and {Amaro-Seoane}, Pau and {Arca Sedda}, Manuel and {Auclair}, Pierre and {Babak}, Stanislav and {Baghi}, Quentin and {Baibhav}, Vishal and {Baker}, Tessa and {Bayle}, Jean-Baptiste and {Berry}, Christopher and {Berti}, Emanuele and {Boileau}, Guillaume and {Bonetti}, Matteo and {Brito}, Richard and {Buscicchio}, Riccardo and {Calcagni}, Gianluca and {Capelo}, Pedro R. and {Caprini}, Chiara and {Caputo}, Andrea and {Castelli}, Eleonora and {Chen}, Hsin-Yu and {Chen}, Xian and {Chua}, Alvin and {Davies}, Gareth and {Derdzinski}, Andrea and {Domcke}, Valerie Fiona and {Doneva}, Daniela and {Dvorkin}, Irna and {Mar{\'\i}a Ezquiaga}, Jose and {Gair}, Jonathan and {Haiman}, Zoltan and {Harry}, Ian and {Hartwig}, Olaf and {Hees}, Aurelien and {Heffernan}, Anna and {Husa}, Sascha and {Izquierdo-Villalba}, David and {Karnesis}, Nikolaos and {Klein}, Antoine and {Korol}, Valeriya and {Korsakova}, Natalia and {Kupfer}, Thomas and {Laghi}, Danny and {Lamberts}, Astrid and {Larson}, Shane and {Le Jeune}, Maude and {Lewicki}, Marek and {Littenberg}, Tyson and {Madge}, Eric and {Mangiagli}, Alberto and {Marsat}, Sylvain and {Vilchez}, Ivan Martin and {Maselli}, Andrea and {Mathews}, Josh and {van de Meent}, Maarten and {Muratore}, Martina and {Nardini}, Germano and {Pani}, Paolo and {Peloso}, Marco and {Pieroni}, Mauro and {Pound}, Adam and {Quelquejay-Leclere}, Hippolyte and {Ricciardone}, Angelo and {Rossi}, Elena Maria and {Sartirana}, Andrea and {Savalle}, Etienne and {Sberna}, Laura and {Sesana}, Alberto and {Shoemaker}, Deirdre and {Slutsky}, Jacob and {Sotiriou}, Thomas and {Speri}, Lorenzo and {Staab}, Martin and {Steer}, Dani{\`e}le and {Tamanini}, Nicola and {Tasinato}, Gianmassimo and {Torrado}, Jesus and {Torres-Orjuela}, Alejandro and {Toubiana}, Alexandre and {Vallisneri}, Michele and {Vecchio}, Alberto and {Volonteri}, Marta and {Yagi}, Kent and {Zwick}, Lorenz},
        title = "{LISA Definition Study Report}",
      journal = {arXiv e-prints},
         year = 2024,
        month = feb,
          eid = {arXiv:2402.07571},
        pages = {arXiv:2402.07571},
          doi = {10.48550/arXiv.2402.07571},
archivePrefix = {arXiv},
       eprint = {2402.07571},
 primaryClass = {astro-ph.CO},
       adsurl = {https://ui.adsriegerabs.harvard.edu/abs/2024arXiv240207571C}
}

@ARTICLE{Supersize_2025ApJS,
       author = {{Pfeifle}, Ryan W. and {Weaver}, Kimberly A. and {Secrest}, Nathan J. and {Rothberg}, Barry and {Patton}, David R.},
        title = "{Super-size Me: The Big Multi-AGN Catalog (The Big MAC) Data Release 1: The Source Catalog}",
      journal = {\apjs},
         year = 2025,
        month = nov,
       volume = {281},
       number = {1},
          eid = {25},
        pages = {25},
          doi = {10.3847/1538-4365/adf845},
archivePrefix = {arXiv},
       eprint = {2411.12799},
 primaryClass = {astro-ph.GA},
       adsurl = {https://ui.adsabs.harvard.edu/abs/2025ApJS..281...25P}
}

@ARTICLE{Habouz_Vo_2025,
       author = {{Puerto-S{\'a}nchez}, Clara and {Habouzit}, M{\'e}lanie and {Volonteri}, Marta and {Ni}, Yueying and {Foord}, Adi and {Angl{\'e}s-Alc{\'a}zar}, Daniel and {Chen}, Nianyi and {Guetzoyan}, Paloma and {Dav{\'e}}, Romeel and {Di Matteo}, Tiziana and {Dubois}, Yohan and {Koss}, Michael and {Rosas-Guevara}, Yetli},
        title = "{Large-scale dual AGN in large-scale cosmological hydrodynamical simulations}",
      journal = {\mnras},
         year = 2025,
        month = jan,
       volume = {536},
       number = {3},
        pages = {3016-3040},
          doi = {10.1093/mnras/stae2763},
archivePrefix = {arXiv},
       eprint = {2411.15297},
 primaryClass = {astro-ph.CO},
       adsurl = {https://ui.adsabs.harvard.edu/abs/2025MNRAS.536.3016P}
}

@ARTICLE{Penil2025P22,
       author = {{Pe{\~n}il}, P. and {Ajello}, M. and {Buson}, S. and {Dom{\'\i}nguez}, A. and {Westernacher-Schneider}, J.~R. and {Rico}, A. and {Adhikari}, S. and {Zrake}, J.},
        title = "{Search for periodic variability in {\ensuremath{\gamma}}-ray blazars Using Fermi-LAT}",
      journal = {\mnras},
         year = 2025,
        month = aug,
       volume = {541},
       number = {4},
        pages = {2955-2977},
          doi = {10.1093/mnras/staf1108},
archivePrefix = {arXiv},
       eprint = {2211.01894},
 primaryClass = {astro-ph.HE},
       adsurl = {https://ui.adsabs.harvard.edu/abs/2025MNRAS.541.2955P}
}

@article{wen2022transformers,
  title={Transformers in time series: A survey},
  author={Wen, Qingsong and Zhou, Tian and Zhang, Chaoli and Chen, Weiqi and Ma, Ziqing and Yan, Junchi and Sun, Liang},
  journal={arXiv preprint arXiv:2202.07125},
  year={2022}
}

@article{shin2025enhancing,
  title={Enhancing Channel-Independent Time Series Forecasting via Cross-Variate Patch Embedding},
  author={Shin, Donghwa and Zhang, Edwin},
  journal={arXiv preprint arXiv:2505.12761},
  year={2025}
}

@article{vaswani2017attention,
  title={Attention is all you need},
  author={Vaswani, Ashish and Shazeer, Noam and Parmar, Niki and Uszkoreit, Jakob and Jones, Llion and Gomez, Aidan N and Kaiser, {\L}ukasz and Polosukhin, Illia},
  journal={Advances in neural information processing systems},
  url={https://proceedings.neurips.cc/paper_files/paper/2017/file/3f5ee243547dee91fbd053c1c4a845aa-Paper.pdf},
  volume={30},
  year={2017}
}

@article{Hammad_2023sbd,
    author = "Hammad, A. and Moretti, S. and Nojiri, M.",
    title = "{Multi-scale cross-attention transformer encoder for event classification}",
    eprint = "2401.00452",
    archivePrefix = "arXiv",
    primaryClass = "hep-ph",
    doi = "10.1007/JHEP03(2024)144",
    journal = "JHEP",
    volume = "03",
    pages = "144",
    year = "2024"
}

@article{Hammad_2024cae,
    author = "Hammad, A. and Nojiri, Mihoko M.",
    title = "{Streamlined jet tagging network assisted by jet prong structure}",
    eprint = "2404.14677",
    archivePrefix = "arXiv",
    primaryClass = "hep-ph",
    doi = "10.1007/JHEP06(2024)176",
    journal = "JHEP",
    volume = "06",
    pages = "176",
    year = "2024"
}

@ARTICLE{Opt_Rad2025ApJ,
       author = {{Kiehlmann}, S. and {de la Parra}, P.~V. and {Sullivan}, A.~G. and {Synani}, A. and {Liodakis}, I. and {Mr{\'o}z}, P. and {N{\ae}ss}, S.~K. and {Readhead}, A.~C.~S. and {Begelman}, M.~C. and {Blandford}, R.~D. and {Chatziioannou}, K. and {Ding}, Y. and {Graham}, M.~J. and {Harrison}, F. and {Homan}, D.~C. and {Hovatta}, T. and {Kulkarni}, S.~R. and {Lister}, M.~L. and {Maiolino}, R. and {Max-Moerbeck}, W. and {Molina}, B. and {O'Dea}, C.~P. and {Pavlidou}, V. and {Pearson}, T.~J. and {Aller}, M.~F. and {Lawrence}, C.~R. and {Lazio}, T.~J.~W. and {O'Neill}, S. and {Prince}, T.~A. and {Ravi}, V. and {Reeves}, R.~A. and {Tassis}, K. and {Vallisneri}, M. and {Zensus}, J.~A.},
        title = "{PKS 2131‑021{\textemdash}Discovery of Strong Coherent Sinusoidal Variations from Radio to Optical Frequencies: Compelling Evidence for a Blazar Supermassive Black Hole Binary}",
      journal = {\apj},
         year = 2025,
        month = may,
       volume = {985},
       number = {1},
          eid = {59},
        pages = {59},
          doi = {10.3847/1538-4357/adc567},
archivePrefix = {arXiv},
       eprint = {2407.09647},
 primaryClass = {astro-ph.HE},
       adsurl = {https://ui.adsabs.harvard.edu/abs/2025ApJ...985...59K}
}

@ARTICLE{Radio2025arXiv251023103M,
       author = {{Molina}, B. and {Mr{\'o}z}, P. and {De la Parra}, P.~V. and {Readhead}, A.~C.~S. and {Surti}, T. and {Aller}, M.~F. and {Scargle}, J.~D. and {Reeves}, R.~A. and {Aller}, H. and {Begelman}, M.~C. and {Blandford}, R.~D. and {Ding}, Y. and {Graham}, M.~J. and {Harrison}, F. and {Hovatta}, T. and {Liodakis}, I. and {Lister}, M.~L. and {Max-Moerbeck}, W. and {Pavlidou}, V. and {Pearson}, T.~J. and {Ravi}, V. and {Sullivan}, A.~G. and {Synani}, A. and {Tassis}, K. and {Tremblay}, S.~E. and {Zensus}, J.~A.},
        title = "{A Search for Supermassive Black Hole Binary Candidates in 46-Year Radio Light Curves of 83 Blazars}",
      journal = {arXiv e-prints},
         year = 2025,
        month = oct,
          eid = {arXiv:2510.23103},
        pages = {arXiv:2510.23103},
          doi = {10.48550/arXiv.2510.23103},
archivePrefix = {arXiv},
       eprint = {2510.23103},
 primaryClass = {astro-ph.HE},
       adsurl = {https://ui.adsabs.harvard.edu/abs/2025arXiv251023103M}
}

@ARTICLE{RedheadHarmonic2026ApJ,
       author = {{Readhead}, A.~C.~S. and {Aller}, M.~F. and {Sullivan}, A.~G. and {Blandford}, R.~D. and {Mr{\'o}z}, P. and {De la Parra}, P.~V. and {Molina}, B. and {Most}, E.~R. and {Lister}, M.~L. and {Synani}, A. and {Aller}, H. and {Begelman}, M.~C. and {Ding}, Y. and {Graham}, M.~J. and {Harrison}, F. and {Hovatta}, T. and {Liodakis}, I. and {Max-Moerbeck}, W. and {Pavlidou}, V. and {Pearson}, T.~J. and {Ravi}, V. and {Reeves}, R.~A. and {Surti}, T. and {Tassis}, K. and {Tremblay}, S.~E. and {Zensus}, J.~A.},
        title = "{Compelling Evidence for a Harmonic in the Light Curve of the Supermassive Black Hole Binary Candidate PKS J1309+1154}",
      journal = {\apjl},
         year = 2026,
        month = jan,
       volume = {996},
       number = {2},
          eid = {L39},
        pages = {L39},
          doi = {10.3847/2041-8213/ae2656},
archivePrefix = {arXiv},
       eprint = {2511.09409},
 primaryClass = {astro-ph.HE},
       adsurl = {https://ui.adsabs.harvard.edu/abs/2026ApJ...996L..39R}
}

@ARTICLE{Sharma_2024ApJ,
       author = {{Sharma}, Ajay and {Banerjee}, Anuvab and {Das}, Avik Kumar and {Mandal}, Avijit and {Bose}, Debanjan},
        title = "{Detection of a Transient Quasiperiodic Oscillation in {\ensuremath{\gamma}}-Rays from Blazar PKS 2255-282}",
      journal = {\apj},
         year = 2024,
        month = nov,
       volume = {975},
       number = {1},
          eid = {56},
        pages = {56},
          doi = {10.3847/1538-4357/ad7391},
archivePrefix = {arXiv},
       eprint = {2408.13052},
 primaryClass = {astro-ph.HE},
       adsurl = {https://ui.adsabs.harvard.edu/abs/2024ApJ...975...56S}
}

@ARTICLE{Tantery_2025,
       author = {{Tantry}, Javaid and {Sharma}, Ajay and {Shah}, Zahir and {Iqbal}, Naseer and {Bose}, Debanjan},
        title = "{Study of multi-wavelength variability, emission mechanism and quasi-periodic oscillation for transition blazar S5 1803+784}",
      journal = {Journal of High Energy Astrophysics},
         year = 2025,
        month = jul,
       volume = {47},
          eid = {100372},
        pages = {100372},
          doi = {10.1016/j.jheap.2025.100372},
archivePrefix = {arXiv},
       eprint = {2503.20379},
 primaryClass = {astro-ph.HE},
       adsurl = {https://ui.adsabs.harvard.edu/abs/2025JHEAp..4700372T}
}

@article{Sarkar_2021, title={Multiwaveband quasi-periodic oscillation in the blazar 3C 454.3}, volume={501}, ISSN={1365-2966}, url={http://dx.doi.org/10.1093/mnras/staa3211}, DOI={10.1093/mnras/staa3211}, number={1}, journal={Monthly Notices of the Royal Astronomical Society}, publisher={Oxford University Press (OUP)}, author={Sarkar, Arkadipta and Gupta, Alok C and Chitnis, Varsha R and Wiita, Paul J}, year={2021}, month=oct, pages={50–61} }

@article{Penil_2025_curved_jet, title={Transient quasiperiodic oscillations of Fermi-LAT blazars under the curved jet model}, volume={700}, ISSN={1432-0746}, url={http://dx.doi.org/10.1051/0004-6361/202555599}, DOI={10.1051/0004-6361/202555599}, journal={Astronomy \& Astrophysics}, publisher={EDP Sciences}, author={Peñil, P. and Otero-Santos, J. and Banerjee, A. and Buson, S. and Rico, A. and Ajello, M. and Adhikari, S.}, year={2025}, month=aug, pages={A208} }

@article{Penil_2025_Transients, title={Transient quasiperiodic oscillations of Fermi-LAT blazars under the curved jet model}, volume={700}, ISSN={1432-0746}, url={http://dx.doi.org/10.1051/0004-6361/202555599}, DOI={10.1051/0004-6361/202555599}, journal={Astronomy \& Astrophysics}, publisher={EDP Sciences}, author={Peñil, P. and Otero-Santos, J. and Banerjee, A. and Buson, S. and Rico, A. and Ajello, M. and Adhikari, S.}, year={2025}, month=aug, pages={A208} }

@ARTICLE{Welsh1999,
       author = {{Welsh}, W.~F.},
        title = "{On the Reliability of Cross-Correlation Function Lag Determinations in Active Galactic Nuclei}",
      journal = {\pasp},
         year = 1999,
        month = nov,
       volume = {111},
       number = {765},
        pages = {1347-1366},
          doi = {10.1086/316457},
archivePrefix = {arXiv},
       eprint = {astro-ph/9911112},
 primaryClass = {astro-ph},
       adsurl = {https://ui.adsabs.harvard.edu/abs/1999PASP..111.1347W}
}

@ARTICLE{2019ChevalierOptical,
       author = {{Chevalier}, J. and {Sanchez}, D.~A. and {Serpico}, P.~D. and {Lenain}, J.-P. and {Maurin}, G.},
        title = "{Variability studies and modelling of the blazar PKS 2155-304 in the light of a decade of multi-wavelength observations}",
      journal = {\mnras},
         year = 2019,
        month = mar,
       volume = {484},
       number = {1},
        pages = {749-759},
          doi = {10.1093/mnras/stz027},
archivePrefix = {arXiv},
       eprint = {1901.01743},
 primaryClass = {astro-ph.HE},
       adsurl = {https://ui.adsabs.harvard.edu/abs/2019MNRAS.484..749C}
}

@ARTICLE{PMW1978,
       author = {{Stellingwerf}, R.~F.},
        title = "{Period determination using phase dispersion minimization.}",
      journal = {\apj},
         year = 1978,
        month = sep,
       volume = {224},
        pages = {953-960},
          doi = {10.1086/156444},
       adsurl = {https://ui.adsabs.harvard.edu/abs/1978ApJ...224..953S}
}

@ARTICLE{CWT1998,
       author = {{Torrence}, Christopher and {Compo}, Gilbert P.},
        title = "{A Practical Guide to Wavelet Analysis.}",
      journal = {Bulletin of the American Meteorological Society},
         year = 1998,
        month = jan,
       volume = {79},
       number = {1},
        pages = {61-78},
          doi = {10.1175/1520-0477(1998)079<0061:APGTWA>2.0.CO;2},
       adsurl = {https://ui.adsabs.harvard.edu/abs/1998BAMS...79...61T}
}

@ARTICLE{2019SKtime,
       author = {{L{\"o}ning}, Markus and {Bagnall}, Anthony and {Ganesh}, Sajaysurya and {Kazakov}, Viktor and {Lines}, Jason and {Kir{\'a}ly}, Franz J.},
        title = "{sktime: A Unified Interface for Machine Learning with Time Series}",
      journal = {arXiv e-prints},
         year = 2019,
        month = sep,
          eid = {arXiv:1909.07872},
        pages = {arXiv:1909.07872},
          doi = {10.48550/arXiv.1909.07872},
archivePrefix = {arXiv},
       eprint = {1909.07872},
 primaryClass = {cs.LG},
       adsurl = {https://ui.adsabs.harvard.edu/abs/2019arXiv190907872L}
}

@inproceedings{Pytorch2017,
  title={Automatic differentiation in PyTorch},
  author={Paszke, Adam and Gross, Sam and Chintala, Soumith and Chanan, Gregory and Yang, Edward and DeVito, Zachary and Lin, Zeming and Desmaison, Alban and Antiga, Luca and Lerer, Adam},
  booktitle={NIPS Workshop},
  url={https://openreview.net/pdf?id=BJJsrmfCZ},
  year={2017}
}

@ARTICLE{2015MNRASDo,
       author = {{Do{\u{g}}an}, Suzan and {Nixon}, Chris and {King}, Andrew and {Price}, Daniel J.},
        title = "{Tearing up a misaligned accretion disc with a binary companion}",
      journal = {\mnras},
         year = 2015,
        month = may,
       volume = {449},
       number = {2},
        pages = {1251-1258},
          doi = {10.1093/mnras/stv347},
archivePrefix = {arXiv},
       eprint = {1502.05410},
 primaryClass = {astro-ph.HE},
       adsurl = {https://ui.adsabs.harvard.edu/abs/2015MNRAS.449.1251D}
}

@ARTICLE{2014ApJFarris,
       author = {{Farris}, Brian D. and {Duffell}, Paul and {MacFadyen}, Andrew I. and {Haiman}, Zoltan},
        title = "{Binary Black Hole Accretion from a Circumbinary Disk: Gas Dynamics inside the Central Cavity}",
      journal = {\apj},
         year = 2014,
        month = mar,
       volume = {783},
       number = {2},
          eid = {134},
        pages = {134},
          doi = {10.1088/0004-637X/783/2/134},
archivePrefix = {arXiv},
       eprint = {1310.0492},
 primaryClass = {astro-ph.HE},
       adsurl = {https://ui.adsabs.harvard.edu/abs/2014ApJ...783..134F}
}

@ARTICLE{SciPy2020,
       author = {{Virtanen}, Pauli and {Gommers}, Ralf and {Oliphant}, Travis E. and {Haberland}, Matt and {Reddy}, Tyler and {Cournapeau}, David and {Burovski}, Evgeni and {Peterson}, Pearu and {Weckesser}, Warren and {Bright}, Jonathan and {van der Walt}, St{\'e}fan J. and {Brett}, Matthew and {Wilson}, Joshua and {Millman}, K. Jarrod and {Mayorov}, Nikolay and {Nelson}, Andrew R.~J. and {Jones}, Eric and {Kern}, Robert and {Larson}, Eric and {Carey}, C.~J. and {Polat}, {\.I}lhan and {Feng}, Yu and {Moore}, Eric W. and {VanderPlas}, Jake and {Laxalde}, Denis and {Perktold}, Josef and {Cimrman}, Robert and {Henriksen}, Ian and {Quintero}, E.~A. and {Harris}, Charles R. and {Archibald}, Anne M. and {Ribeiro}, Ant{\^o}nio H. and {Pedregosa}, Fabian and {van Mulbregt}, Paul and {SciPy 1.  0 Contributors}},
        title = "{SciPy 1.0: fundamental algorithms for scientific computing in Python}",
      journal = {Nature Medicine},
         year = 2020,
        month = feb,
       volume = {17},
        pages = {261-272},
          doi = {10.1038/s41592-019-0686-2},
archivePrefix = {arXiv},
       eprint = {1907.10121},
 primaryClass = {cs.MS},
       adsurl = {https://ui.adsabs.harvard.edu/abs/2020NatMe..17..261V}
}

@ARTICLE{2022Dorigo-PGz,
       author = {{Dorigo Jones}, J. and {Johnson}, S.~D. and {Muzahid}, Sowgat and {Charlton}, J. and {Chen}, H.-W. and {Narayanan}, A. and {Sameer} and {Schaye}, J. and {Wijers}, N.~A.},
        title = "{Improving blazar redshift constraints with the edge of the Ly {\ensuremath{\alpha}} forest: 1ES 1553+113 and implications for observations of the WHIM}",
      journal = {\mnras},
         year = 2022,
        month = jan,
       volume = {509},
       number = {3},
        pages = {4330-4343},
          doi = {10.1093/mnras/stab3331},
archivePrefix = {arXiv},
       eprint = {2111.06927},
 primaryClass = {astro-ph.GA},
       adsurl = {https://ui.adsabs.harvard.edu/abs/2022MNRAS.509.4330D}
}

@ARTICLE{Bell2011,
       author = {{Bell}, M.~E. and {Tzioumis}, T. and {Uttley}, P. and {Fender}, R.~P. and {Ar{\'e}valo}, P. and {Breedt}, E. and {McHardy}, I. and {Calvelo}, D.~E. and {Jamil}, O. and {K{\"o}rding}, E.},
        title = "{X-ray and radio variability in the low-luminosity active galactic nucleus NGC 7213}",
      journal = {\mnras},
         year = 2011,
        month = feb,
       volume = {411},
       number = {1},
        pages = {402-410},
          doi = {10.1111/j.1365-2966.2010.17692.x},
archivePrefix = {arXiv},
       eprint = {1009.2430},
 primaryClass = {astro-ph.GA},
       adsurl = {https://ui.adsabs.harvard.edu/abs/2011MNRAS.411..402B}
}

@ARTICLE{Neill-2022,
       author = {{O'Neill}, S. and {Kiehlmann}, S. and {Readhead}, A.~C.~S. and {Aller}, M.~F. and {Blandford}, R.~D. and {Liodakis}, I. and {Lister}, M.~L. and {Mr{\'o}z}, P. and {O'Dea}, C.~P. and {Pearson}, T.~J. and {Ravi}, V. and {Vallisneri}, M. and {Cleary}, K.~A. and {Graham}, M.~J. and {Grainge}, K.~J.~B. and {Hodges}, M.~W. and {Hovatta}, T. and {L{\"a}hteenm{\"a}ki}, A. and {Lamb}, J.~W. and {Lazio}, T.~J.~W. and {Max-Moerbeck}, W. and {Pavlidou}, V. and {Prince}, T.~A. and {Reeves}, R.~A. and {Tornikoski}, M. and {Vergara de la Parra}, P. and {Zensus}, J.~A.},
        title = "{The Unanticipated Phenomenology of the Blazar PKS 2131-021: A Unique Supermassive Black Hole Binary Candidate}",
      journal = {\apjl},
         year = 2022,
        month = feb,
       volume = {926},
       number = {2},
          eid = {L35},
        pages = {L35},
          doi = {10.3847/2041-8213/ac504b},
archivePrefix = {arXiv},
       eprint = {2111.02436},
 primaryClass = {astro-ph.HE},
       adsurl = {https://ui.adsabs.harvard.edu/abs/2022ApJ...926L..35O}
}

@ARTICLE{Scargle1981,
       author = {{Scargle}, J.~D.},
        title = "{Studies in astronomical time series analysis. I - Modeling random processes in the time domain}",
      journal = {\apjs},
         year = 1981,
        month = jan,
       volume = {45},
        pages = {1-71},
          doi = {10.1086/190706},
       adsurl = {https://ui.adsabs.harvard.edu/abs/1981ApJS...45....1S}
}

@ARTICLE{Feigelson2018,
       author = {{Feigelson}, Eric D. and {Babu}, G. Jogesh and {Caceres}, Gabriel A.},
        title = "{Autoregressive Times Series Methods for Time Domain Astronomy}",
      journal = {Frontiers in Physics},
         year = 2018,
        month = aug,
       volume = {6},
          eid = {80},
        pages = {80},
          doi = {10.3389/fphy.2018.00080},
archivePrefix = {arXiv},
       eprint = {1901.08003},
 primaryClass = {astro-ph.IM},
       adsurl = {https://ui.adsabs.harvard.edu/abs/2018FrP.....6...80F}
}

@inproceedings{seabold2010statsmodels,
  title={statsmodels: Econometric and statistical modeling with python},
  author={Seabold, Skipper and Perktold, Josef},
  booktitle={9th Python in Science Conference},
  year={2010},
}

@ARTICLE{Covino2020,
       author = {{Covino}, Stefano and {Landoni}, Marco and {Sandrinelli}, Angela and {Treves}, Aldo},
        title = "{Looking at Blazar Light-curve Periodicities with Gaussian Processes}",
      journal = {\apj},
         year = 2020,
        month = jun,
       volume = {895},
       number = {2},
          eid = {122},
        pages = {122},
          doi = {10.3847/1538-4357/ab8bd4},
archivePrefix = {arXiv},
       eprint = {2004.10763},
 primaryClass = {astro-ph.HE},
       adsurl = {https://ui.adsabs.harvard.edu/abs/2020ApJ...895..122C}
}

@ARTICLE{Penil2024-period22yr,
       author = {{Adhikari}, S. and {Pe{\~n}il}, P. and {Westernacher-Schneider}, J.~R. and {Dom{\'\i}nguez}, A. and {Ajello}, M. and {Buson}, S. and {Rico}, A. and {Zrake}, J.},
        title = "{Constraining the PG 1553+113 Binary Hypothesis: Interpreting Hints of a New, 22 yr Period}",
      journal = {\apj},
         year = 2024,
        month = apr,
       volume = {965},
       number = {2},
          eid = {124},
        pages = {124},
          doi = {10.3847/1538-4357/ad310a},
archivePrefix = {arXiv},
       eprint = {2307.11696},
 primaryClass = {astro-ph.HE},
       adsurl = {https://ui.adsabs.harvard.edu/abs/2024ApJ...965..124A}
}

@article{THEODOSIOU2011STL,
title = {Forecasting monthly and quarterly time series using STL decomposition},
journal = {International Journal of Forecasting},
volume = {27},
number = {4},
pages = {1178-1195},
year = {2011},
issn = {0169-2070},
doi = {https://doi.org/10.1016/j.ijforecast.2010.11.002},
url = {https://www.sciencedirect.com/science/article/pii/S0169207011000070},
author = {Marina Theodosiou}
}

@article{Hyndman2008,
 title={Automatic Time Series Forecasting: The forecast Package for R},
 volume={27},
 url={https://www.jstatsoft.org/index.php/jss/article/view/v027i03},
 doi={10.18637/jss.v027.i03},
 number={3},
 journal={Journal of Statistical Software},
 author={Hyndman, Rob J. and Khandakar, Yeasmin},
 year={2008},
 pages={1–22}
}

@book{HyndmanForecastingBook2021,
title = "Forecasting: Principles and Practice",
author = "Rob Hyndman and G. Athanasopoulos",
year = "2021",
language = "English",
publisher = "OTexts",
address = "Australia",
edition = "3rd",
}

@misc{Hashad2026Zenodo,
  author       = {Hashad, M. A. and Hammad, A. and EL-Zant, A.},
  title        = {Potential periodic signals in blazars significance forecasting and deep learning},
  year         = {2026},
  version      = {v1.2.0},
  publisher    = {Zenodo},
  doi          = {10.5281/zenodo.20672481},
  url          = {https://doi.org/10.5281/zenodo.20672481}
}
\bibliographystyle{aasjournal}



\end{document}